\documentclass[prb,aps,groupedaddress,twocolumn]{revtex4-1}
\usepackage{graphicx}% Include figure files
\usepackage{dcolumn}% Align table columns on decimal point
\usepackage{bm}% bold math
\usepackage{color}

\usepackage{epstopdf}

\usepackage{amsmath, amssymb, graphics, setspace}

\newcommand{\mathsym}[1]{{}}
\newcommand{\unicode}[1]{{}}

\usepackage{titlesec}
\usepackage{bm}
\usepackage{comment}
\usepackage{verbatim}

\usepackage{graphicx}
\usepackage{tabularx}

\usepackage{color}
\usepackage[colorlinks,bookmarks=false,citecolor=blue,linkcolor=red,urlcolor=blue]{hyperref}

\definecolor{darkred}{rgb}{0.7,0.0,0.0}
\definecolor{darkblue}{rgb}{0,0.02,0.45}

\definecolor{darkgreen}{rgb}{0.02,0.45,0.0}

\definecolor{violet}{rgb}{0.8,0.2,0.6}

\def\be{\begin{equation}}
\def\ee{\end{equation}}
\def\bea{\begin{eqnarray}}
\def\eea{\end{eqnarray}}

\def\vec{\mathbf}

\def\mc{\mathcal}

\usepackage{times}

\begin{document}
%\title{Magnetic excitations in hyperhoneycomb Kitaev system}
\title{ Magnetic structure and excitation spectrum of the hyperhoneycomb Kitaev magnet $\beta$-Li$_2$IrO$_3$}

\author{Samuel Ducatman}
\affiliation{School of Physics and Astronomy, University of Minnesota, Minneapolis,
MN 55116, USA}

\author{Ioannis Rousochatzakis}
\affiliation{School of Physics and Astronomy, University of Minnesota, Minneapolis,
MN 55116, USA}

\author{Natalia  B. Perkins}

\affiliation{School of Physics and Astronomy, University of Minnesota, Minneapolis,
MN 55116, USA}

\begin{abstract}
We present a theoretical study of the static and dynamical properties of the three-dimensional, hyperhoneycomb Kitaev magnet $\beta$-Li$_2$IrO$_3$. We argue that the observed incommensurate order can be understood in terms of a long-wavelength twisting of a nearby commensurate period-3 state, with the same key qualitatively features. The period-3 state shows very different structure when either the Kitaev interaction $K$ or the off-diagonal exchange anisotropy $\Gamma$ is dominant. A comparison of the associated static spin structure factors with reported scattering expoeriments in zero and finite fields gives strong evidence that $\beta$-Li$_2$IrO$_3$ lies in the regime of dominant Kitaev coupling, and that  the Heisenberg exchange $J$ is much weaker than both $K$ and $\Gamma$. Our predictions for the magnon excitation spectra, the dynamical spin structure factors and their polarization dependence provide additional distinctive fingerprints that can be checked experimentally.
\end{abstract}
\maketitle

\section{Introduction}
\vspace*{-0.3cm}
Transition-metal-based insulators  with partially filled $4d$ and $5d$ shells on tri-coordinated lattices have recently attracted a lot of interest as a novel platform for quantum spin liquid physics.~\cite{Jackeli2009,Jackeli2010,BookCao,Balents2014,Rau2016,Trebst2017,Hermanns2017,Winter2017} Most of the materials studied so far are based on Ir$^{4+}$ or Ru$^{3+}$ ions, which are characterized by effective, $J_{\rm eff}\!=\!1/2$ pseudospin degrees of freedom. Due to the strong spin orbit coupling (SOC) and the edge-sharing IrO$_6$ or RuCl$_6$ octahedra structure, the dominant exchange interactions between the pseudospins are Ising-like, $S_i^\alpha S_j^\alpha$, with the quantization axis $\alpha$ depending on the spatial orientation of the bond $(ij)$.~\cite{Jackeli2009,Jackeli2010} When acting alone, this so-called Kitaev anisotropy gives rise to exactly solvable, quantum spin liquid phases.~\cite{Kitaev2006,Mandal2009,Kimchi2014,Hermanns2016}

As it turns out, however, the Kitaev spin liquids are very fragile against various perturbations that are present in real materials, and indeed all Kitaev materials known so far eventually order magnetically at low enough temperatures.~\cite{Singh2010, Singh2012,Liu2011,Johnson2015,Williams2016,Modic2014,Biffin2014a,Biffin2014b,Takayama2015} Thus, despite  recent developments showing that external perturbations, such as pressure~\cite{Takayama2015,Breznay2017,Haskel2017,Tsirlin2018} or magnetic field,~\cite{Baek2017,Yadav2016,Modic2017,Sears2017,Zheng2017} may lead to spin liquid behavior, it is still crucial to understand the role of the most relevant perturbations in the existing materials, to map out the corresponding instabilities, and identify their distinctive experimental fingerprints.\cite{Winter2017}

In this context, we study the static and dynamic properties of the three-dimensional (3D) hyperhoneycomb iridate $\beta$-Li$_2$IrO$_3$.~\cite{Takayama2015,Biffin2014a,Ruiz2017} This magnet shows, below $T_N\!=\!37$ K, a non-coplanar incommensurate modulation  with counter-rotating moments. Interestingly, the main features of this peculiar phase manifest in two more tri-coordinated iridates, the 3D stripy-honeycomb $\gamma$-Li$_2$IrO$_3$~\cite{Biffin2014b,Modic2014} and the layered honeycomb $\alpha$-Li$_2$IrO$_3$,~\cite{Williams2016}  suggesting that the minimal microscopic description is similar in these compounds.

Indeed, Lee {\it et al}~\cite{Lee2015,Lee2016} have proposed that the experimentally observed order of $\beta$-Li$_2$IrO$_3$ (and $\gamma$-Li$_2$IrO$_3$) can be explained within a minimal model with three types of nearest-neighbor (NN) interactions: the Kitaev coupling $K$, the isotropic Heisenberg exchange $J$, and the symmetric portion of the off-diagonal exchange anisotropy, the so-called $\Gamma$ interaction.~\cite{Katukuri2014,Rau2014,Lee2015,Lee2016,KimKimKee2016,IoannisGamma} 
In this minimal, $J$-$K$-$\Gamma$ model, an incommensurate spiral order arises in a large region of the parameter space where $K\!<\!0$, $J\!>\!0$ and $\Gamma\!<\!0$, and has the main qualitative features observed experimentally.~\cite{Biffin2014a} Namely, it describes a counter-rotating modulation that belongs to the observed irreducible representation, the propagation vector ${\bf Q}$ is along the orthorhombic ${\bf a}$-axis, and the ratio $h\!=\!\frac{Q}{(2\pi/a)}$, where $a$ is the lattice constant along ${\bf a}$, varies smoothly around the observed value $h\!=\!0.57$. 
A complementary picture for the counter-rotating moments arises in the context of the so-called $J$-$K$-$I_c$ model,~\cite{Kimchi2015} and some approximate, one-dimensional (1D) single-chain models.~\cite{Kimchi2015,Kimchi2016} Both pictures agree in that the realization of counter-rotating moments in $\beta$-Li$_2$IrO$_3$ requires a ferromagnetic (FM) NN Kitaev interaction.

The motivation of the present study is to better understand the nature of the incommensurate phase of $\beta$-Li$_2$IrO$_3$ and to compute its spin-wave excitation spectrum using the minimal $J$-$K$-$\Gamma$ model. The main challenge in computing the spectrum is that in strongly-anisotropic magnets, such as $\beta$-Li$_2$IrO$_3$, a generic incommensurate configuration cannot be described by a single-${\bf Q}$ modulation, but instead  by a linear combination of a large number of harmonic wave-vectors ${\bf Q}$.~\cite{Z2Ioannis2016,Becker2015,Sizyuk2014,Lee2015,Janssen2016,Chern2016} In contrast to commensurate modulations, these states are {\it inhomogeneous} in the sense that the magnitudes of the local fields exerted at the magnetic sites  from their neighboring spins have  a non-trivial distribution. 
The simplest way to see this is via the so-called Luttinger-Tisza approach~\cite{LT,Bertaut,Litvin,Kaplan} which, by construction, targets the minimum energy configurations that are {\it homogeneous}, with the magnitude of the local field being the same everywhere. 
And it so happens~\cite{Lee2015,Lee2016} that, in the incommensurate region of interest, the minimum energy configurations obtained from the Luttinger-Tisza approach do not satisfy the spin-length constraint for all sites, and therefore the true minima correspond to inhomogeneous modulations that break translational symmetry  in a non-trivial way. And, unless we are sitting at special parameter points of high (continuous) symmetry,~\cite{Kimchi2016} the semiclassical expansion around such states contains umklapp magnon scattering processes, which lead to an intractable, spin-wave Hamiltonian matrix of infinite size.

To circumvent this obstacle we exploit the idea that such inhomogeneous states typically represent a long-wavelength twisting of a nearby commensurate state. We believe that irrespectively of the way this twisting is taking place (e.g., via soliton-like `discommensurations' of various types~\cite{DesGennes1975,McMillan1976,Bak1982,Schaub1985,Abrikosov1957,Wright89,Bogdanov1989,Roessler2006,Z2Ioannis2016}), it is reasonable to assume that the magnetic structure and correlations at short distances follow to some extent the ones of the nearby commensurate state. 
So a first step in computing the excitation spectrum of $\beta$-Li$_2$IrO$_3$ is to search for the simplest nearby commensurate state with counter-rotating moments, the same irreducible representation and similar periodicity with the one observed experimentally. Ideally, the excitation spectrum of the commensurate state should follow closely the spectrum of the actual structure above a low-energy cutoff, which is set by the  perturbations that drive the system from the commensurate to the observed incommensurate state.~\footnote{A characteristic example where this has been demonstrated is the excitation spectrum of the skyrmionic chiral Mott insulator Cu$_2$SeO$_3$.~\cite{Judit,Mike,JuditINS2,JuditINS3} The spectrum of this system is determined by the isotropic Heisenberg interactions over a wide energy bandwidth of about 600 K,~\cite{Judit,Mike} down to a low energy cut-off of the order of 1-2 meV. Below this cut-off, the spectrum is affected by the much weaker Dzyaloshinskii-Moriya interactions,~\cite{JuditINS2,JuditINS3} and the precise changes reflect the long-wavelength twisting of the ferrimagnetic-like order parameter into mesoscopic skyrmions~\cite{Oleg}}

Our Landau-Lifshitz-Gilbert simulations~\cite{Serpico2001,Mochizuki2010,Choi2013,ChoiPRB2013} and analytical considerations show that the experimentally relevant parameter region hosts two such  commensurate states, with $h\!=\!2/3$, one in the region of dominant $K$ and the other in the region of dominant $\Gamma$. These states, which are called `$K$-state' and `$\Gamma$-state' in the following, show qualitatively different magnetic structures. Most notably, the $K$-state has six spin sublattices and contains FM spin dimers, while the $\Gamma$-state has ten sublattices and contains antiferromagnetic (AF) dimers.

Importantly, the static structure factors of both $K$- and $\Gamma$-states comprise, in addition to the dominant Fourier component at ${\bf Q}\!=\!(2/3,0,0)$, a weak uniform canting component with ${\bf Q}\!=\!0$. The latter reflects the deviation of the each state from an ideal 120$^\circ$-pattern realized at $J\!\to\!0^+$. In particular, the ${\bf Q}\!=\!0$ component of the $K$-state is in {\it full agreement} with the ${\bf Q}\!=\!0$ Bragg peaks observed in recent experiments in a field.~\cite{Ruiz2017} 
In conjunction with previous measurements at zero field,~\cite{Biffin2014a} the results signify that $\beta$-Li$_2$IrO$_3$ lies in the regime of dominant Kitaev coupling, and that $J$ is much weaker than both $K$ and $\Gamma$, consistent with {\it ab initio} calculations.~\cite{KimKimKee2016,Katukuri2016}

Furthermore,  the two commensurate structures can be understood in terms of a simple, single-chain Hamiltonian $\mc{H}_c$, which differs from other single-chain models proposed previously.~\cite{Kimchi2015,Kimchi2016} Specifically,  both $K$- and $\Gamma$-states can be shown to arise by simply `tiling' the minima of $\mc{H}_c$ to the whole 3D lattice. This important property is actually also shared by the so-called $120^\circ$ state of the layered honeycomb model.~\cite{Rau2014}

It is also noteworthy that the  boundary line between the $K$- and $\Gamma$-state begins at a hidden, isotropic SO(3) point, $K\!=\!\Gamma$ and $J\!=\!0$, which is related to a 24-sublattice duality transformation, similar to the ones found previously in many other anisotropic models.~\cite{Khaliullin2005,Jackeli2010,Z2Ioannis2016,ioannisK1K2,Chaloupka2015,Kimchi2014a,Kimchi2015} The proximity to this point  marks a non-trivial evolution of the excitation spectra and dynamical structural factors, especially as we cross the boundary from one commensurate phase to the other.

The remaining part of the paper is organized as follows. We begin with the general aspects of the lattice structure (Sec.~\ref{sec:structure}), the minimal $J$-$K$-$\Gamma$ model (Sec.~\ref{sec:model}) and the hidden SO(3) point (Sec.~\ref{sec:SO3}). We then proceed in Sec.~\ref{sec:TwoMainRegions} to analyze in detail the two commensurate phases (Secs.~\ref{sec:Kdom} and \ref{sec:Gdom}),  their characteristic, nearly-120$^\circ$ pattern (Sec.~\ref{sec:Angles}), the close relation to the special line $J=0$ of the phase diagram (Sec.~\ref{sec:SpecialLine}) and the insights from the analysis of the single-chain Hamiltonian $\mc{H}_c$ (Sec.~\ref{sec:Hc}). The analysis of the associated static spin structure factors are  presented in Sec.~\ref{sec:SofQTheory}. We then present our results for the quadratic spin-wave spectrum, the evolution of the spin-gap at the center of the Brillouin zone (BZ), and the dynamic spin structure factor $\mc{S}({\bf Q},\omega)$ (Sec.~\ref{sec:LSW}). We conclude with a general discussion of our results in Sec.~\ref{sec:concl}. Technical details and other auxiliary information are provided in App.~\ref{App:A}-\ref{app:D}.

\begin{figure}[!t]
\includegraphics[width=0.99\columnwidth]{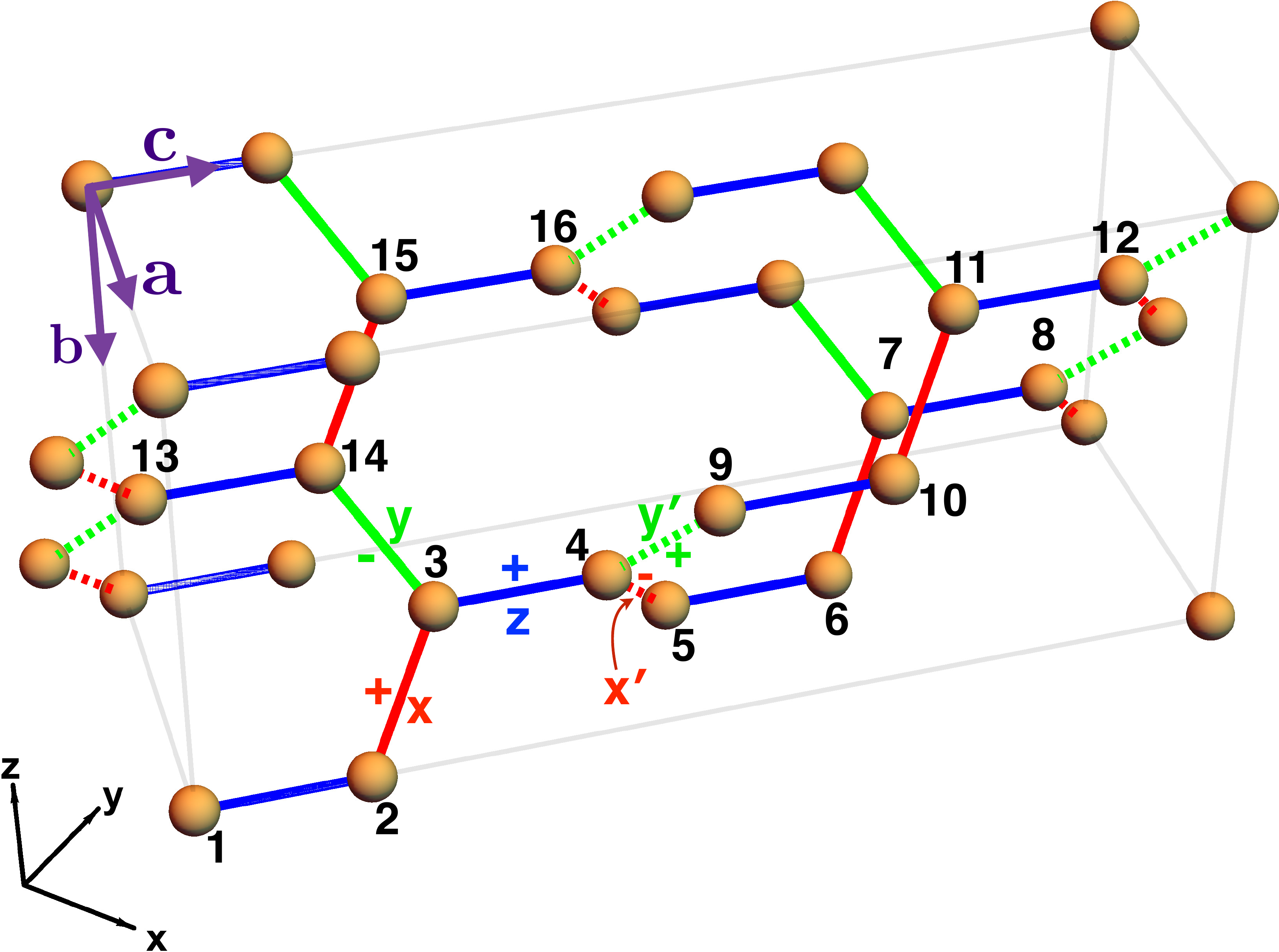}
\caption{Lattice structure and orthorhombic unit cell of $\beta$-Li$_2$IrO$_3$. The bonds are color coded according to the type $t$ of the Kitaev interaction: red, green and blue for $t\!=\!(x,x^\prime)$, $(y,y^\prime)$ and $z$, respectively. The $\pm$ signs denote the sign of $\sigma_t\!=\!\pm1$ in Eq.~(\ref{Ham1}).}\label{fig:betalattice}
\end{figure}

\vspace*{-0.3cm}
\section{Structure and magnetic interactions}\label{sec:StructureAndModel}
\vspace*{-0.3cm}

\subsection{Main aspects of the hyperhoneycomb lattice}\label{sec:structure}
\vspace*{-0.3cm}
The Ir$^{4+}$ ions of $\beta$-Li$_2$IrO$_3$ sit at the vertices of a 3D hyperhoneycomb lattice (see Fig.~\ref{fig:betalattice}), which has a primitive unit cell of four Ir ions. The more convenient, orthorhombic unit cell contains four primitive cells and thus 16 Ir ions. The positions of 4 sites of the  primitive unit cell of Fig.~\ref{fig:betalattice} are
\begin{eqnarray}\label{primitive}
\begin{array}{l}
\vec{r}_1 = (0,0,0), \,
\vec{r}_2 = (0,0,\frac{1}{6}), \\
\vec{r}_3 = (-\frac{1}{4},-\frac{1}{4},\frac{1}{4}),\,
\vec{r}_4 =  (-\frac{1}{4},-\frac{1}{4},\frac{5}{12}),
\end{array}
\end{eqnarray}
where all distances are measured in terms of  fractions of the orthorhombic lattice vectors $\vec{a}$, $\vec{b}$, and $\vec{c}$. The orthorhombic structural unit cell contains 4 primitive cells which can be obtained from the primitive unit cell using translations with  lattice vectors $\boldsymbol{\rho}_p$  given by
\begin{eqnarray}\label{rho}
\begin{array}{l}
\boldsymbol{\rho}_0 = (0,0,0) ,\,
\boldsymbol{\rho}_1 = (\frac{-1}{2},0,\frac{1}{2}),\, 
\\
\boldsymbol{\rho}_2 =  (0,-\frac{1}{2},\frac{1}{2}),\,
\boldsymbol{\rho}_3 = (-\frac{1}{2},-\frac{1}{2},0 ).
\end{array}
\end{eqnarray}
Thus, the  positions of Ir sites shown in Fig. \ref{fig:betalattice}
 labeled  $\vec{r}_5$-$\vec{r}_8$, $\vec{r}_9$-$\vec{r}_{12}$  and $\vec{r}_{13}$-$\vec{r}_{16}$ can be obtained by adding, correspondingly,  $\boldsymbol{\rho}_1$, $\boldsymbol{\rho}_2$, and $\boldsymbol{\rho}_3$ to sites $\vec{r}_1$ through $\vec{r}_4$.  

The Ir$^{4+}$ ions form zigzag chains stacked along the  $\mathbf{c}$-axis and directed alternatively along $({\bf a}$+${\bf b})$ and $({\bf a}$-${\bf b})$. These chains are shown respectively by solid and dashed lines in Fig.~\ref{fig:betalattice}. 
There are five types  of NN bonds, labeled by $t\!=\!x$, $y$, $x^\prime$, $y^\prime$ and $z$  in Fig.~\ref{fig:betalattice}. The zigzag chains running along $({\bf a}$+${\bf b})$ consist of alternating $x$ and $y$ bonds, while those running along $({\bf a}$-${\bf b})$ consist of alternating $x^\prime$ and $y^\prime$ bonds. Adjacent zigzag chains are connected by $z$-bonds, which are all directed along the ${\bf c}$-axis. 

The crystal structure is invariant under the $\pi$-rotations $C_{2{\bf c}}$ around the ${\bf c}$-axes that pass through the $z$-bonds. These rotations map $x$-bonds to $y$-bonds and $x^\prime$-bonds to $y^\prime$-bonds. In addition, there is a $\pi$-rotation symmetry $C_{2{\bf a}}$ around the ${\bf a}$-axis that passes through the middle of the $z$-bonds. This operation maps $x$-bonds to $y^\prime$-bonds and $y$-bonds to $x^\prime$-bonds. The Cartesian axes $\hat{{\bf x}}$, $\hat{{\bf y}}$ and $\hat{{\bf z}}$ that enter the spin Hamiltonian below, are defined by (see Fig.~\ref{fig:betalattice}):
\be
\hat{{\bf x}}\!=\!(\hat{{\bf a}}+\hat{{\bf c}})/\sqrt{2}, ~~
\hat{{\bf y}}\!=\!(\hat{{\bf c}}-\hat{{\bf a}})/\sqrt{2},~~ 
\hat{{\bf z}}\!=\!-\hat{{\bf b}} \,.
\ee  
In the following we shall use square and round brackets to represent vectors in the Cartesian and orthorhombic global frames, respectively.

\vspace*{-0.3cm}
\subsection{The minimal pseudospin-1/2 $J$-$K$-$\Gamma$ Hamiltonian}\label{sec:model}
\vspace*{-0.3cm}
The Ir$^{4+}$ ions of $\beta$-Li$_2$IrO$_3$ are described by a spin-orbit entangled doublet and have effective moments of about 1.7$\mu_B$,\cite{Biffin2014a,Takayama2015,Ruiz2017} which  is very close to the expected value for ideal pseudospins $J_{\rm eff}\!=\!1/2$. Hereafter, we  shall  label these pseudospins  by $S$. 

The edge-sharing IrO$_6$ octrahedra and tri-coordinated lattice structure give rise to dominant Kitaev interactions, which for two spins, ${\bf S}_i$ and ${\bf S}_j$, occupying a NN bond of type $t$, take the Ising-like form $K S_i^{\alpha_t}S_j^{\alpha_t}$, where $K$ is the coupling constant, and the Cartesian component $\alpha_{x}\!=\!\alpha_{x^\prime}\!=\!x$, $\alpha_{y}\!=\!\alpha_{y^\prime}\!=\!y$, and $\alpha_{z}\!=\!z$. Besides the dominant Kitaev interactions, there are other appreciable couplings that are  allowed by symmetry and cannot be ignored. The most important ones among the NN interactions are the Heisenberg exchange $J$ and the symmetric, off-diagonal portion of the exchange anisotropy, the so-called $\Gamma$-coupling.~\cite{Katukuri2014,Rau2014,Lee2015,Lee2016,KimKimKee2016,IoannisGamma} Altogether the minimal Hamiltonian with only NN couplings can be written as a sum over bonds of different types $t$:\cite{Lee2015,Lee2016}
\bea\label{Ham1}
\begin{array}{c}
\mc{H} =\sum_t \sum_{\langle ij\rangle \in t}  \mc{H}^{(t)}_{ij}, ~~t\in\{x,y,z,x^\prime,y^\prime\}, \\[1em]
\mc{H}^{(t)}_{ij}\!=\!
J\, \vec{S}_i\!\cdot\!\vec{S}_j 
\!+\! K\,S_j^{\alpha_t} S_j^{\alpha_t}
\!+\! \sigma_t \Gamma (S_i^{\beta_t} S_j^{\gamma_t}\!+\! S_i^{\gamma_t} S_j^{\beta_t})\,,
\end{array}
\eea
where 
$(\beta_t,\gamma_t,\sigma_t)=(y,z,1)$ for $t=x$, 
$(y,z,-1)$ for $t=x^\prime$, 
$(z,x,-1)$ for $t=y$, 
$(z,x,1)$ for $t=y^\prime$, 
and $(x,y,1)$ for $t=z$. 
Note in particular the alternation in the sign of the prefactor $\sigma_t$ along a given $xy$- or $x^\prime y^\prime$-zigzag chain,~\cite{Lee2015} see Fig.~\ref{fig:betalattice}. This alternation is required by the symmetries $C_{2{\bf c}}$ and $C_{2{\bf a}}$ mentioned above,  which, in spin space, map $[S^x,S^y,S^z]$ to $[S^y,S^x,-S^z]$ and $[-S^y,-S^x,-S^z]$, respectively. 

Following Ref.~[\onlinecite{Rau2014}], we parametrize the full parameter space of the model (\ref{Ham1}) in terms of two parameters $\phi$ and $r$: 
\bea
J=\sin r\cos\phi,~~
K=\sin r\sin\phi,~~
\Gamma= \text{sgn}(\Gamma) \cos r,
\eea
where  $\phi\!\in\![0,2\pi)$ and  $r\!\in\![0,\frac{\pi}{2}]$. For a given sign of $\Gamma$, this range of parameters can be visualized as a disc of radius $\pi/2$, with $\phi$ and $r$ denoting the azimuthal angle and the distance from the center of the disc, respectively.
The center of the disc ($r\!=\!0$) corresponds to the pure $\Gamma$ model ($J\!=\!K\!=\!0$) studied in Ref.~[\onlinecite{IoannisGamma}], while the circle $r\!=\!\pi/2$ corresponds to the $J$-$K$ model ($\Gamma\!=\!0$) studied in Ref.~[\onlinecite{Lee2014}].
The full $J$-$K$-$\Gamma$-model was studied in detail in Refs.~[\onlinecite{Kimchi2014,Lee2015,Lee2016}].
In the present study, we shall focus entirely on the parameter region that is believed~\cite{Lee2015,Lee2016} to host the counter-rotating, non-coplanar phase found experimentally.~\cite{Biffin2014a} This is the shaded region shown in Fig.~\ref{fig:LTphasediagram}, which occupies a significant part of the fourth quadrant of the disc, where $K\!<\!0$ and $J\!>\!0$, and in addition $\Gamma\!<\!0$.~\cite{Lee2015,KimKimKee2016}

\vspace*{-0.3cm}
\subsection{Hidden isotropic SO(3) point}\label{sec:SO3}
\begin{figure}[!t]
\includegraphics[width=0.85\columnwidth]{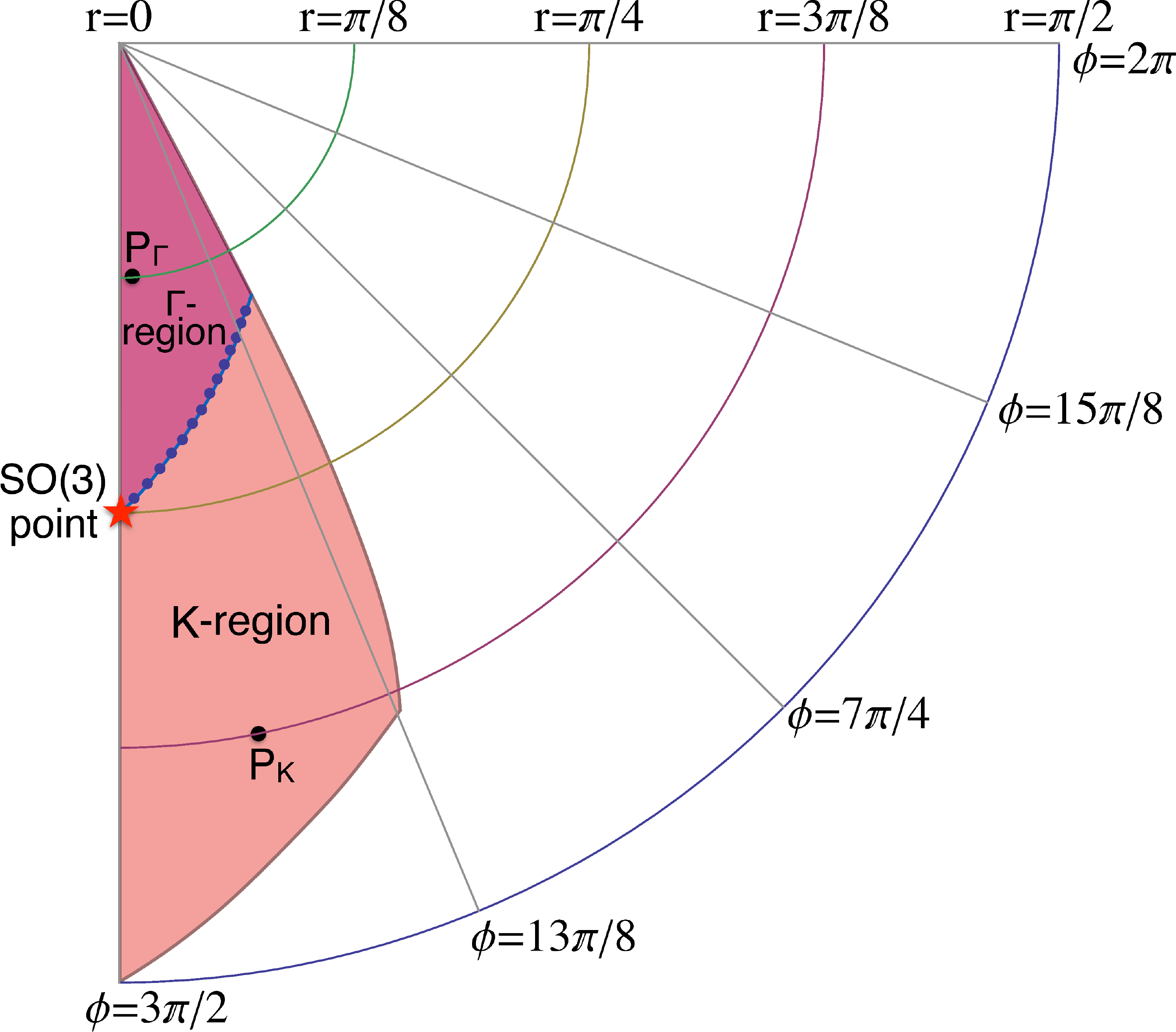}
\caption{The shaded region shows the portion of the parameter space $(r,\phi)$ that is believed~\cite{Lee2015,Lee2016} to be relevant for $\beta$-Li$_2$IrO$_3$. The two main subregions discussed in the text, the `$K$-region' and the `$\Gamma$-region', are separated  by the boundary which begins at the hidden, isotropic SO(3) point (red star). On this boundary, the classical energies of the $K$-state Eq.~(\ref{eq:EnK}) and of the  $\Gamma$-state Eq.~(\ref{eq:EnG}) are degenerate. The two representative points shown, $P_K$ and $P_\Gamma$, correspond to $(r,\phi)=(\frac{3\pi}{8},\frac{25\pi}{16})$ and $(\frac{\pi}{8},\frac{97\pi}{64})$, respectively. For the phases outside the shaded region (including the white region shown), see detailed analysis in Refs.~[\onlinecite{Lee2015}] and [\onlinecite{Lee2016}].}\label{fig:LTphasediagram}
\end{figure}

\vspace*{-0.3cm}
The boundary of the shaded region of Fig.~\ref{fig:LTphasediagram} includes a hidden, isotropic SO(3) point at $(r,\phi)\!=\!(\frac{\pi}{4},\frac{3\pi}{2})$, where $K\!=\!\Gamma$ and $J\!=\!0$. To show this, we follow Chaloupka and Khaliullin,~\cite{Chaloupka2015} and generate local transformations that map this special point to a dual point where $K'\!=\!\Gamma'\!=\!0$ and $J'\!=\!-K$. For an isolated $xy$-chain, this can be achieved by the six-sublattice decomposition represented schematically as
\be\label{eq:XYchain}
\includegraphics[width=2.65in]{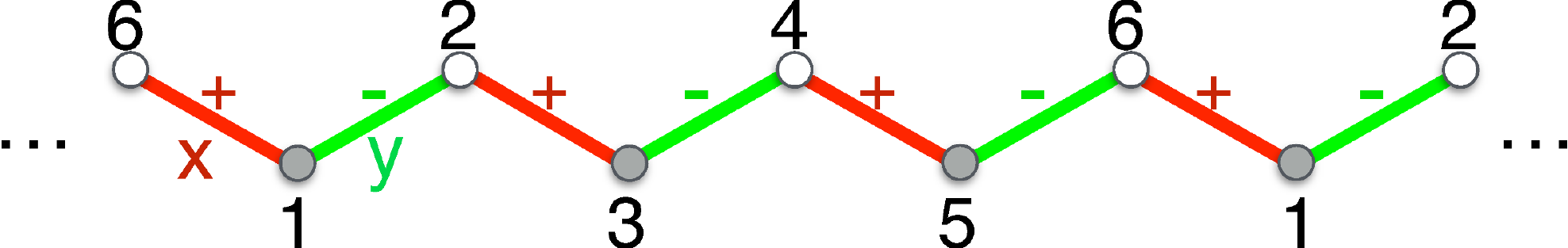}
\ee 
and the following local transformations:
\be\label{eq:6sub}
\begin{array}{c |  rrr}
\text{site index}~ j & S_j^x & S_j^y & S_j^z \\
\hline
1 & S_1^x{^\prime}& S_1^y{^\prime}&S_1^z{^\prime} \\
2 &S_2^z{^\prime}&-S_2^y{^\prime}& S_2^x{^\prime}\\
3 &-S_3^z{^\prime}&-S_3^x{^\prime}&S_3^y{^\prime}\\
4 &S_4^y{^\prime}&S_4^x{^\prime}&-S_4^z{^\prime} \\
5 &-S_5^y{^\prime}&S_5^z{^\prime}&-S_5^x{^\prime} \\
6 &-S_6^x{^\prime}&-S_6^z{^\prime}&-S_6^y{^\prime}
\end{array}
\ee
Let us demonstrate the duality for one bond only, e.g., the $y$-bond $(1,2)$ of (\ref{eq:XYchain}). Using (\ref{eq:6sub}) with $J\!=\!0$ and $K\!=\!\Gamma$ we get:
\be
K (S_1^y S_2^y - S_1^x S_2^z-S_1^z S_2^x) \to
-K{\bf S}_1^\prime \cdot {\bf S}_2^\prime \,,
\ee
which is a Heisenberg coupling with $J'\!=\!-K$, which is antiferromagnetic for negative $K$. (A similar SO(3) point with ferromagnetic $J'$ occurs at the point $K\!=\!\Gamma\!>\!0$, $J\!=\!0$). 
We can proceed in a similar way to generate the corresponding transformation rules along the neighboring chains. It then follows that for the entire lattice, the transformation has 24 sublattices in total and not just 6. But each separate chain hosts only 6 sublattices.

\vspace*{-0.3cm}
\section{The two main subregions of interest and the associated commensurate local minima}\label{sec:TwoMainRegions}
\vspace*{-0.3cm}
In this section and the next we treat the problem classically and, without loss of generality, set the spin length to $S\!=\!1$. 

In the shaded region of Fig.~\ref{fig:LTphasediagram}, the minimum of the classical energy computed with the Luttinger-Tisza (LT) approach~\cite{LT,Bertaut,Litvin,Kaplan} takes place at a wavevector ${\bf Q}\!=\!2\pi h \hat{{\bf a}}$, where $h$ is equal to $\frac{2}{3}$ at $\phi\!=\!\frac{3\pi}{2}$ and shows a weak decrease with increasing $\phi$, but does not depend strongly on $r$. The experimental value~\cite{Biffin2014a} $h\!=\!0.57$ is included in this region. Some details of the analysis are presented in App.~\ref{App:A},  see also Ref.~[\onlinecite{Lee2015,Lee2016}]. 
As mentioned above, the incommensurate solutions delivered by the Luttinger-Tisza method do not satisfy the spin-length constraint at each site, meaning that the actual ground states are inhomogeneous, featuring a distribution of local mean fields with more than one value (unlike the states delivered by the Luttinger-Tisza method).

\begin{figure*}[!t]
\includegraphics[width=0.45\textwidth]{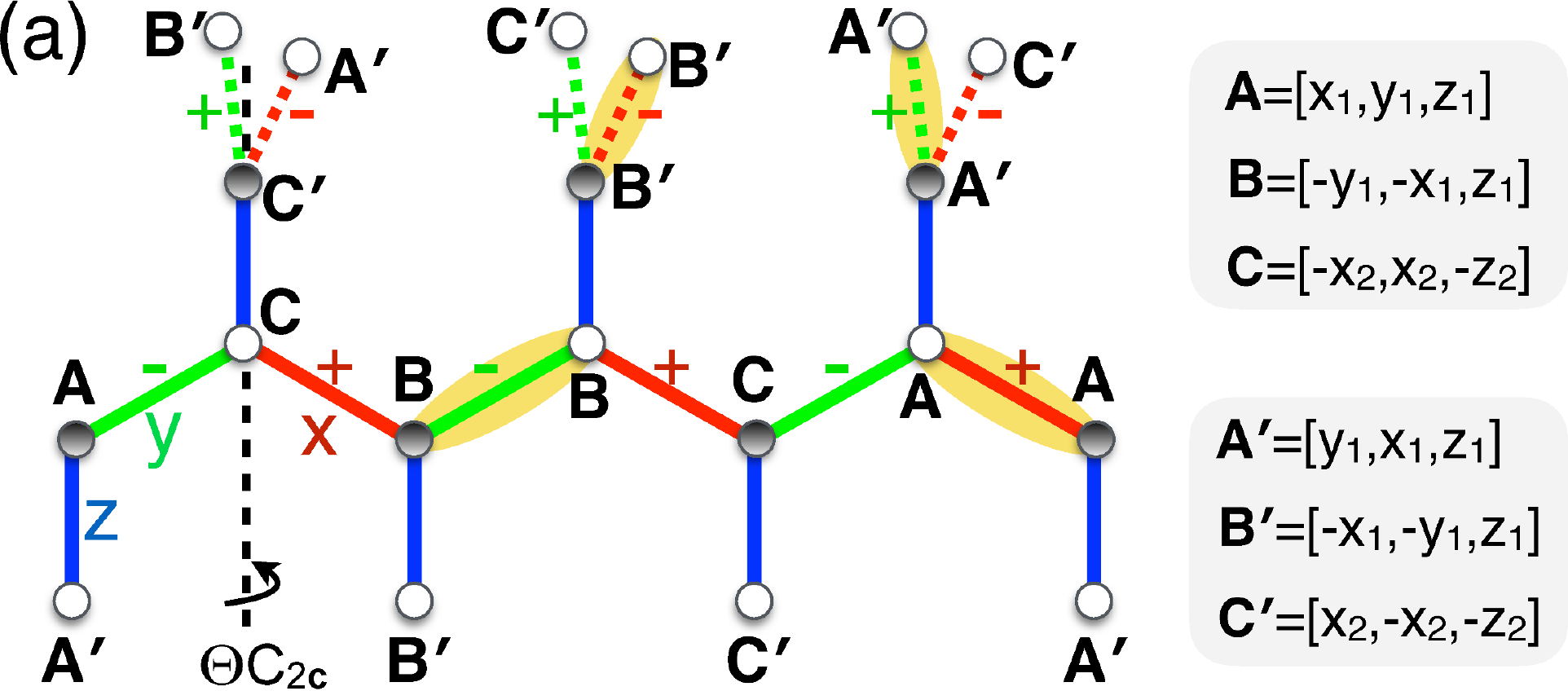}
\hspace{1cm}
\includegraphics[width=0.45\textwidth]{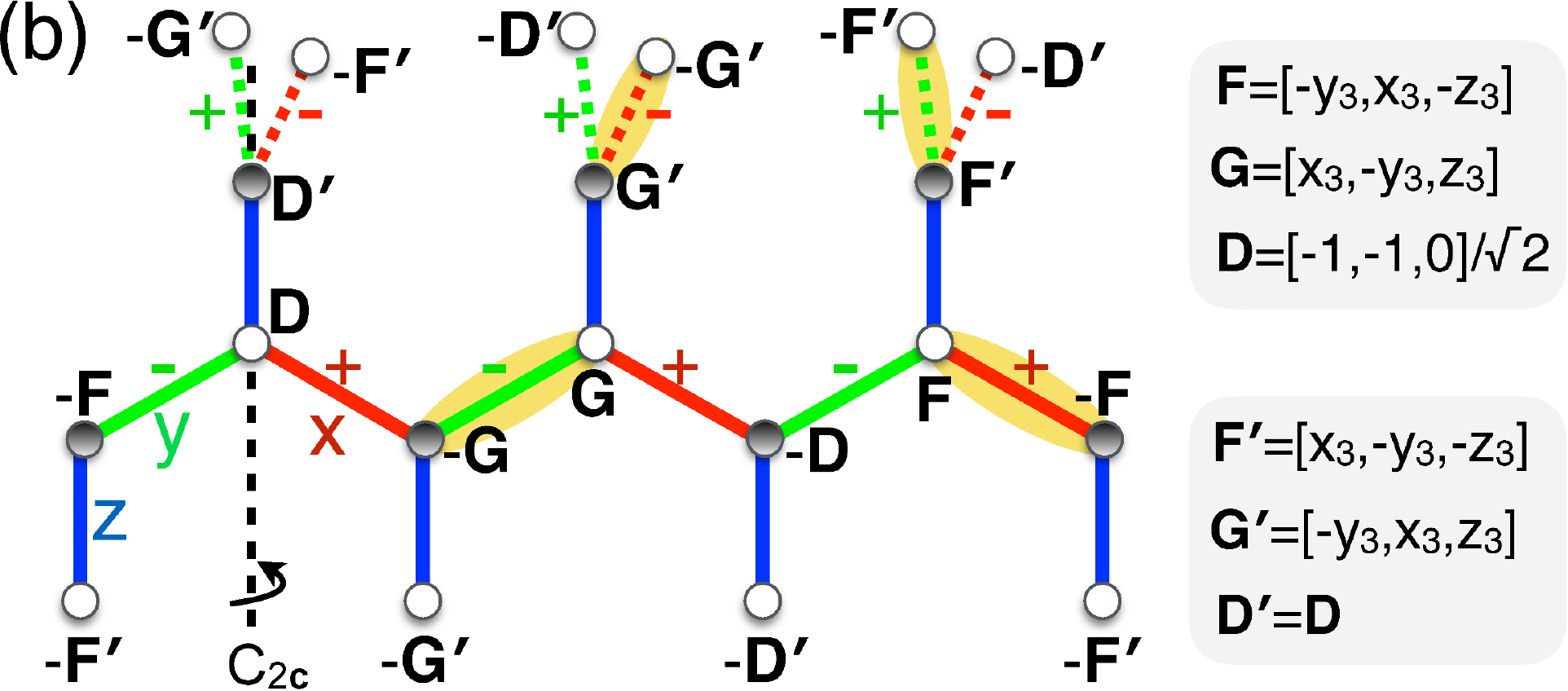}
\caption{Schematic structure of the (a) $K$-state and  (b) $\Gamma$-state discussed in Secs.~\ref{sec:Kdom} and \ref{sec:Gdom}. The spins on $xy$-  and $x'y'$ zigzag chains are denoted, correspondingly, by ${\bf A},{\bf B},{\bf C}$ and  ${\bf A}^\prime,{\bf B}',{\bf C}'$  in the $K$-state  and ${\bf D},{\bf F},{\bf G}$ and  ${\bf D}',{\bf F}',{\bf G}'$  in the $\Gamma$-state. The side panels show the corresponding Cartesian components.}
\label{fig:KdomGdom}
\end{figure*}

To get further insights into the structure of the actual ground states, we used overdamped dynamics simulations based on the Landau-Lifshitz-Gilbert (LLG) equations,~\cite{Serpico2001,Mochizuki2010,Choi2013,ChoiPRB2013}  see details in App.~\ref{App:B}. The results show that the minima inside the shaded region of Fig.~\ref{fig:LTphasediagram} correspond to two types of commensurate phases described by a wavevector ${\bf Q}\!=\!(\frac{2}{3},0,0)$. %
These two states, dubbed `$h\!=\!2/3$ phases', are inhomogeneous and feature two distinct values of local mean fields, repeated with periodicity-3. The magnitude of the spin-spin correlations between even (odd) spins along any of the zigzag chains also alternates between  two different values with the same periodicity, see App.~\ref{App:B} and detailed analysis below. 

These LLG results were obtained on a cluster with 240$\times$2$\times$2 orthorhombic unit cells with periodic boundary conditions and $N\!=\!15360$ spins.
Although we have examined other finite-size clusters as well, we were not able to find other solutions with larger periods. This is likely due to the inherent difficulty of single-spin update algorithms (such as LLG) to deliver states with non-trivial incommensurate structures, especially when dealing with almost flat energy landscapes.
Still, the mere fact that the minima from the Luttinger-Tisza method sit at an incommensurate point as soon as we depart from the line $\phi=3\pi/2$, suggests that the system will develop some kind of long-wavelength deformation of the commensurate $h\!=\!2/3$ phases found here. 
We do expect however that the $h\!=\!2/3$ phases will survive as global minima in a finite window close to $\phi\!=\!3\pi/2$ due to the lattice cutoff. Otherwise, these phases are only local minima, as shown explicitly by the fact that all computed local torques practically vanish (see App.~\ref{App:B}) in the whole shaded region of Fig.~\ref{fig:LTphasediagram}.

The stability regions of the two $h\!=\!2/3$ phases are indicated in Fig.~\ref{fig:LTphasediagram} by `$K$-region' (light red) and `$\Gamma$-region' (purple), corresponding, respectively, to dominant Kitaev or $\Gamma$ interactions. 
The boundary between the two regions begins at the hidden SO(3) point $(r,\phi)=(\frac{\pi}{4},\frac{3\pi}{2})$  discussed above. In the following, we shall describe the main features of the two $h\!=\!2/3$ states in detail.

\vspace*{-0.3cm}
\subsection{$K$-state}\label{sec:Kdom}
\begin{figure*}[!t]
\includegraphics[width=1\textwidth]{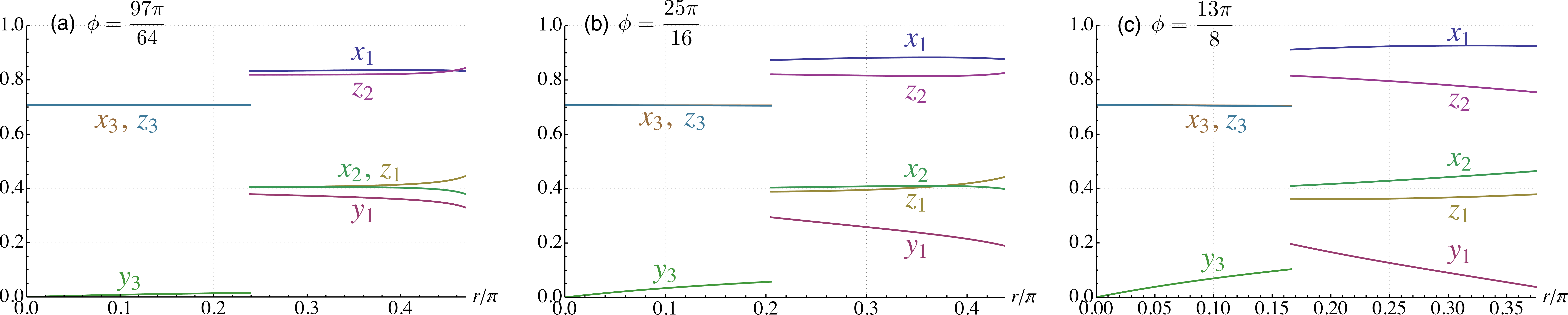}
\caption{Evolution of the components $x_1$, $y_1$, $z_1$, $x_2$, $z_2$ of Eq.~(\ref{eq:ABC}), and the components $x_3$, $y_3$ and $z_3$ of Eq.~(\ref{eq:FGD}), as we change the parameter $r$ for (a) $\phi\!=\!1.515625\pi$, (b) $\phi\!=\!1.5625\pi$, and (c) $\phi\!=\!1.625\pi$. Data are shown up to the point $r_c(\phi)$ where we exit the $K$-region, see Fig.~\ref{fig:LTphasediagram}.}
\label{fig:x1y1etc}
\end{figure*}

\vspace*{-0.3cm}
The $K$-state is shown schematically in Fig.~\ref{fig:KdomGdom}\,(a) and consists of six sublattices, ${\bf A}$, ${\bf B}$, ${\bf C}$, ${\bf A}^\prime$, ${\bf B}^\prime$, and ${\bf C}^\prime$. Their {\it Cartesian} components are given by 
\be\label{eq:ABC}
\begin{array}{ll}
{\bf A}=[x_1,y_1,z_1], &
{\bf A}^\prime=[y_1,x_1,z_1],\\
{\bf B}=[-y_1,-x_1,z_1],&
{\bf B}^\prime=[-x_1,-y_1,z_1],\\
{\bf C}=[-x_2,x_2,-z_2],&
{\bf C}^\prime=[x_2,-x_2,-z_2]\,,
\end{array}
\ee
and depend on three independent, {\it positive} real numbers, $x_1$, $y_1$ and $x_2$ (Note that $z_1\!=\![1-x_1^2-y_1^2]^\frac{1}{2}$ and $z_2\!=\![1-2x_2^2]^\frac{1}{2}$, due to the spin-length constraint). These numbers can be found by minimizing the energy.  Fig.~\ref{fig:x1y1etc} shows the resulting values along three special lines in parameter space, see also Fig.~\ref{fig:CartesianCompsAlongChain}\,(a) in App.~\ref{app:C} for the modulation of the spin components along a single $xy$-chain.

Each chain has a three-sublattice structure, with sublattices ${\bf A}$, ${\bf B}$ and ${\bf C}$ on $xy$-chains, and sublattices ${\bf A}^\prime$, ${\bf B}^\prime$ and ${\bf C}^\prime$ on $x^\prime y^\prime$-chains. In each given chain, spins sitting on even sites modulate in a counter-rotating manner from those sitting on odd sites. On the $xy$-chains, for example, the odd sites (gray circles) modulate in a ${\bf A, B, C, A, B} \cdots$ pattern, while the even sites (white circles) show a ${\bf C, B, A, C, B}\cdots$ pattern, and similarly for the modulation along the $x^\prime y^\prime$-chains.

As shown in Fig.~\ref{fig:ABC-FGD} and will be discussed in detail in Sec.~\ref{sec:Angles} below, the sublattices $\{{\bf A}, {\bf B}, {\bf C}\}$ form an almost ideal coplanar 120$^\circ$-structure and the same is true for the sublattices $\{{\bf A}', {\bf B}', {\bf C}'\}$.  
The deviation from the ideal 120$^\circ$-structure is small and their nature can be seen by examining the following vectors in the {\it orthorhombic} frame
\be\label{eq:ABCcanting}
\begin{array}{c}
{\bf A}\!+\!{\bf B}\!+\!{\bf C}\!=\!\left(\sqrt{2}(x_1\!-\!y_1\!-\!x_2),-2z_1\!+\!z_2,0 \right),\\
{\bf A}'\!+\!{\bf B}'\!+\!{\bf C}'\!=\!\left(-\sqrt{2}(x_1\!-\!y_1\!-\!x_2),-2z_1\!+\!z_2,0 \right).
\end{array}
\ee

So, the structure features an in-plane AF canting along ${\bf a}$ and an out-of-plane FM canting along ${\bf b}$. 
The AF canting is proportional to the quantity $x_1\!-\!y_1\!-\!x_2$, and alternates in sign from the primed to unprimed chains. In particular, we will see below in Sec.~\ref{sec:SofQ} that this AF canting is of the so-called zig-zag type.
The out-of-plane FM canting, on the other hand, is uniform and is proportional to $2z_1\!-\!z_2$. As a result the $K$-state has a total magnetization along the ${\bf b}$-axis. 
Now, both canting components are numerically very small in the entire shaded region of Fig.~\ref{fig:LTphasediagram}. For example, the total magnetization per site, $(2 z_1$-$z_2)/3$, is about $0.002$ at $P_K$. See also the evolution of the quantities $M_a'\propto x_1-y_1-x_2$ and $M_b'\propto 2 z_1-z_2$ plotted in Fig.~\ref{fig:SofQcomponents} below, along different paths in parameter space.

Another important feature of the $K$-state is the presence of FM dimers. Each $xy$-chain features alternating (${\bf A}{\bf A}$) and (${\bf B}{\bf B}$) FM dimers, separated by spins pointing along ${\bf C}$, and similarly for the $x^\prime y^\prime$-chains.  
Furthermore, the $K$-state respects two important symmetries. The first is $\Theta C_{2{\bf c}}$, which consists of a $\pi$-rotation in spin and real space around the black dashed line of Fig.~\ref{fig:KdomGdom}\,(a), followed by the time-reversal operation $\Theta$. This symmetry gives the following relations: $C_y\!=\!-C_x$ and $[B_x,B_y,B_z]\!=\![-A_y,-A_x,A_z]$ and similarly for the $x^\prime y^\prime$-chains, $C_y^\prime\!=\!-C_x^\prime$ and $[B_x^\prime,B_y^\prime,B_z^\prime]\!=\![-A_y^\prime,-A_x^\prime,A_z^\prime]$. 
The second symmetry operation is $\Theta C_{2{\bf a}}$, which consists of a $\pi$-rotation around the ${\bf a}$-axis  passing through the middle of the $z$-bonds, followed by $\Theta$. This symmetry maps the configuration of an $xy$-chain to that of a neighboring $x^\prime y^\prime$-chain, and gives, for example, $[A_x^\prime,A_y^\prime,A_z^\prime]\!=\![A_y,A_x,A_z]$. 

This relation between the Cartesian components of ${\bf A}$ and ${\bf A}^\prime$ (and similarly for ${\bf B}$ and ${\bf B}^\prime$ or ${\bf C}$ and ${\bf C}^\prime$) plays a special role for the energy contributions from the $z$-bonds, which are always of the type (${\bf A}{\bf A}^\prime$),  (${\bf B}{\bf B}^\prime$) or (${\bf C}{\bf C}^\prime$). 
Indeed, the fact that the $x$ and $y$ components get swapped between the two sites sharing the $z$-bonds, while the $z$-components remain the same follow the recipes described in Refs.~[\onlinecite{IoannisGamma},\onlinecite{IoannisK}] for minimizing the $\Gamma$- and $K$-coupling, individually (see also discussion in Sec.~\ref{sec:SpecialLine}) Here both interactions are present, and the negative energy contributions from the two couplings are $\Gamma (x_1^2+y_1^2)$ and $K z_1^2$ for  (${\bf A}{\bf A}^\prime$)  and   (${\bf B}{\bf B}^\prime$) bonds,   and similarly  $\Gamma (2\,x_2^2)$ and $K z_2^2$ for  (${\bf C}{\bf C}^\prime$) bonds.  
Given that $x_1^2\!+\!y_1^2\!+\!z_1^2\!=\!1$, the spin arrangement on the $z$-bonds is then essentially a compromise between the two anisotropic couplings. 
The contributions from $K$ and $\Gamma$ from all other type of bonds are always negative.

\begin{figure}[!b]
\includegraphics[width=0.7\columnwidth]{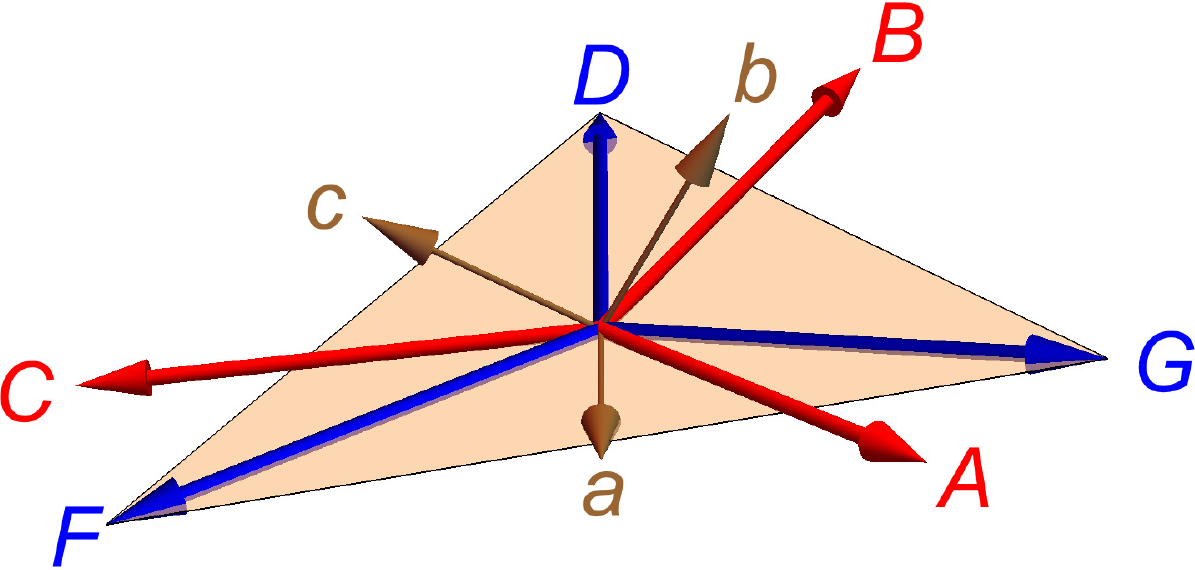}
\caption{Direction of the vectors ${\bf A}$, ${\bf B}$, ${\bf C}$ and ${\bf F}$, ${\bf G}$, ${\bf D}$ at one of the points on the boundary between the $K$- and $\Gamma$-states ($\phi=25\pi/16$, $r=0.41065\pi/2$). ${\bf a}$, ${\bf b}$, and ${\bf c}$ are the orthorhombic lattice vectors.}
\label{fig:ABC-FGD}
\end{figure}  

The  {\it Cartesian} components of the local fields are given by
\bea
\begin{array}{c}
{\bf h}_A \!=\! [h_A^x,h_A^y, h_A^z],~~
{\bf h}_B \!=\! [-h_A^y, -h_A^x, h_A^z],\\
{\bf h}_C \!=\! [h_C^x, h_C^x, h_C^z],\\
{\bf h}_{A^\prime} \!=\! [h_A^y, h_A^x, h_A^z],~~
{\bf h}_{B^\prime} \!=\! [-h_A^x, -h_A^y, h_A^z],\\
{\bf h}_{C^\prime} \!=\! [-h_C^x, -h_C^x, h_C^z],
\end{array}
\eea
where 
\be
\begin{array}{l}
h_A^x= J(x_1-x_2+y_1)+K x_1+\Gamma (x_1+z_2),\\
h_A^y=J (x_1+x_2+y_1)+K x_2+\Gamma (y_1+z_1),\\
h_A^z=J (2z_1-z_2) + K z_1 +\Gamma (x_2+y_1),\\
h_C^x= J(x_1+x_2-y_1)-K y_1-\Gamma (x_2+z_1),\\
h_C^z=J (2z_1- z_2)-K z_2-2 \Gamma x_1\,.
\end{array}
\ee
The magnitudes of the local fields obey the relations 
\bea
\begin{array}{c}
h_B =  h_{B^\prime} = h_{A^\prime}  = h_A,~~
h_{C^\prime} = h_C \neq h_A \,.
\end{array}
\eea
It follows that the distribution of the local field magnitudes contains only two distinct values. This also means that the $K$-state cannot be obtained by the standard version of the Luttinger-Tisza method, but only by an appropriate generalization of it.~\cite{LyonsKaplan1960,CairoB} 
Also, the total energy per site is equal to 
\be\label{eq:EnK}
E_K/N = - (2 h_{A}+h_{C})/6.
\ee
Finally, we note that the $K$-state appears visually similar to the  $\overline{\text{SP}}_{b^-}$ phase found by Lee and Kim (see Fig.~7\,(c) of Ref.~[\onlinecite{Lee2015}]), 
which is stabilized outside (below) the shaded region of Fig.~\ref{fig:LTphasediagram} (we have confirmed this numerically). However, the $\overline{\text{SP}}_{b^-}$ phase propagates along the ${\bf b}$-axis and not along the ${\bf a}$-axis, and as a result, the associated irreducible representation differs from that of the $K$-state, see also below.

\vspace*{-0.3cm}
\subsection{$\Gamma$-state}\label{sec:Gdom}
\vspace*{-0.3cm}
The $\Gamma$-state appears visually similar (and is most likely the same) with the $\overline{\text{SP}}_{a^-}$ phase found by Lee and Kim (see Fig.~6\,(c) of Ref.~[\onlinecite{Lee2015}]).
It is shown schematically in Fig.~\ref{fig:KdomGdom}\,(b) and consists of ten sublattices, $\pm{\bf F}$, $\pm{\bf G}$, $\pm{\bf D}$, $\pm{\bf F}^\prime$ and $\pm{\bf G}^\prime$. Their {\it Cartesian} components are given by 
\be\label{eq:FGD}
\begin{array}{ll}
{\bf F}=[-y_3,x_3,-z_3], &  {\bf F}^\prime=[x_3,-y_3,-z_3],\\
{\bf G}=[x_3,-y_3,z_3], &  {\bf G}^\prime=[-y_3,x_3,z_3],\\
{\bf D}=-\frac{1}{\sqrt{2}}[1,1,0]=-{\bf c}, &  {\bf D}^\prime={\bf D},\\
\end{array}
\ee
which depend on two independent {\it positive} real numbers, $x_3$ and $y_3$ (Note that $z_3\!=\![1-x_3^2-y_3^2]^{1/2}$ due to the spin-length constraint).
These numbers can again be obtained by minimizing the energy. The resulting numerical values are shown in Fig.~\ref{fig:x1y1etc} along three special lines in parameter space, see also Fig.~\ref{fig:CartesianCompsAlongChain}\,(b) in App.~\ref{app:C} for the modulation of the components along a single $xy$-chain.

Here, the spin structure along each zigzag chain requires six sublattices because the configuration on the odd sites features the time-reversed version of the configuration on the even sites. 
However, the odd sites again modulate  in a counter-rotating manner from the even sites, as in the $K$-state. On the $xy$-chains, for example, the odd sites (gray circles) modulate in a $-{\bf F}, -{\bf G}, -{\bf D}, -{\bf F}, -{\bf G}\cdots$ pattern, while the even sites (white circles) show a ${\bf D, G, F, D, G}\cdots$ pattern. The pattern on even sites is furthermore coplanar and the same is true for the pattern on the odd sites, but the two respective planes do not coincide in general. So the state is globally non-coplanar. 
 
 \begin{figure*}[!t]
\includegraphics[width=0.99\textwidth]{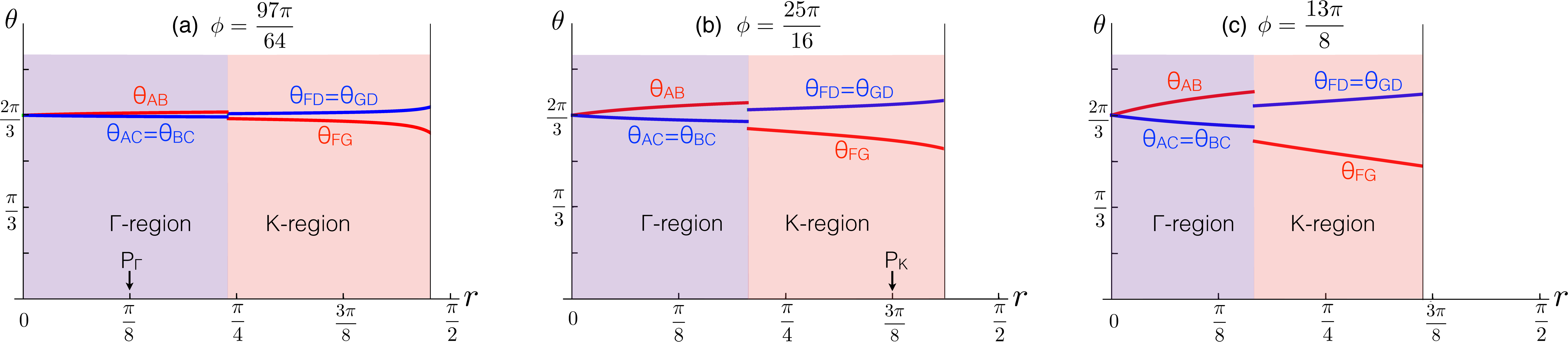}
\caption{Evolution of various angles between different sublattices as we change the parameter $r$ for (a) $\phi\!=\!1.515625\pi$, %\frac{97\pi}{64}$, 
(b) $\phi\!=\!1.5625\pi$, %\frac{25\pi}{16}$ 
and (c) $\phi\!=\!1.625\pi$. %\frac{13\pi}{8}$. 
Data are shown up to the point $r_c(\phi)$ where we exit the $K$-region, see Fig.~\ref{fig:LTphasediagram}.}
\label{fig:angles}
\end{figure*}

Similarly to the sublattices $\{{\bf A}, {\bf B}, {\bf C}\}$ of the $K$-state, the sublattices $\{{\bf F}, {\bf G}, {\bf D}\}$ also form an almost ideal 120$^\circ$-structure, see Fig.~\ref{fig:ABC-FGD}. Here the deviation from the ideal 120$^\circ$-structure features only an in-plane AF canting since ${\bf F}$, ${\bf G}$ and ${\bf D}$ are coplanar. Specifically, in the {\it orthorhombic} frame, 
\be\label{eq:FGDcanting}
{\bf F}\!+\!{\bf G}\!+\!{\bf D}\!=\!{\bf F}'\!+\!{\bf G}'\!+\!{\bf D}'\!=\!(0,0,\sqrt{2}(x_3\!-\!y_3)\!-\!1).
\ee
So, the in-plane AF canting away from the ideal 120$^\circ$-structure is now along the {\bf c}-axis, and is proportional to the quantity $x_3-y_3-\frac{1}{\sqrt{2}}$. We will see below in Sec.~\ref{sec:SofQ} that this canting is of the so-called stripy type.
 
Next, in contrast to the $K$-state that contains FM dimers, the $\Gamma$-state contains AF dimers. Each $xy$-chain features, for example, alternating $({\bf F},-{\bf F})$ and $(-{\bf G},{\bf G})$ dimers, separated by spins pointing along ${\bf D}$  and $-{\bf D}$, and similarly for the $x^\prime y^\prime$-chains.
Furthermore, the $\Gamma$-state  is invariant under $C_{2{\bf c}}$ (and not $\Theta C_{2{\bf c}}$, which is the reason why the $z$-component of ${\bf D}$ vanishes), and under $\Theta C_{2{\bf a}}$ symmetries (like the $K$-state).  As before, the latter symmetry maps the spin configuration of $xy$-chain to that in the neighboring  $x^\prime y^\prime$-chain.

Turning to the energetics, the $z$-bonds are always of the type (${\bf D}{\bf D}^\prime$),  (${\bf F}{\bf F}^\prime$) or  (${\bf G}{\bf G}^\prime$) and their time-reversed versions, respectively. The contribution to the energy from the $K$- and $\Gamma$-couplings on the (${\bf F}{\bf F}^\prime$) or  (${\bf G}{\bf G}^\prime$)  bonds are equal to $K z_3^2$ and $\Gamma (x_3^2+y_3^2)$ (both negative), and so the spin arrangement on these bonds is again a compromise between the two anisotropic couplings. On the (${\bf D}{\bf D}^\prime$) bonds, the corresponding contributions are $0$ and $\Gamma$, respectively. So the $\Gamma$-state maximizes the energy gain from the $\Gamma$-coupling on 1/3 of the $z$-bonds. This is also partly the reason why this state is stabilized for dominant $\Gamma$. 
For the other types of bonds, the anisotropic contributions to the energy are again all negative, as in the $K$-state.

The  {\it Cartesian} components of the local fields are given by
\be
\begin{array}{c}
{\bf h}_G = [h_G^x,h_G^y, h_G^z],~
{\bf h}_F = [h_G^y, h_G^x,- h_G^z],\\
{\bf h}_D = {\bf h}_{D'} = [h_D^x, h_D^x, 0],\\
{\bf h}_{G'} = [h_G^y, h_G^x,h_G^z],~
{\bf h}_{F'} = [h_G^x,h_G^y, -h_G^z],
\end{array}
\ee
where the independent components are 
\be
\begin{array}{l}
h_G^x = -J(x_3+y_3-\frac{1}{\sqrt{2}})+\frac{K}{\sqrt{2}}+\Gamma (x_3+z_3),\\
h_G^y = J (x_3+y_3+\frac{1}{\sqrt{2}})+(K-\Gamma) y_3, \\
h_G^z = K z_3+\Gamma  (x_3+\frac{1}{\sqrt{2}})\,.
\end{array}
\ee
The magnitudes of the local fields are 
\be
h_F = h_{F^\prime} = h_{G^\prime} = h_G \neq h_D\,.
\ee
So, there are two distinct local field magnitudes as in the $K$-state. Finally, the total energy per site is equal to 
\be\label{eq:EnG}
E_\Gamma/N = -(2 h_{G}+h_{D})/6.
\ee

\vspace*{-0.3cm} 
\subsection{The nearly 120$^\circ$ pattern of the $K$- and $\Gamma$-state}\label{sec:Angles}
\vspace*{-0.3cm}
As mentioned above and shown explicitly in Fig.~\ref{fig:ABC-FGD}, the $K$- and $\Gamma$-states feature a distinctive nearly 120$^\circ$ pattern. Figure~\ref{fig:angles} shows the evolution of the angles between the spins along a single zigzag chain, as a function of the parameter $r$, for three representative values of $\phi$: (a) $\frac{97\pi}{64}$, (b) $\frac{25\pi}{16}$, and (c) $\frac{13\pi}{8}$. 
The angles $\theta_{AB}$ (between ${\bf A}$ and ${\bf B}$ sublattices inside the $K$-region) and $\theta_{FG}$ (between ${\bf F}$ and ${\bf G}$ sublattices inside the $\Gamma$-region) are shown by red lines, while the corresponding angles $\theta_{AC}\!=\!\theta_{BC}$ ($K$-region) and $\theta_{FD}\!=\!\theta_{GD}$ ($\Gamma$-region) are shown by blue lines. The results are shown up to the critical value $r=r_c(\phi)$, where we exit from the experimental relevant region of interest.  
We see  that for dominant $\Gamma$ interaction (i.e., small $r$) both $\theta_{FG}$ and $\theta_{FD}$ are almost equal to 120$^\circ$. The deviation from 120$^\circ$ is particularly small for the smallest value of $\phi\!=\!\frac{97\pi}{64}$, and it only slightly increases for bigger $\phi$. Thus, in this parameter range the magnetic ground state can be described by a collection of zigzags in which each half of the chain is almost a 120$^\circ$ coplanar spiral. The spins in the other half rotate in the opposite direction, as discussed above.
Note that as we approach the line $\phi=3\pi/2$ all angles shown in Fig.~\ref{fig:angles} tend to 120$^\circ$, irrespective of the value of $r$. This is because, as $\phi\!\to\!(3\pi/2)^+$ all the dot products 
\small
\be\begin{array}{l}
{\bf A}\!\cdot\!{\bf B}\!=\!z_1^2\!-\!2x_1y_1,~
{\bf B}\!\cdot\!{\bf C}\!=\!{\bf C}\cdot{\bf A}\!=\!(y_1\!-\!x_1)x_2\!-\!z_1z_2,\\
{\bf F}\!\cdot\!{\bf G}\!=\!-2x_3y_3\!-\!z_3^2,~
{\bf G}\!\cdot\!{\bf D}\!=\!{\bf D}\cdot{\bf F}\!=\!(y_3\!-\!x_3)/\sqrt{2},
\end{array}\ee
\normalsize
tend to $-1/2$, because in that limit, 
\be\label{eq:3pio2}
\begin{array}{c}
\{x_1,y_1,z_1,x_2,z_2\}\!\to\!\{\frac{2}{\sqrt{6}},\frac{1}{\sqrt{6}},\frac{1}{\sqrt{6}},\frac{1}{\sqrt{6}},\frac{2}{\sqrt{6}}\},\\
\{x_3,y_3,z_3\}\!\to\!\{\frac{1}{\sqrt{2}},0,\frac{1}{\sqrt{2}}\},
\end{array}
\ee
see Fig.~\ref{fig:x1y1etc} and Sec.~\ref{sec:SpecialLine} below. 
With increasing $\phi$ (increasing $J$), the deviation from this ideal 120$^\circ$ pattern increases.

\vspace*{-0.3cm} 
\subsection{The transition between $K$- and $\Gamma$ states}\label{sec:Boundary}
\vspace*{-0.3cm}
Fig.~\ref{fig:angles} shows in addition that when we cross the boundary between the $K$- and $\Gamma$-states, there is a discontinuous jump between $\theta_{FG}$ and $\theta_{AB}$ and between $\theta_{FD}$ and $\theta_{AC}$. This shows that the transition between the two states is of first order, which is further confirmed by the qualitatively different ${\bf Q}=0$ Fourier components of the two states (see Sec.~\ref{sec:SofQ} below). 
This is  also demonstrated in Fig.~\ref{fig:ABC-FGD} which shows the qualitatively different structures of $\{{\bf A},{\bf B},{\bf C}\}$ and $\{{\bf F},{\bf G},{\bf D}\}$ sublattices on the boundary between the $K$- and $\Gamma$-states ($\phi=25\pi/16$, $r=0.41065\pi/2$).

Note that the discontinuous jumps become smaller and smaller as we approach the hidden SO(3) point $(r,\phi)\!=\!(\frac{\pi}{4},\frac{3\pi}{2})$, discussed above. The reason is that at this special point the $K$- and $\Gamma$-state become members of the symmetry-related SO(3) degeneracy in the rotated frame. So apart from their global orientation (in the rotated frame),  at this special point, the two states are indistinguishable from each other, with the same relative angles between different sublattices.

\vspace*{-0.3cm}
\subsection{The special line $\phi=3\pi/2$}\label{sec:SpecialLine}
\begin{figure}[!t]
\includegraphics[width=0.35\textwidth]{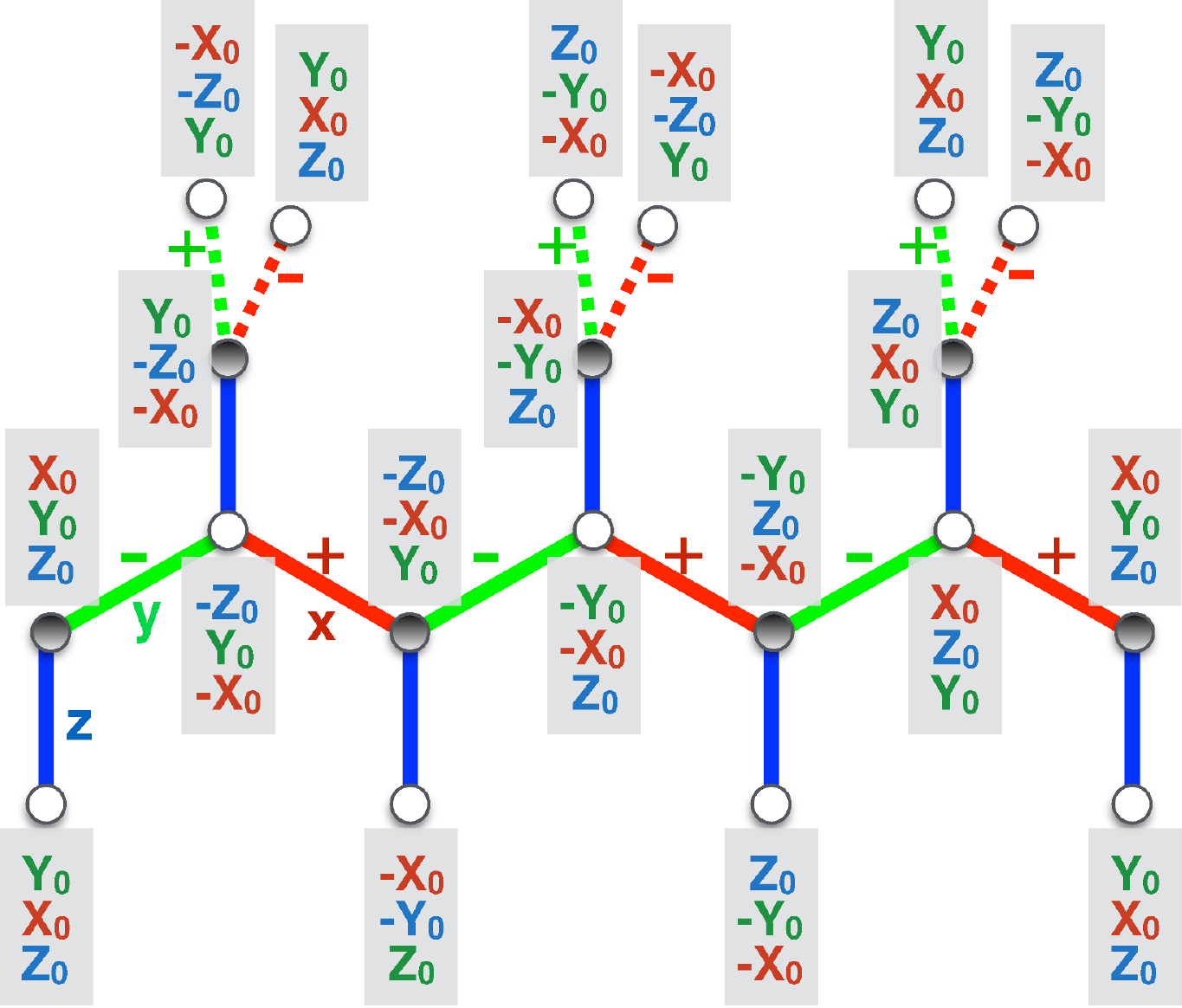}
\caption{The two-parameter family of classical ground states along the special line $\phi=3\pi/2$ of Fig.~\ref{fig:LTphasediagram}, where $\Gamma\!<\!0$ and $K\!<\!0$. The vectors shown at each site are the Cartesian components.}
\label{fig:3pio2}
\end{figure}

\vspace*{-0.3cm}
It turns out that many of the properties of the two commensurate phases described above descend from the structure of the classical ground state manifold along the special line $\phi=3\pi/2$, where $J=0$. 
To understand the structure of this manifold, we combine the two recipes described in Refs.~[\onlinecite{IoannisGamma},\onlinecite{IoannisK}] for minimizing the $\Gamma$- and $K$-coupling, individually. To this end, we consider one of the two building blocks of the structure, which contains the bonds labeled by $x$, $y$ and $z$ in Fig.~\ref{fig:betalattice} (the second building block contains the bonds labeled by $x'$, $y'$ and $z$ and the analysis is similar),
\be\label{gr:3bonds}
\includegraphics[width=0.85in]{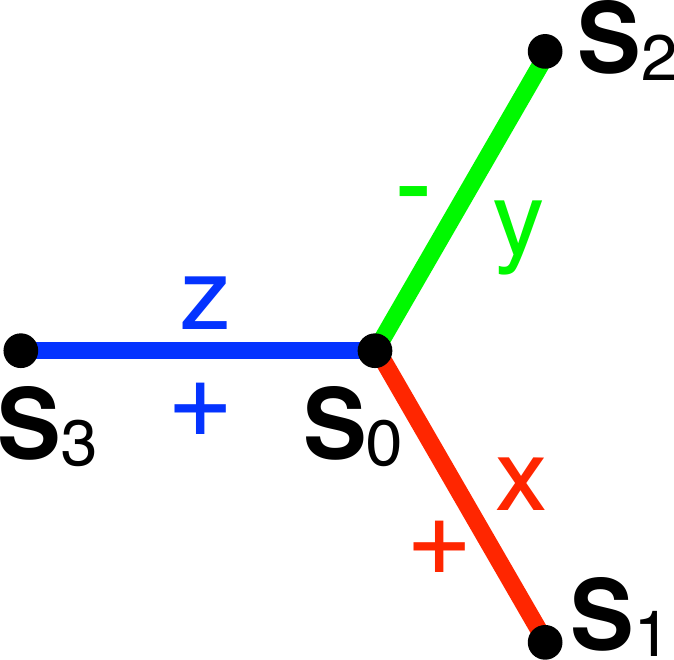}
\ee 
where $\vec{S}_i$ denotes the pseudospin 1/2 at site $i$.  
To find the minimum, we begin by aligning the central spin $\vec{S}_0$ along an arbitrary direction $[X_0,Y_0,Z_0]$ in the Cartesian frame. Then we go to one of the neighboring sites, say the site $\vec{S}_1$ of (\ref{gr:3bonds}), which shares a $x$-type of bond with $S_0$. The interaction between the two sites is of the form $\Gamma (S_0^yS_1^z+S_0^zS_1^y)+K S_0^xS_1^x$, and both $K$ and $\Gamma$ are negative. To satisfy this coupling we take $\vec{S}_1=[X_0,Z_0,Y_0]$, i.e. we copy the $x$ component and switch the $y$ and $z$ components relative to $\vec{S}_0$. Similarly, the site $\vec{S}_2$ of (\ref{gr:3bonds}) shares a $y$-type of bond with $\vec{S}_0$, and their mutual coupling is now of the form $-\Gamma (S_0^zS_2^x+S_0^xS_2^z)+K S_0^yS_2^y$. To satisfy this coupling we now take $\vec{S}_2=[-Z_0,Y_0,-X_0]$, i.e. we copy the $y$ component and switch the $x$ and $y$ components relative to $\vec{S}_0$, and at the same time we multiply with minus one, because the $\Gamma$ coupling has an extra minus sign on the $y$-type of bonds. Finally, for the site $\vec{S}_3$ of (\ref{gr:3bonds}) we take $\vec{S}_3=[Y_0,X_0,Z_0]$. We can then proceed to the neighboring sites of $\vec{S}_1$, $\vec{S}_2$ and $\vec{S}_3$ following the same recipe, until we cover the whole lattice. The resulting magnetic structure is shown in Fig.~\ref{fig:3pio2} and corresponds to a continuum, two-parameter family of states associated with the direction of the initial central spin $\vec{S}_0$.

We next show that these configurations saturate the lower energy bound set by the minimum eigenvalue $\lambda_{\text{min}}=(K+2\Gamma)S^2/2$ of the Luttinger-Tisza matrix, and are therefore ground states. 
Indeed, collecting all energy contributions from the three types of bonds of the cluster shown in (\ref{gr:3bonds}) gives
\be
E_{\includegraphics[width=0.1in]{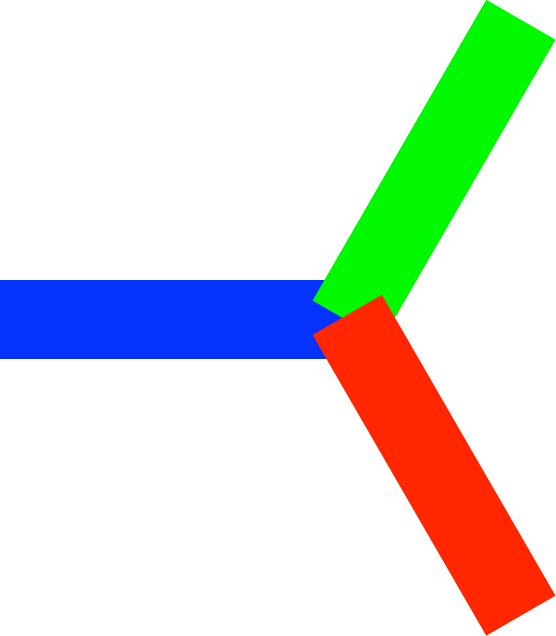}}
=(K+2\Gamma) (X_0^2+Y_0^2+Z_0^2)=(K+2\Gamma)S^2.
\ee
This result is the same for all such clusters in the 3D structure. So the total energy per site of the resulting configurations is $E/N=\frac{1}{2}(K+2\Gamma)S^2$, which coincides with $\lambda_{\text{min}}$. 

The $\mc{S}^2$ degeneracy associated with the two-parameter family of states shown in Fig.~\ref{fig:3pio2} is accidental everywhere along the line $\phi=3\pi/2$, except at $K=\Gamma$ where the degeneracy is related to the hidden SO(3) symmetry discussed above.

Next, we examine the fate of this degeneracy as we include an infinitesimal coupling $J$. Within the above manifold of states, this coupling gives an energy contribution 
\be\label{eq:ObD}
E_J \propto J (X_0 -Y_0 -Z_0)^2 + \text{constant},
\ee
This expression is invariant under the symmetry $C_{2{\bf c}}$ and $\Theta C_{2{\bf c}}$.
According to (\ref{eq:ObD}), a positive $J$ selects the submanifold of states with $X_0=Y_0+Z_0$.

Turning now to the $K$-state, the minimization of $E_K$ of Eq.~(\ref{eq:EnK}) at $\phi=3\pi/2$ delivers not one but a continuous family of states, described by  $z_1=x_2=y_1$, see Eq.~(\ref{eq:CondK}) in App.~\ref{app:D}. These states belong to the $\mc{S}^2$ manifold at $\phi=3\pi/2$. An infinitesimal positive $J$ will select the state with $x_1=y_1+z_1$ (the condition $X_0=Y_0+Z_0$ above for the sublattice ${\bf A}$ of the $K$-state), which gives $[x_1,y_1,z_1]\to\frac{1}{\sqrt{6}}[2,1,1]$.

For the $\Gamma$-state, the minimization of $E_\Gamma$ of Eq.~(\ref{eq:EnG}) for $\phi=3\pi/2$ delivers one solution only, with $x_3=z_3=\frac{1}{\sqrt{2}}$, see Eq.~(\ref{eq:CondG}) in App.~\ref{app:D}. This solution is  also a member of the $\mc{S}^2$ manifold, and in addition already satisfies the condition $X_0=Y_0+Y_0$ for the sublattice ${\bf F}$ (which has $[X_0,Y_0,Z_0]=[-y_3,x_3,-z_3]=[0,1,-1]/\sqrt{2}$).

At this point it is useful to digress a little and discuss what happens for negative $J$. The reason we wish to do this is that the available {\it ab initio} calculations~\cite{KimKimKee2016,Tsirlin2018} deliver a negative $J$ rather than a positive $J$ that we consider here. Eq.~(\ref{eq:ObD}) shows why a negative $J$ is not consistent with experimental data: A negative $J$ lifts the $\mc{S}^2$ degeneracy completely and selects a state with $[X_0,Y_0,Y_0]=\frac{1}{\sqrt{3}}[1,-1,-1]$. Based on Fig.~\ref{fig:3pio2}, in this state the spins of the unprimed chains point along $[1\bar{1}\bar{1}]$, while the spins of the primed chains point along $[\bar{1}1\bar{1}]$. So the state comprises two FM subsystems, and has a finite total magnetization along the ${\bf z}$-axis. (This is also the state denoted by `FM-SZ$_{\text{FM}}$' in Fig.~5 (a) of Ref.~[\onlinecite{Lee2015}].) Clearly, this state is not compatible with the observed counter-rotating state, and therefore we can safely conclude that, within the $J$-$K$-$\Gamma$ model description of $\beta$-Li$_2$IrO$_3$, the Heisenberg coupling $J$ must be antiferromagnetic.

\vspace*{-0.3cm}
\subsection{The single-chain Hamiltonian $\mc{H}_c$}\label{sec:Hc}
\vspace*{-0.3cm} 
Let us now analyze a central property that is shared by both $K$- and $\Gamma$-states, namely that both states are invariant under the operation $\Theta C_{2{\bf a}}$. According to this property, the global structure of the states arises simply by `tiling' the spin configuration of a single $xy$-chain to the whole lattice using the appropriate rotation $\Theta C_{2{\bf a}}$. This raises the question of whether there exists a single-chain Hamiltonian whose classical minima coincide with the actual configuration on $xy$-chains. Indeed, the structure of the system allows to split the Hamiltonian into a sum over single-chain Hamiltonians $\mc{H}_c$ for $xy$-chains and $\mc{H}_{c^\prime}$ for $x^\prime y^\prime$-chains, 
\be
\mc{H}=\sum_{xy\text{-chains}~c} \mc{H}_c+\sum_{x^\prime y^\prime\text{-chains}~c^\prime} \mc{H}_{c^\prime}\,,
\ee
where $\mc{H}_c$ and $\mc{H}_{c^\prime}$ include half of the inter-chain couplings, which reside on $z$-bonds. Schematically, $\mc{H}_c$ takes the form
\be\label{eq:Hc}
\includegraphics[width=1.6in]{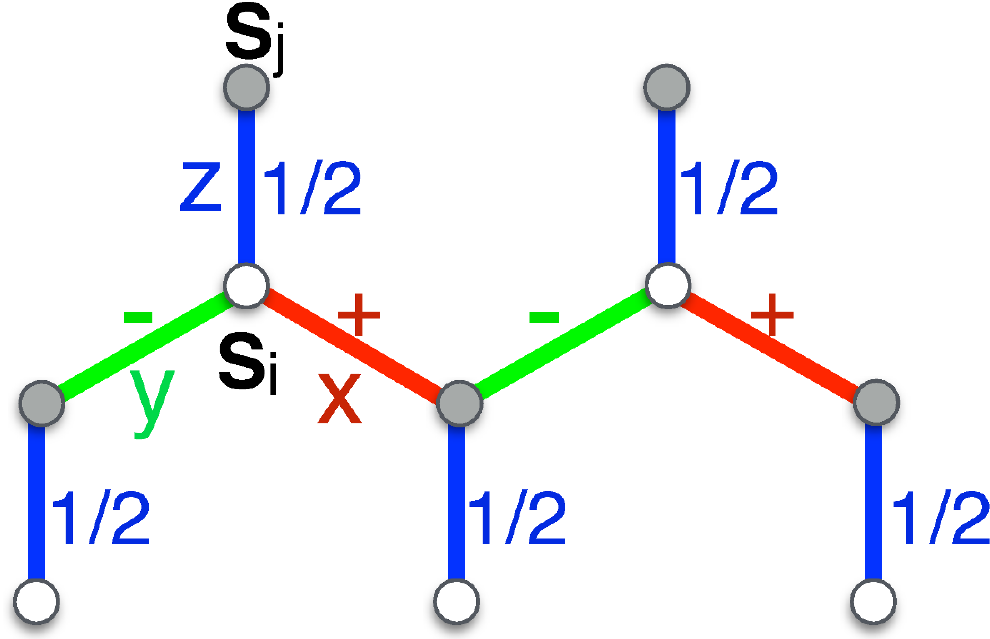}
\ee 
where the factors of $1/2$ on the vertical, $z$-bonds indicate that $J$, $K$ and $\Gamma$ should be replaced with $J/2$, $K/2$ and $\Gamma/2$, respectively. Note that this Hamiltonian differs from the single-chain Hamiltonian of Kimchi {\it et al.},\cite{Kimchi2015,Kimchi2016} which includes only the couplings on the $x$- and $y$-bonds.

Now, suppose we have found a minimum energy configuration of $\mc{H}_c$, with energy $E_c$. From this configuration we can then generate a minimum energy configuration of $\mc{H}_{c^\prime}$, with the same energy $E_c$, by simply applying the operation $\Theta C_{2{\bf a}}$. In addition, since the sites $i$  and $j$, sharing a $z$-bond are mapped to each other by this operation, we must have $[S_i^x,S_i^y,S_i^z]\!=\![S_j^y,S_j^x,S_j^z]$. This relation is satisfied in both the $K$-state and the $\Gamma$-state.

The crucial point is whether the single-chain minimum can be tiled in the whole lattice or not. The answer depends on the form of the state,  on the connectivity and on the loop-structure of the lattice. If the answer is yes, then clearly the generated state saturates the global energy minimum and is therefore a classical ground state. 
According to the above, the $K$-state and the $\Gamma$-state belong to this family of solutions, and it is plausible that all classical ground states of the shaded region of Fig.~\ref{fig:LTphasediagram} (plus other states of the phase diagram as well) belong to this family too. 
This suggests that solving the much simpler single-chain Hamiltonian $\mc{H}_c$ may be the route to deducing the detailed structure of the phases inside the shaded region of Fig.~\ref{fig:LTphasediagram}, which is the experimentally relevant region for $\beta$-Li$_2$IrO$_3$. In particular, this approach can help clarifying whether this region consists, e.g., of a cascade of first-order transitions between commensurate phases with different periodicity. Such a detailed investigation is however out of the scope of the present paper.

We should  also comment on the similarity  between our  periodicity-3 states and the 120$^{\circ}$-phase of the $J$-$K$-$\Gamma$-model on the 2D honeycomb lattice  which appears at the same region of the parameter space  (see Fig. 2\,(f) of Ref.~[\onlinecite{Rau2014}]).  Here again the magnetic structure can be tiled by the spin configuration of a single $xy$-chain.  Similarly to the $\Gamma$-state,
 the 120$^{\circ}$-phase  of the 2D honeycomb lattice contains AF dimers.
 
\begin{figure*}[!t]
\includegraphics[width=1\textwidth]{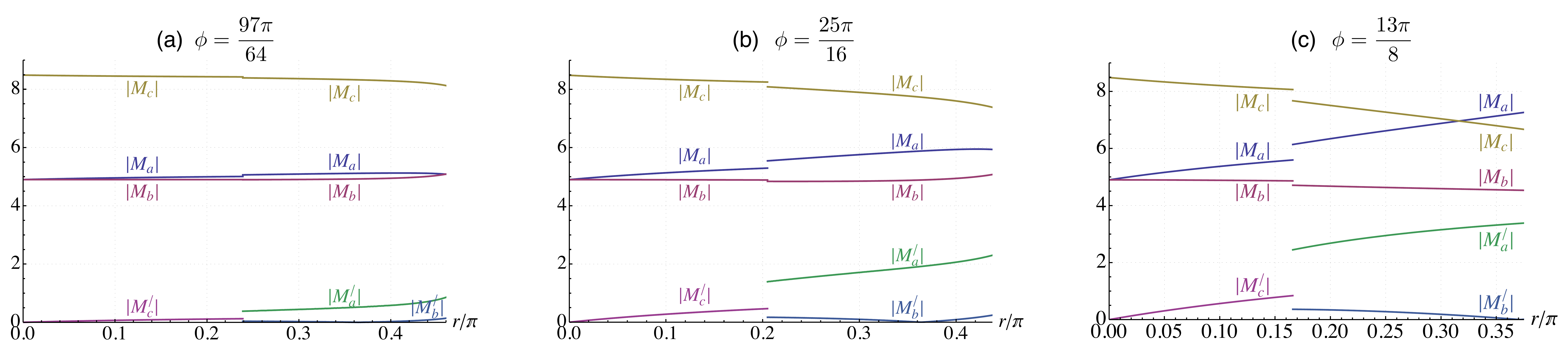}
\caption{Evolution of the absolute values of the various components of the static structure factor as a function of $r$, as we cross the boundary between the $\Gamma$- and $K$-states, for (a) $\phi\!=\!1.515625\pi$, (b) $\phi\!=\!1.5625\pi$, and (c) $\phi\!=\!1.625\pi$.}
\label{fig:SofQcomponents}
\end{figure*}

\vspace*{-0.3cm}
\section{Static spin-spin structure factor}\label{sec:SofQ}
\vspace*{-0.3cm}
\subsection{Theoretical results}\label{sec:SofQTheory}
\vspace*{-0.3cm}
Next, we analyze the static spin structure factors of both $K$- and $\Gamma$-states and compare with the irreducible representation reported experimentally. To this end, we follow Ref.~[\onlinecite{Biffin2014a}] and introduce the four-component vector $\vec{M}^T_{\bf Q}=(\vec{M}_{{\bf Q},1},\vec{M}_{{\bf Q},2},\vec{M}_{{\bf Q},3},\vec{M}_{{\bf Q},4})$, where
\begin{eqnarray}\label{M}
\vec{M}_{\vec{Q},n}&=&\sum_{\vec{R}}\sum_{p=0}^3 \vec{S}(\vec{R},\boldsymbol{\rho_p}, \vec{r}_n) e^{i \vec{Q} \cdot ( \vec{R}+ \boldsymbol{\rho_p} + \vec{r}_n)} 
\end{eqnarray}
are the Fourier transforms of the magnetic moments at the four sites $n=1$-$4$ of the primitive cell, ${\bf Q}$ belongs to the orthorhombic BZ, $\vec{R}$ runs over the orthorhombic unit cells, and $\vec{r}_n$ and $ \boldsymbol{\rho_p}$ are given by Eqs. (\ref{primitive}) and (\ref{rho}), respectively. 
The four component vector $\vec{M}_{\bf Q}$ can be expressed in terms of the symmetry basis vectors:
\bea
F=\left[\begin{array}{r}\nonumber
1\\1\\1\\1
\end{array}
\right]\!,\,
A=\left[\begin{array}{r}
1\\-1\\-1\\1
\end{array}
\right]\!,\,
C=\left[\begin{array}{r}
1\\1\\-1\\-1
\end{array}
\right],\!\,
G=\left[\begin{array}{r}
1\\-1\\1\\-1
\end{array}
\right],
\eea
which describe the ferromagnetic (F), N\'eel (A), stripy (C) and zig-zag order (G), respectively.
 
The zero-field scattering experiments of Ref.~[\onlinecite{Biffin2014a}] have detected a Fourier component of the static structure factor with ${\bf Q}=(0.57,0,0)$, belonging to the $\Gamma_4$ irreducible representation, with
\be\label{MK}
\vec{M}_{(0.57,0,0)}= (i M_a A, \,iM_b C,\, M_c F).
\ee
To compare with our theoretical results we must first note that the site labeling in Ref.~[\onlinecite{Biffin2014a}] is different from ours. The primitive unit cell used there contains the following 4 sites (denoted with a superscript $c$):
\begin{eqnarray}
\begin{array}{l}
\vec{r}_{1}^c = (\frac{1}{8},\frac{1}{8},z),\,
\vec{r}_{2}^c = (\frac{1}{8},\frac{5}{8},\frac{3}{4}-z)  \\ 
\\
\vec{r}_{3}^c = (\frac{3}{8},\frac{3}{8},1-z) ,\,
\vec{r}_{4}^c = (\frac{3}{8},\frac{7}{8},\frac{1}{4}+z), 
\end{array}
\end{eqnarray}
where $z=\frac{17}{24}$. Therefore, there is the following mapping between the notations of the sites belonging to  the primitive unit cell given in Ref.~[\onlinecite{Biffin2014a}] and our labeling of the sites presented in Fig.~\ref{fig:betalattice}: $\vec{r}_{1}^c\rightarrow \vec{r}_{4}$, $\vec{r}_{2}^c \rightarrow \vec{r}_{11}$, $\vec{r}_{3}^c \rightarrow \vec{r}_{1}$ and $\vec{r}_{4}^c \rightarrow \vec{r}_{10}$.  So in order to effectively compare the basic states describing our period-3 orders to the ones used in Ref.~[\onlinecite{Biffin2014a}], we have relabeled the sites of our magnetic unit cell  accordingly. 
 
Let us summarize our findings. Since  both $K$- and $\Gamma$-states are characterized by periodicity-3,  we have three momenta to consider: $ {\bf Q}=(0,0,0), (1/3,0,0)$ and $(2/3,0,0)$ and expect three Bragg peaks in general. However,   the  Fourier components of the magnetic structure are non-zero only  at $ (0,0,0)$ and $(2/3,0,0)$. For  the $K$-state  we find 
\bea\label{MK}
\begin{array}{c}
\vec{M}_{(2/3,0,0)}= (i M_a A, \,iM_b C,\, M_c F),\\
\vec{M}_{(0,0,0)}= (M'_a G, \,M'_b F,\, 0),
\end{array}
\eea
with
\be\begin{array}{c}
M_a= i (x_1+2x_2-y_1), M_b=- i (z_1+z_2), \\
M_c= i \sqrt{3}(x_1+y_1),\\
M_a'=-2(x_1-y_1-x_2), M_b'=(2z_1-z_2)~,
\end{array}\ee
while for the $\Gamma$-state we find 
\bea\label{MGamma}
\begin{array}{c}
\vec{M}_{(2/3,0,0)}\!=\!(i M_a A, \,iM_b C,\, M_c F),\\
\vec{M}_{(0,0,0)}\!=\! (0, \,0,\, M'_c  C),
\end{array}
\eea
with
\be\begin{array}{c}
M_a\!=\!\sqrt{3}w(x_3\!+\!y_3),~ M_b\!=\!-\sqrt{3} w z_3,\\
M_c\!=\!w(x_3\!-\!y_3\!+\!\sqrt{2}),~ M_c'\!=\!-2(x_3\!-\!y_3\!-\!\frac{1}{\sqrt{2}})~,
\end{array}
\ee
and $w\!=\!e^{i\frac{\pi}{3}}$. 
So we find that both $K$- and $\Gamma$-states contain two Fourier components, one at ${\bf Q}=(2/3,0,0)$ and another at ${\bf Q}=0$. 
The latter which has not been observed so far in zero-field (see below), reflects the canting structure of the two states out of the perfect 120-degrees coplanar state. In particular, the amplitudes $M_a'$ and $M_b'$ of the $K$-state are proportional, respectively, to the in-plane zig-zag and out-of-plane FM canting of the $\{{\bf A},{\bf B},{\bf C}\}$ sublattices, see Eq.~(\ref{eq:ABCcanting}). Similarly, the amplitude $M_c'$ of the $\Gamma$-state tracks the in-plane stripy canting of the $\{{\bf F},{\bf G},{\bf D}\}$ sublattices along the {\bf c}-axis, see Eq.~(\ref{eq:FGDcanting}). As mentioned above then, the ${\bf Q}=0$ components of the structure are ramifications of the Heisenberg exchange coupling $J$ and vanish as we approach the line $\phi=3\pi/2$. 

Figure~\ref{fig:SofQcomponents} shows the evolution of the various components of the structure factor as we cross the boundary between the two phases, for three values of $\phi$ and varying $r$.
The three components corresponding to ${\bf Q}=(2/3,0,0)$, $M_a$, $M_b$ and $M_c$, change very slightly with $r$ and $\phi$. 
In particular the ratios between them is consistent with the reported relative ratios $M_{a}:M_{b}:M_{c}=0.45:0.56:1$ that give the best fit to the azimuthal  intensity dependence in Ref.~[\onlinecite{Biffin2014a}].

Turning to the components corresponding to ${\bf Q}=0$, these are generally much smaller than $M_a$, $M_b$ and $M_c$. In particular, they all tend to zero as we approach the line $\phi=3\pi/2$, and this is true irrespective of the value of $r$. This behavior reflects the small deviation of the magnetic structure from the ideal 120$^\circ$-pattern, discussed above. 

\vspace*{-0.3cm}
\subsection{Comparison to experiments}\label{sec:CompWithExp}
\vspace*{-0.3cm}
The ${\bf Q}=(2/3,0,0)$ components have all the qualitative features observed experimentally.~\cite{Biffin2014a} Indeed, defining $A_a\!=\!M_a A$, $C_b\!=\!M_b C$ and $F_c\!=\!M_c F$, the structure corresponding to $\vec{Q}=(2/3,0,0)$ transforms as ($iA_a$, $iC_b$, $F_c$), consistent with the irreducible representation $\Gamma_4$ found experimentally.~\cite{Biffin2014a}  This agreement gives strong support to the idea exploited here that the observed incommensurate order with ${\bf Q}=(0.57,0,0)$ must be some type of long-wavelength deformation of the ${\bf Q}=(2/3,0,0)$ order.  

Let us now turn to the ${\bf Q}=0$ components, which consist of a FM canting along ${\bf b}$ axis ($M_b'$) and a zig-zag canting along ${\bf a}$ axis ($M_a'$) for the $K$-state, or a stripy canting along ${\bf c}$ axis($M_c'$) for the $\Gamma$-state. 
First of all, the fact that the ${\bf Q}=0$ components were not seen in the zero-field scattering experiments of Ref.~[\onlinecite{Biffin2014a}] may well signify that the corresponding amplitudes are too weak to be observed, and that the system is close to the line $\phi=3\pi/2$ (i.e., $J$ is much weaker than both $\Gamma$ and $K$). 
On the other hand, the ${\bf Q}=0$ components found here for the $K$-state, i.e. the components $M_a'$ and $M_b'$, are precisely the ones reported in the more recent~\cite{Ruiz2017} scattering experiments under a magnetic field along the ${\bf b}$-axis. This agreement signifies that $\beta$-Li$_2$IrO$_3$ lies inside the $K$-region of Fig.~\ref{fig:LTphasediagram}.

The experiments of Ref.~[\onlinecite{Ruiz2017}] have in addition revealed that the ${\bf Q}=0$ components $M_a'$ and $M_b'$ grow very fast with the field, at the expense of the incommensurate, finite-${\bf Q}$ components $M_a$, $M_b$ and $M_c$, which decrease very fast with field. 
These findings can be explained by noting that a uniform magnetic field along the ${\bf b}$-axis couples linearly to both $M_a'$ and $M_b'$. The former proceeds via the off-diagonal element $g_{ab}$ of the ${\bf g}$-tensor, which is staggered between the primed and unprimed chains, while the coupling to $M_b'$ proceeds via the uniform diagonal element $g_{bb}$.~\cite{Ruiz2017} A detailed theoretical analysis of the behavior of the $K$-state in the magnetic field will be reported elsewhere.

 \begin{figure*}
\includegraphics[width=0.48\textwidth]{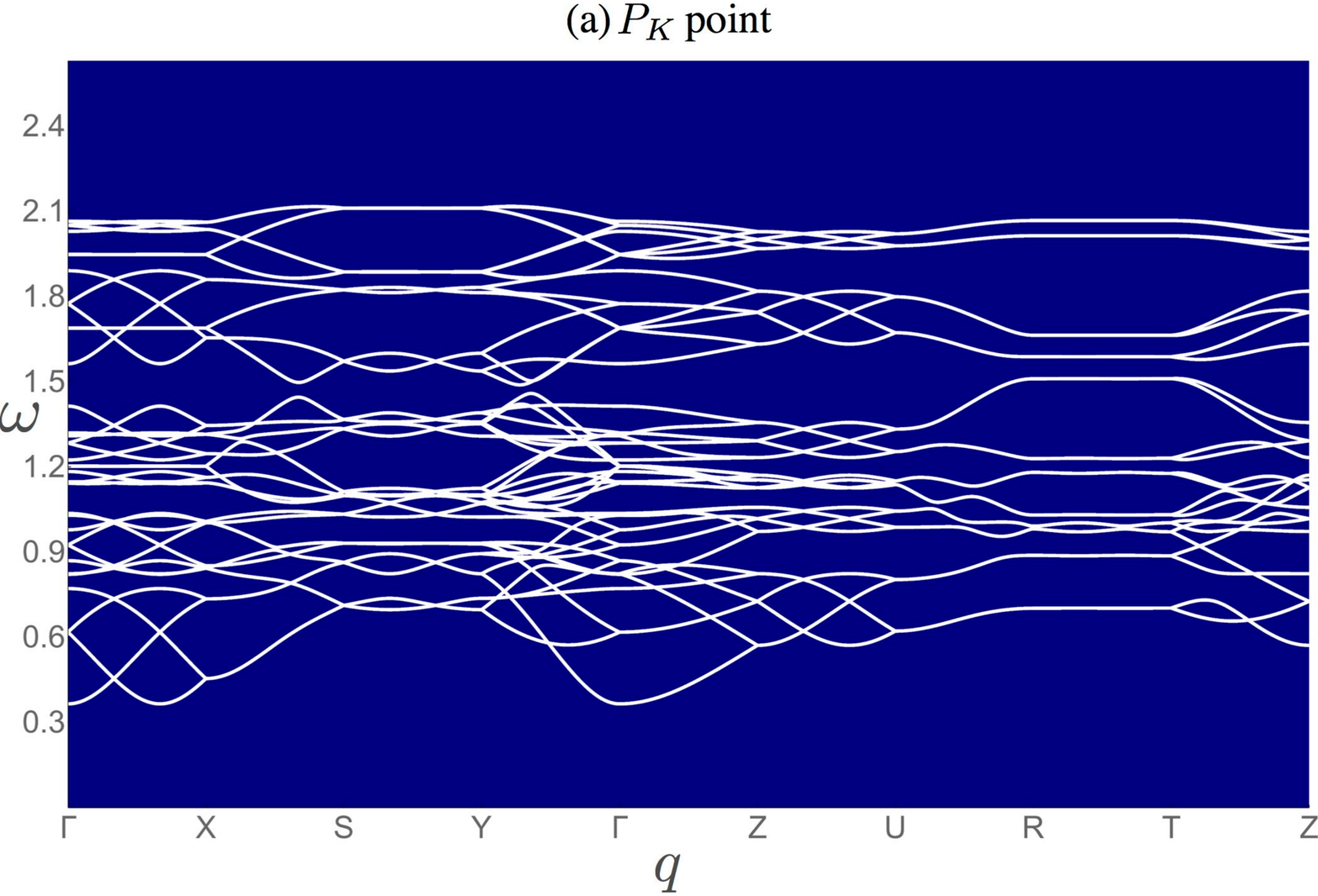}
~~
\includegraphics[width=0.48\textwidth]{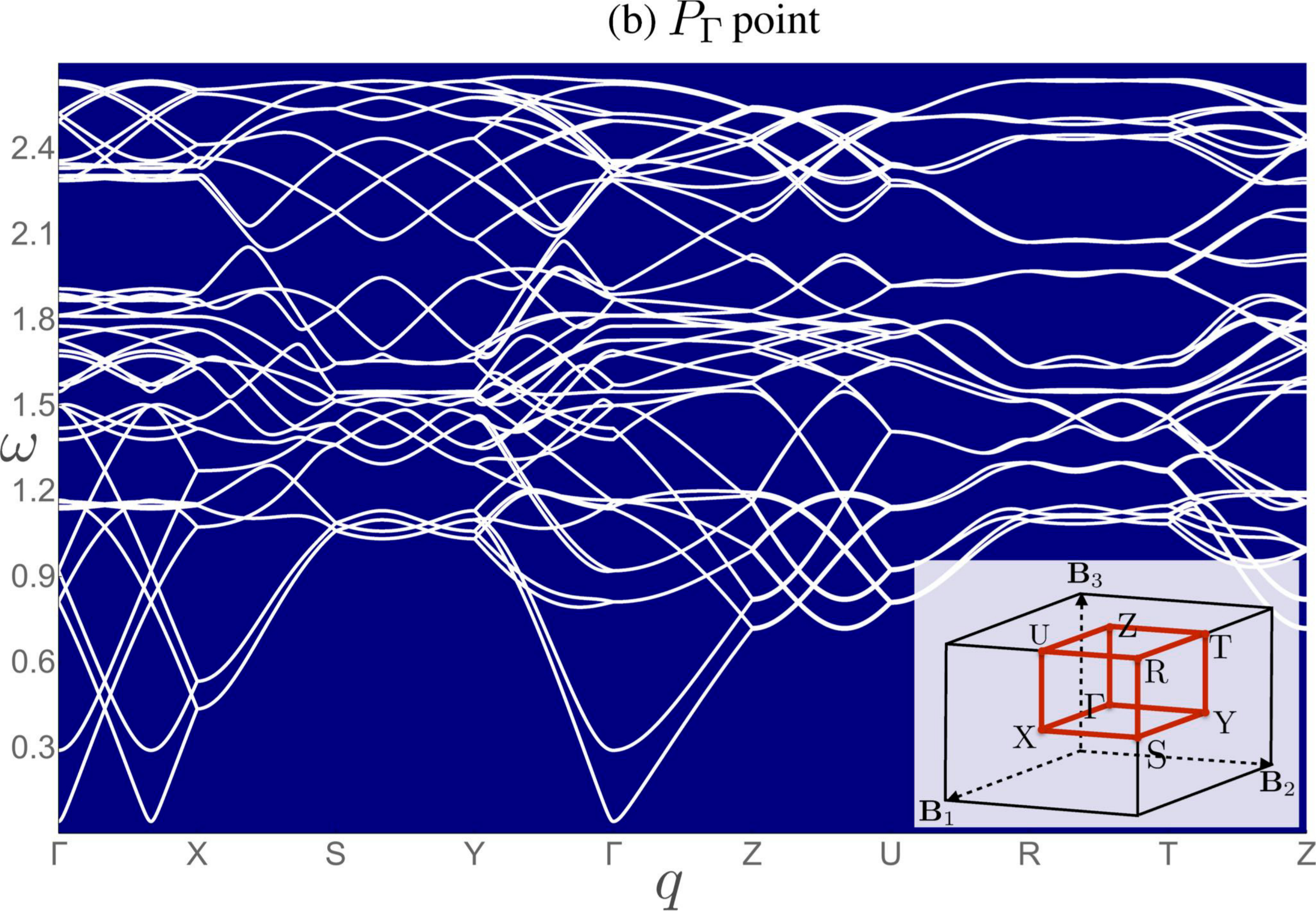}
\vspace*{-0.3cm}
\caption{The LSW spectra computed at  (a) $P_K$ and (b)  $P_{\Gamma}$ points  of Fig.~\ref{fig:LTphasediagram} are shown by white solid lines along the high-symmetry paths. The inset shows the magnetic Brillouin zone (in red) along with the high-symmetry points. The vectors ${\bf B}_1=\frac{2\pi}{a}\hat{{\bf a}}$, ${\bf B}_2=\frac{2\pi}{a}\hat{{\bf b}}$ and ${\bf B}_3=\frac{2\pi}{c}\hat{{\bf c}}$ are the reciprocal vectors of the orthorhombic BZ.}\label{fig:LSWspectra}
\end{figure*}

\vspace*{-0.3cm}
\section{Spin-wave spectra and dynamical spin structure factor}\label{sec:LSW}
\vspace*{-0.5cm}
We now turn to the semiclassical expansion around the above states and restore the spin length to $S\!=\!1/2$.
\vspace*{-0.5cm}
\subsection{Technical details of the semiclassical expansion}\label{sec:Technical}
\vspace*{-0.5cm}
The magnetic excitations and the dynamical spin structure factor for the states discussed above can be computed by employing the standard semiclassical Holstein-Primakoff expansion.~\cite{HP1940} To this end, we make use of an enlarged magnetic unit cell composed of three orthorhombic unit cells along the ${\bf a}$-axis, and thus contains 48 magnetic sites. The spins can then be labeled as $(i,\mu)$, where $i$ labels the enlarged magnetic unit cell and the index $\mu=1$-$48$ labels the spins inside that unit cell. To proceed we introduce local reference frames $(\tilde{{\bf x}},\tilde{{\bf y}},\tilde{{\bf z}})$, in such a way that the local $\tilde{{\bf z}}$ axis coincides with the corresponding direction of the given spin in the classical state around which we expand. 
The components of the spin in the laboratory frame $({\bf a},{\bf b},{\bf c})$ are given by:
\bea\label{rotation}
\begin{array}{l}
S^{a}_{i,\mu}= c_{\theta_{\mu}}c_{\varphi_{\mu}}\,S^{\tilde x}_{i,\mu} - s_{\varphi_{\mu}} S^{\tilde y}_{i,\mu} +s_{\theta_{i,\mu}}c_{\varphi_{\mu}} \,S^{\tilde z}_{i,\mu}, \\
S^{b}_{i,\mu}= c_{\theta_{\mu}} s_{\varphi_{\mu}} \,S^{\tilde x}_{i,\mu} + c_{\varphi_{\mu}}\,S^{\tilde y}_{i,\mu} + s_{\theta_{\mu}}s_{\varphi_{\mu}}\,S^{\tilde z}_{i,\mu},\\
S^{c}_{i,\mu}= -s_{\theta_{\mu}}\, S^{\tilde x}_{i,\mu}+c_{\theta_{\mu}} S^{\tilde z}_{i,\mu},
\end{array}
\eea
where $c_{\varphi}\!\equiv\!\cos\varphi$ and $s_{\varphi}\!\equiv\!\sin\varphi$. Next, we perform the standard Holstein-Primakoff transformation to lowest order~\cite{HP1940}
\bea\label{transf}
\begin{array}{l}
S^{\tilde x}_{i,\mu}\simeq \sqrt{S/2} (a_{i,\mu}+a^\dagger_{i,\mu}),  \\
S^{\tilde y}_{i,\mu}\simeq-i \sqrt{S/2}(a_{i,\mu}-a^\dagger_{i,\mu}), \\ 
S^{\tilde z}_{i,\mu}=S-a^\dagger_{i,\mu} a_{i,\mu},
\end{array}
\eea
where $S=1/2$. We then go into momentum space with 
\be
a_{i,\mu}=\frac{1}{\sqrt{N_m}}\sum_{\bf q} e^{i {\bf q }\cdot{\bf r}_{i,\mu}} a_{\mu,{\bf q}}~,
\ee
where $N_m=N/48$ is the number of magnetic unit cells ($N$ is the total number of sites), ${\bf q}$ belongs to the magnetic BZ, and the position ${\bf r}_{i,\mu}={\bf r}_{i}+{\bf d}_{\mu}$, where ${\bf r}_{i}$ is the origin of the magnetic unit cell and ${\bf d}_{\mu}$ is the position of the sublattice spin $\mu$ inside that unit cell.  Of course, ${\bf r}_{i,\mu}$ can be equivalently rewritten in terms of the vectors ${\bf R}$, $\boldsymbol{\rho_p}$ and $\vec{r}_n$ discussed in the previous section, but here it is more convenient to use the representation in terms of the magnetic BZ.

Replacing in the original spin-Hamiltonian and collecting the quadratic boson terms gives the spin-wave Hamiltonian
\be\label{H2}
H_2 = \frac{S}{2} \sum_{\bf q} x^\dagger_{\bf q} \cdot H_{\bf q} \cdot x_{\bf q}~,
\ee
where the vector 
\be
x_{\bf q} = (a_{1,{\bf q}},\cdots,a_{48,{\bf q}}, a_{1,-{\bf q}}^\dagger, \cdots ,a_{48,-{\bf q}}^\dagger)^T~,
\ee
and the interaction matrix $H_{\bf q}$ has the general form
\be
H_{\bf q} =\left(
\begin{array}{cc}
 Q_{\bf q} & R_{\bf q}  \\ R^*_{-{\bf q} } & Q^*_{-{\bf q} } 
\end{array}
\right).
\ee
To diagonalize the Hamiltonian (\ref{H2}), we use the standard Bogoliubov transformation~\cite{Blaizot}
\bea \label{xq}
x_{\bf q}= \mathcal{T}_{\bf q} \cdot y_{\bf q},
\eea
where $ y_{\bf q}= (b_{1,{\bf q}},...,b_{48,{\bf q}}, b_{1,-{\bf q}}^\dagger, ...,b_{48,-{\bf q}}^\dagger)^T$ represents the vector of Bogoluibov quasiparticles and the transformation matrix $\mathcal{T}_{\bf q}$ takes the general form
\be\label{sq}
\mathcal{T}_{\bf q} =\left(
\begin{array}{cc}
 U_{\bf q} & V^*_{-{\bf q}} \\ V_{\bf q} & U^*_{-{\bf q}}
\end{array}
\right).
\ee
To preserve the bosonic commutation relations, $\mathcal{T}_{\bf q}$ must satisfy $\mathcal{T}^\dagger_{\bf q}\eta\mathcal{T}_{\bf q}=\eta$, where 
$\eta\!=\!\left(\!
\begin{array}{cc}
I & 0 \\ 0 & -I 
\end{array}
\!\right)$  
and $I$ is the $48\times48$ unit matrix.
With these conditions, the entries of the matrix in (\ref{sq}) can then constructed numerically from the eigenvectors of $\eta H_{\bf q}$.~\cite{Blaizot}
After diagonalization we get
\be\label{H2diagonal}
H_2 = \frac{S}{2} \sum_{\bf q} y^\dagger_{\bf q} \cdot \Omega_{\bf q} \cdot y_{\bf q},
\ee
where 
$
\Omega_{\bf q}\!=\!\mathcal{T}_{\bf q}^\dagger H_{\bf q} \mathcal{T}_{\bf q}\!=\!
\left(\!\begin{array}{cc}
    \omega_{\bf q} & 0 \\
    0 & -\omega_{\bf q}
    \end{array}\!\right)
$
and $\omega_{\bf q}$ is the diagonal matrix $\omega_{\bf q} = \mbox{diag}[\omega_{1,\bf q},\,\omega_{2,\bf q},\,...,\,\omega_{48,\bf q}]$.

\begin{figure*}[!t]
\includegraphics[width=0.33\textwidth]{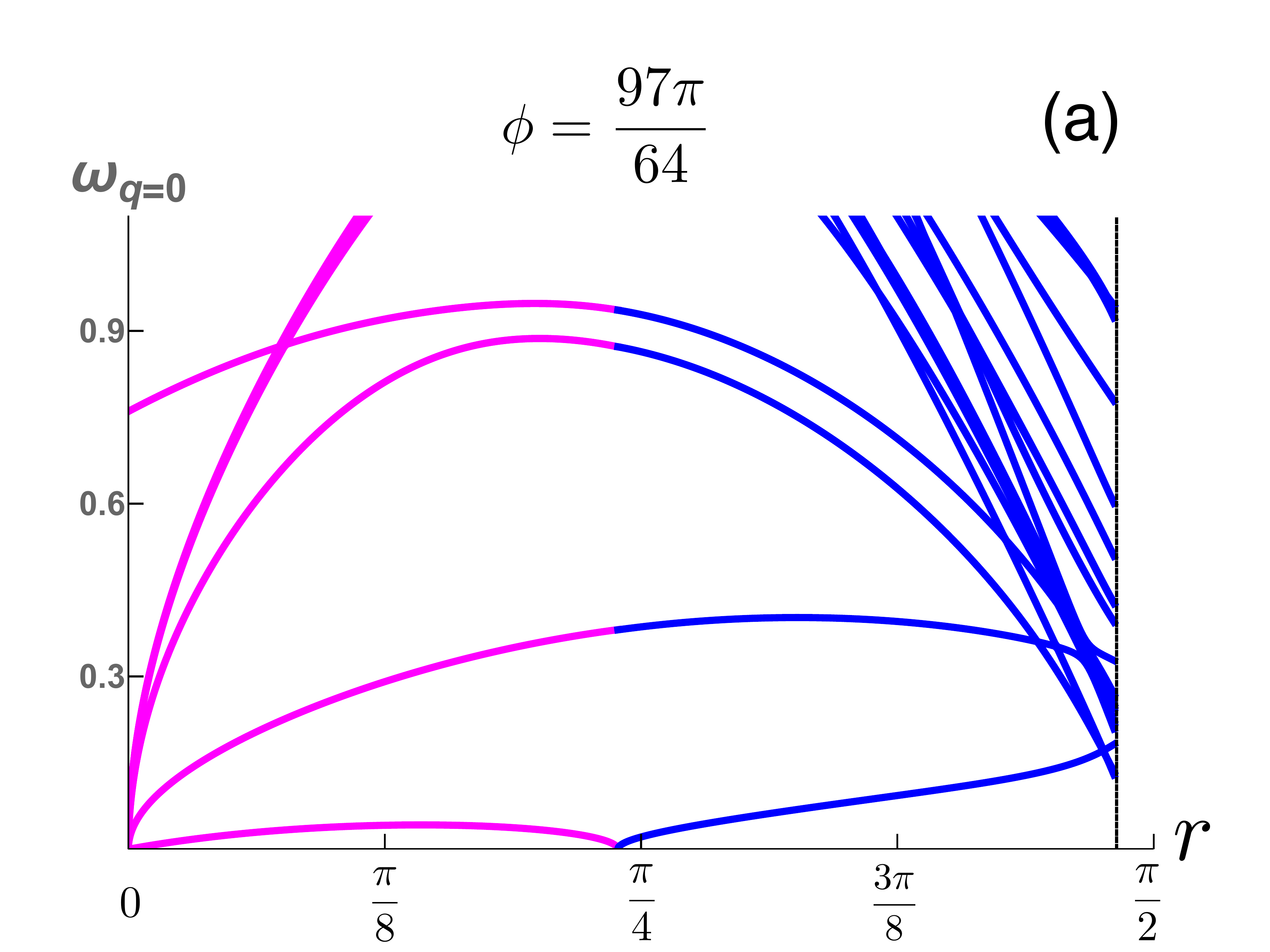}
\includegraphics[width=0.33\textwidth]{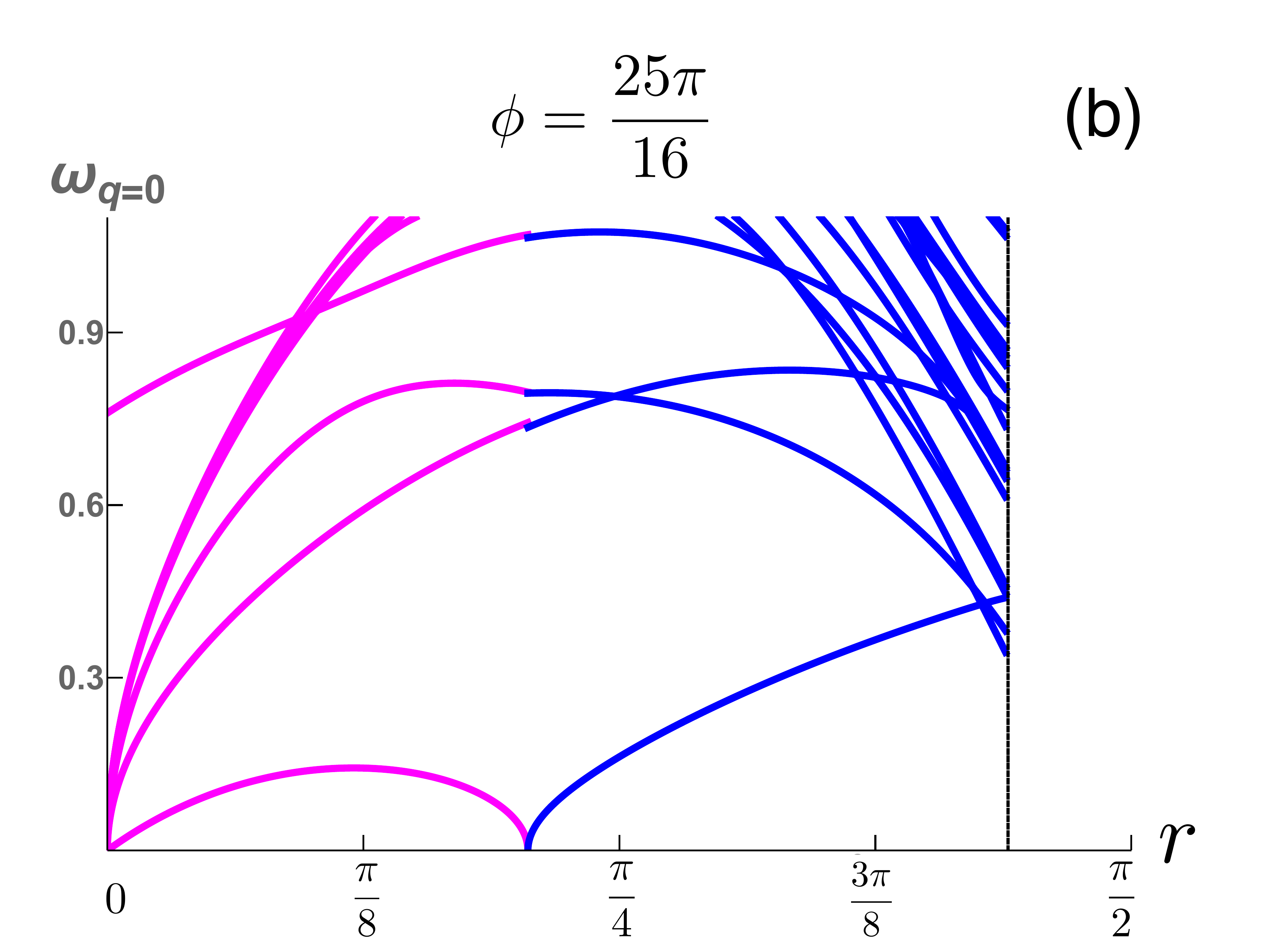}
\includegraphics[width=0.33\textwidth]{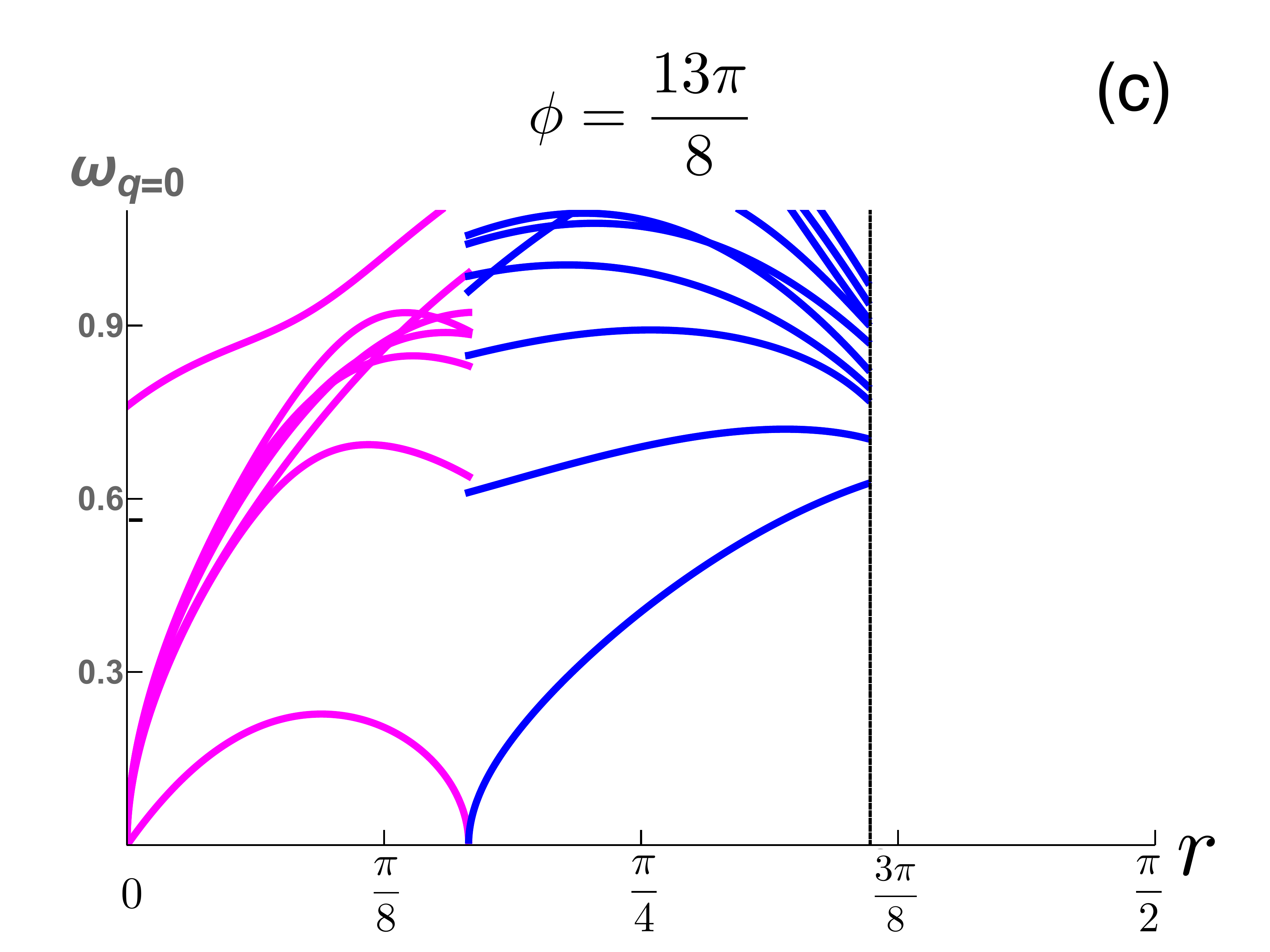}
\caption{
Lowest branches $\omega_{\nu,{\bf q}=0}$ as we change the parameter $r$ for (a) $\phi\!=\!1.515625\pi$, (b) $\phi\!=\!1.5625\pi$ and (c) $\phi\!=\!1.625\pi$. 
Magenta and blue lines show the branches inside the $\Gamma$- and $K$-regions, respectively.
Data are shown up to $r_c(\phi)$ where we exit the $K$-region, see Fig.~\ref{fig:LTphasediagram}.
}
\label{fig:branches}
\end{figure*}

\vspace*{-0.3cm}
\subsection{Spin-wave spectra}
\vspace*{-0.3cm}
Fig.~\ref{fig:LSWspectra} shows the computed magnon branches at the points $P_K$ (a) and $P_{\Gamma}$ (b) of Fig.~\ref{fig:LTphasediagram}. The dispersions are shown along a high-symmetry path within the magnetic Brillouin zone (inset). The LSW spectra for these parameter sets are significantly different. While the LSW spectra are gapped at both $P_K$ and $P_{\Gamma}$ points, the gap is significantly smaller at the latter, i.e. when the $\Gamma$-interaction is dominant.
The reason is that the $P_\Gamma$ point is very close to the line $\phi=3\pi/2$ of Fig.~\ref{fig:LTphasediagram}, along which the ground state of the model has infinite accidental degeneracy.
We should stress however that the difference between the magnon spectra above the $K$- and $\Gamma$-states weakens as the two parameter sets get closer to the boundary line between the two states, see for example the sets of panels (c,d) and (e,f) in Fig.~\ref{fig:intensity} below.
     
To study in more detail the dependence of the spin-wave gap on the parameters of the model,   we show in Fig.~\ref{fig:branches} the dependence of $\omega_{{\bf q}=0}$ on the parameter $r$, for several lowest branches computed for  (a) $\phi=\frac{97\pi}{64}$, (b) $\phi=\frac{25\pi}{16}$ and (c) $\phi=\frac{13\pi}{8}$.  The branches in the $K$-  and  $\Gamma$-states are shown by blue and purple  lines, respectively.   We can see that spin wave excitations are generically gapped, except at $r=0$ and at the boundary between  the $K$- and  $\Gamma$-states. %shown  in Fig.~\ref{fig:LTphasediagram}.
The $r=0$ point is a special  point corresponding to the  pure $\Gamma$-model. This model  is highly frustrated and the classical ground state is macroscopically degenerate.\cite{IoannisGamma} However, this degeneracy is accidental, and the spurious zero modes will be eventually gapped out by spin-wave interactions. 

The gapless excitations along the boundary between the $K$-  and  $\Gamma$-states are also due to accidental degeneracy between the $K$-  and $\Gamma$-states. This degeneracy will also be lifted by spin wave interactions.  It is only at the special SO(3) point, $(r,\phi)=(\pi/4,3\pi/2)$, where the gapless excitations are protected by symmetry. As we discussed above, at this point a 24-sublattice transformation maps the Hamiltonian to a fully SU(2) symmetric Heisenberg model.
We also note that at $\phi=\frac{97\pi}{64}$ the entire spectrum of the $K$-state becomes nearly identical with the spectrum of the $\Gamma$-state at the point where the two states become degenerate. This happens because $\phi=\frac{97\pi}{64}$ is close to the line $\phi=3\pi/2$, and the boundary point is in the vicinity of the SO(3) point. For larger values of $\phi$ [Figs.~\ref{fig:branches}~(b,c)], the two sets of excitations depart from each other, except for the lowest mode where an (accidental) degeneracy remains, as discussed above.

\vspace*{-0.5cm}
\subsection{Dynamical spin structure factor}\label{sec:LSW}
\vspace*{-0.5cm}

\begin{figure*}[!t]
\includegraphics[width=0.33\textwidth]{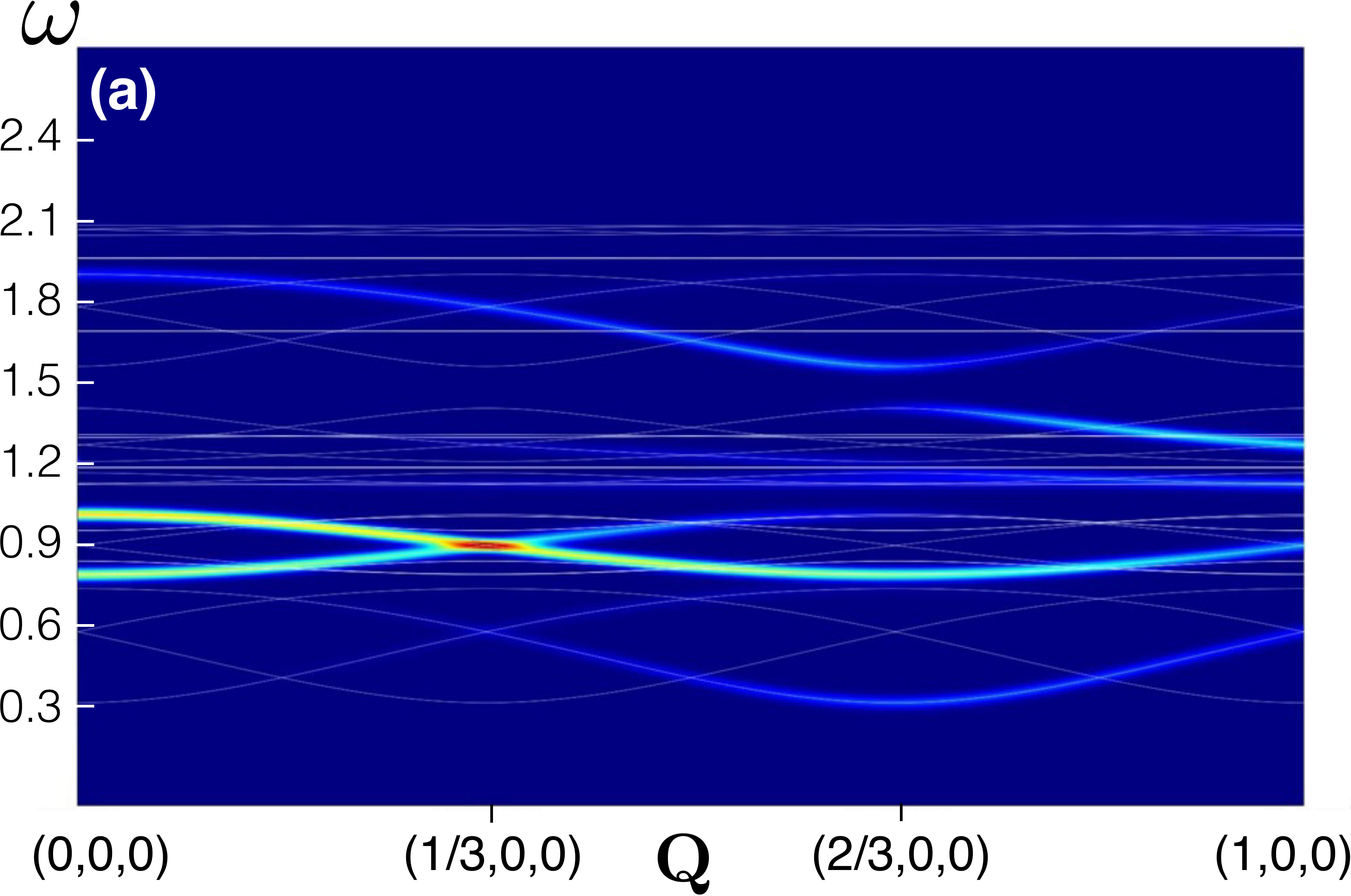}
\includegraphics[width=0.33\textwidth]{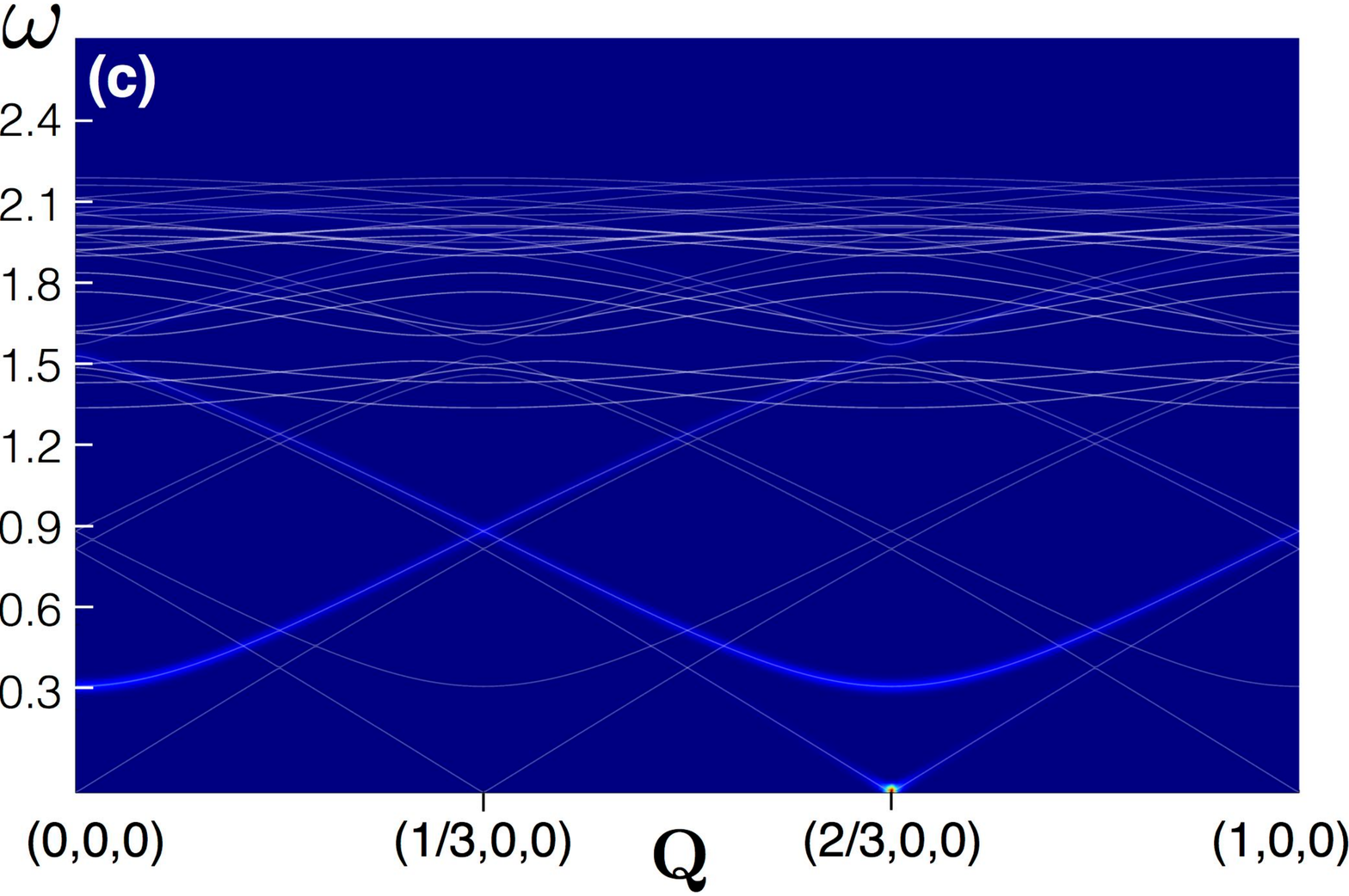}
\includegraphics[width=0.33\textwidth]{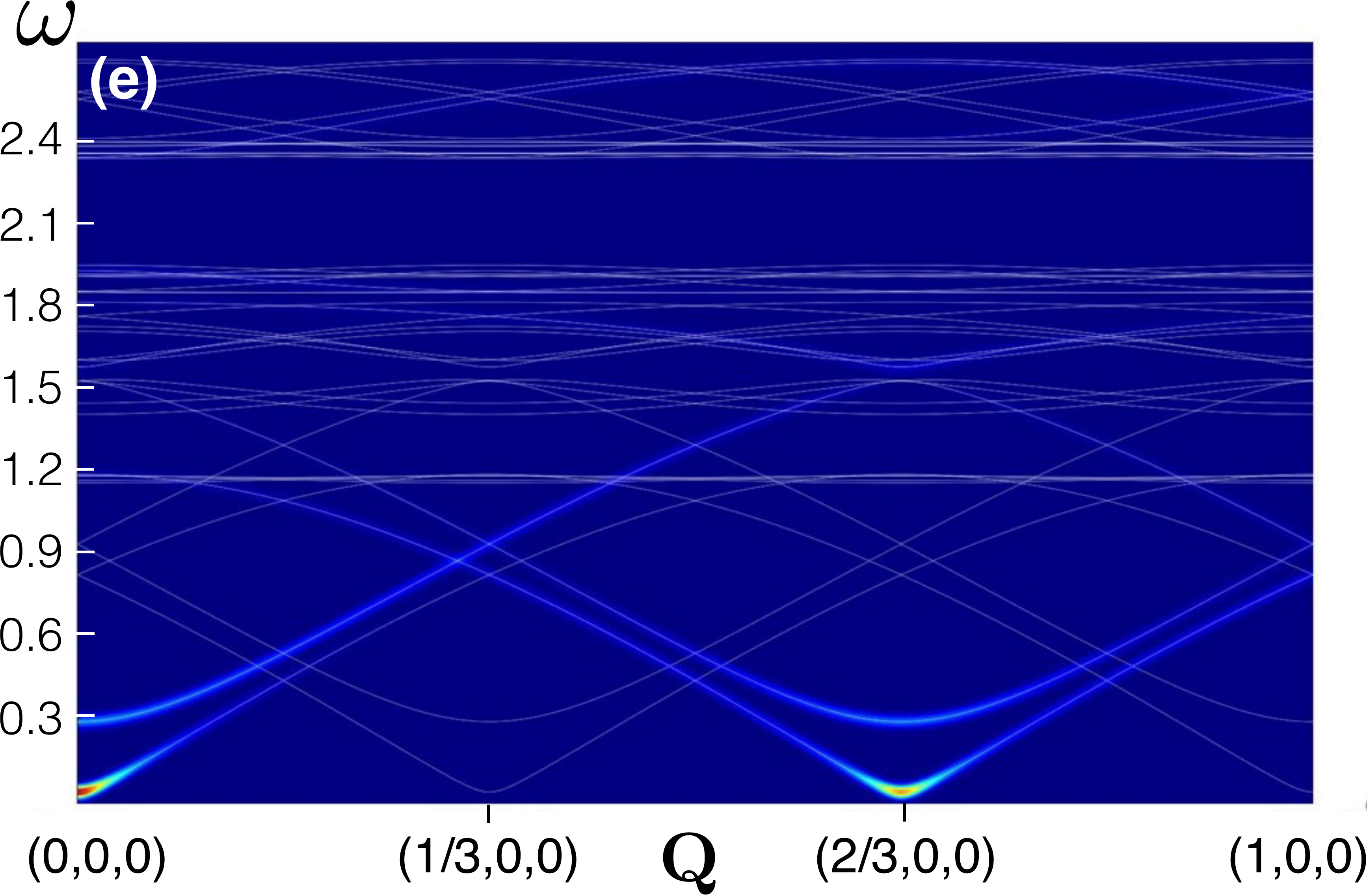}
\includegraphics[width=0.33\textwidth]{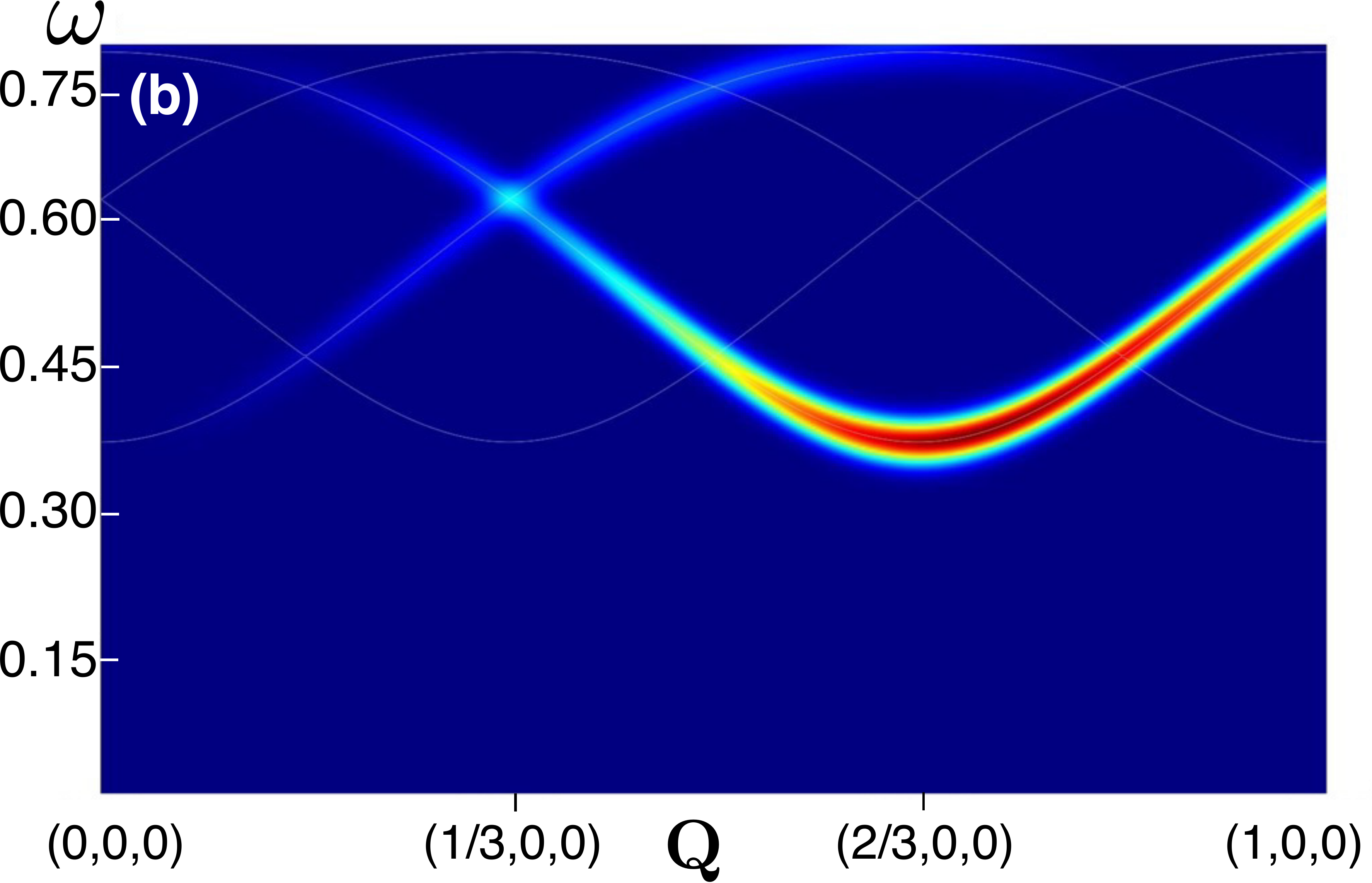}
\includegraphics[width=0.33\textwidth]{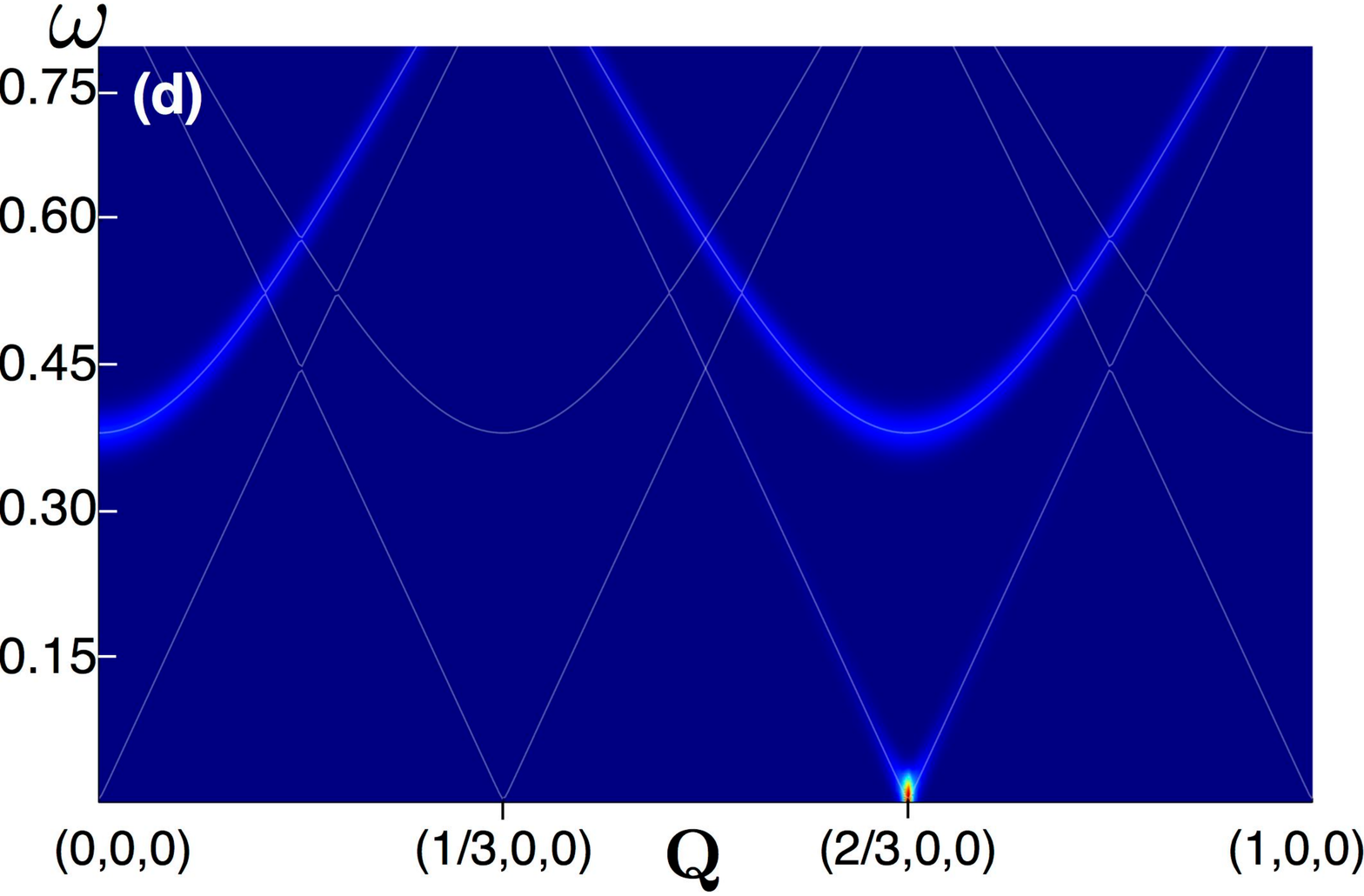}
\includegraphics[width=0.33\textwidth]{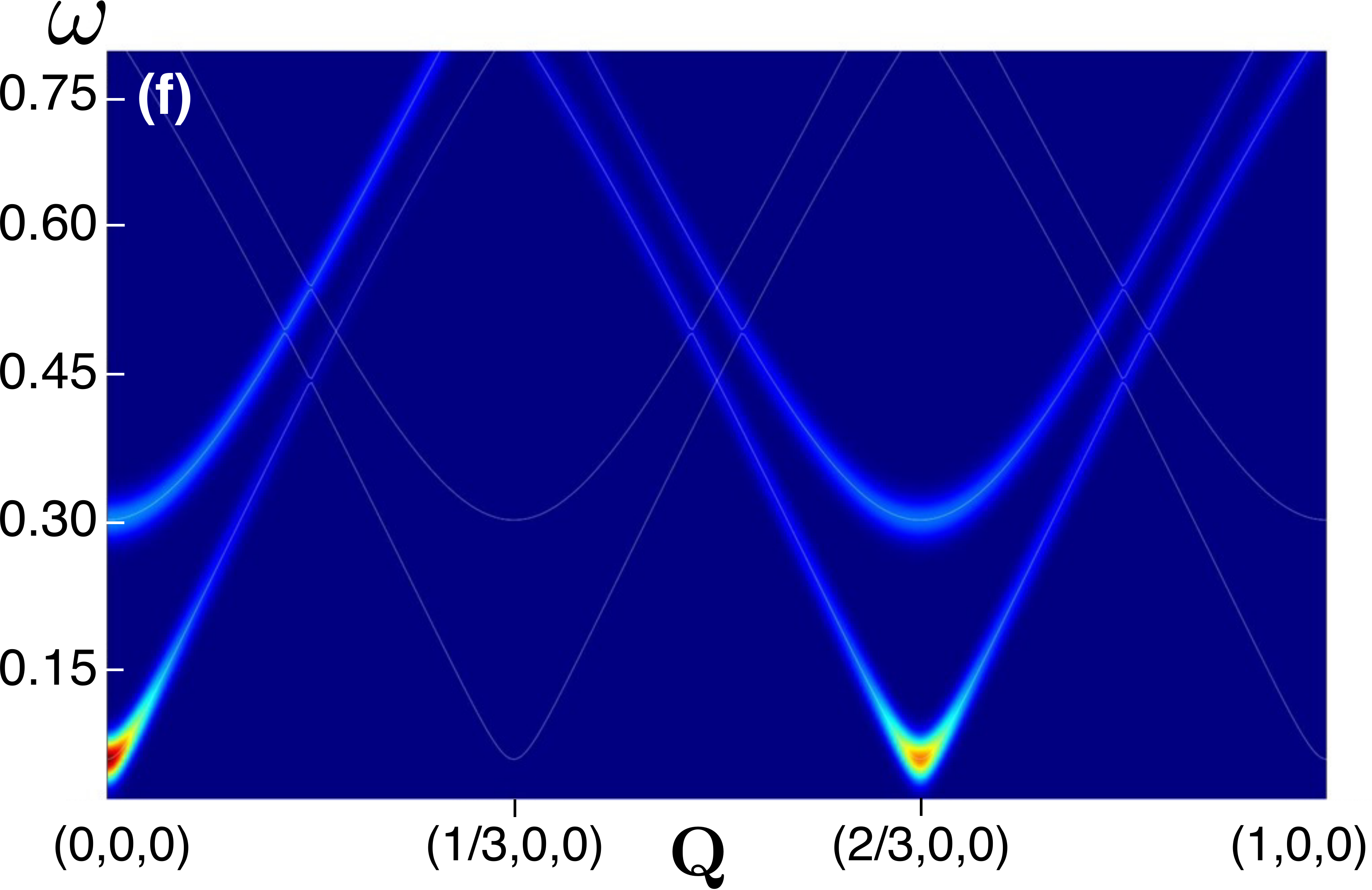}
\caption{Total intensity $I({\bf Q},\omega)$ shown along $\Gamma-X-\Gamma'$ direction of the orthorhombic BZ for:
(a-b) $(r,\phi)=(\frac{3\pi}{8},\frac{25\pi}{16})$, which is the $P_K$ point of Fig.~\ref{fig:LTphasediagram},
(c-d) $(r,\phi)=(0.239\pi,\frac{97\pi}{64})$, which lies inside the $K$-region of Fig.~\ref{fig:LTphasediagram}, but close to the boundary with the $\Gamma$-state,
and 
(e-f) $(r,\phi)=(\frac{\pi}{8},\frac{97\pi}{64})$, which is the $P_\Gamma$ point of Fig.~\ref{fig:LTphasediagram}.
The panels in the lower row show the intensity of the four lowest branches only.
The LSW spectra are shown by white solid
lines.
The colors and the width indicate the magnitude of the intensity after convolving the structure factor with a gaussian of finite width to emulate finite experimental resolution. The color scale  runs from `blue' color corresponding to the minimum to  `red' color corresponding to the maximum of the intensity, and it is  independently normalized for each plot.}\label{fig:intensity}
\end{figure*}

 In this section, we  evaluate the inelastic neutron-scattering cross-section or, equivalently, the intensity $\mathcal{I}({\bf Q}, \omega)$, where $\omega$ is the energy transfer, $\bf{Q}={\bf k_i}-{\bf k_f}$ is the wavevector transfer, and ${\bf k_i}$ and ${\bf k_f}$ are the momenta of the incident and scattered neutron, respectively. The intensity is given by 
\bea\label{intensity}
\begin{array}{l}
I ({\bf Q}, \omega) \!\propto\!  \sum_{\alpha, \beta} (\delta_{\alpha, \beta} -\frac{ Q^{\alpha} Q^{\beta}}{Q^2})S^{\alpha\beta}  ({\bf Q}, \omega),
\end{array}
\eea
where $\alpha,\beta$ run over the orthorhombic axes ${\bf a}$, ${\bf b}$ and ${\bf c}$, and ${\mathcal S}({\bf Q}, \omega)$ is the dynamical  structure factor given by
\be\label{Structurefactor}
\begin{array}{l}
{\mathcal S}^{\alpha \beta}({\bf Q}, \omega) \!=\!\sum_{\mu,\mu'}\!\! \int \frac{dt}{2\pi} e^{i \omega t} \langle S^{\alpha}_{\mu'}(-{\bf Q},t) S^{\beta}_{\mu}({\bf Q},0)\rangle~.
\end{array}
\ee
Here $\mu,\mu'=1$-$48$ are the sublattice indices inside the magnetic unit cell, 
\be
\begin{array}{l}
S^{\beta}_{\mu}({\bf Q},0)=\frac{1}{\sqrt{N_m}}\sum_{i,\mu} e^{-i{\bf Q}\cdot{\bf r}_{i,\mu}} S^{\beta}_{\mu}({\bf r}_{i,\mu},0)~,
\end{array}
\ee
where ${\bf r}_{i,\mu}={\bf r}_i+d_\mu$ are the actual positions of the Ir ions. To proceed we write ${\bf Q}={\bf q}+{\boldsymbol\tau}$, where ${\bf q}$ belongs to the first magnetic BZ and ${\boldsymbol\tau}$ is a reciprocal lattice vector of the magnetic BZ.
The structure factor then reduces to 
\small
\be\label{Structurefactor1}
\mc{S}^{\alpha \beta} ({\bf Q}, \omega) \!=\!\sum_{\mu,\mu'} \!\!\int\! \frac{dt}{2\pi} e^{i \omega t} 
e^{-i{\boldsymbol \tau}\cdot{\bf d}_{\mu\mu '}}\langle S^{\alpha}_{\mu'}(-{\bf q},t) S^{\beta}_{\mu}({\bf q},0)\rangle ,
\ee
\normalsize
where ${\bf d}_{\mu\mu '}={\bf d}_\mu-{\bf d}_{\mu '}$. 
At zero temperature, this can be rewritten as 
\bea\label{G}
\begin{array}{c}
S^{\alpha\beta}  ({\bf Q}, \omega) =
 -\frac{1}{\pi} {\rm Im}[ \sum_{\mu \mu'}e^{-i{\boldsymbol \tau}\cdot{\bf d}_{\mu\mu '}}
 \sum_{\tilde{\alpha}\tilde{\beta}\nu}
 F^{\alpha{\tilde\alpha}}_{\mu'}F^{\beta{\tilde\beta}}_{\mu}
\\
\!\!\!\!\!\!\!\!\!\!\!\!
\times \frac{ \langle 0|S^{{\tilde \alpha}}_{{\mu}',{\bf q}} |\nu\rangle \langle \nu |S^{{\tilde\beta}}_{{\mu},{-\bf q}}|0\rangle}{\omega-\omega_{\nu,{\bf q}}+i\eta}],
\end{array}
\eea
where for each given sublattice the indices $\tilde{\alpha}$ and $\tilde{\beta}$ run over the corresponding local axes $\tilde{\bf x}$ and $\tilde{\bf y}$ (the components involving the $\tilde{\bf z}$ axis do not contribute to leading order), and $ F^{\alpha{\tilde\alpha}}_{\mu}$ are functions of $\theta_{\mu}$ and $\varphi _{\mu}$ defined in Eq.~(\ref{rotation}).
The state labeled by $|0\rangle$ is the vacuum of the Bogoliubov bosons, $|\nu\rangle$ are excited eigenstates of Eq.~(\ref{H2diagonal}), and $\omega_{\nu,{\bf q}}$ are the corresponding eigenenergies. 
The  matrix elements entering to Eq.~(\ref{G}) can be computed  using  the Bogoliubov transformation of Eq.~(\ref{xq}):
\small
\begin{eqnarray}
\begin{array}{l}
\langle 0|S^{\tilde x}_{\mu,{\bf q}}|\nu\rangle   =\sqrt{\frac{S}{2}} (U_{\mu \nu,{\bf q}}+V_{\mu \nu,{\bf q}}),\\
\langle 0|S^{\tilde y}_{\mu,{\bf q}}|\nu\rangle=-i \sqrt{\frac{S}{2}} (U_{\mu \nu,{\bf q}}-V_{\mu \nu,{\bf q}}).
\end{array}
\end{eqnarray}
\normalsize
Figure~\ref{fig:intensity} shows the calculated scattering intensities along the direction $\Gamma-X-\Gamma'$ of the orthorhombic BZ (i.e. for ${\bf Q}\parallel {\bf a}$), for three different points in parameter space: 
The points $P_K$ and $P_\Gamma$ [panels (a,b) and (e,f), respectively], and a point intermediate between the two [panels (c,d)], which lies inside the $K$-region but closer to the boundary line than $P_K$. 

For the $P_K$ point [Fig.~\ref{fig:intensity} (a,b)], the maximum intensity is observed at high energies and around the wavevector ${\bf Q}=(\frac{1}{3},0,0)$.  
At lower energies, most of the intensity is concentrated around the momentum ${\bf Q}=(\frac{2}{3},0,0)$, describing the modulation of the dominant component of the magnetic order, see Fig.~\ref{fig:intensity}~(b). This is also true for the intermediate point that is closer to the boundary line [panels (c,d)]. 
The intensity of the associated soft modes at ${\bf Q}=0$ [panels (a-c) and (d-f)] are much weaker. 
Generally, this is consistent with the fact that the quantities $|M_a'|^2$ and $|M_b'|^2$ are much smaller than $|M_a|^2$, $|M_b|^2$ and $|M_c|^2$, for almost all values of parameters inside the $K$-region (see Fig.~\ref{fig:SofQcomponents}). 
Turning to the results at the $P_\Gamma$ point [panels (e,f)], the maximum of the intensity is observed at the low-energy modes at ${\bf Q}=(\frac{2}{3},0,0)$ and at ${\bf Q}=0$. Note that, unlike the $K$-region, here the intensity of the ${\bf Q}=0$ soft mode is comparable to that of the ${\bf Q}=(\frac{2}{3},0,0)$ soft mode, despite the fact that $|M_c'|^2$ is much smaller than $|M_a|^2$, $|M_b|^2$ and $|M_c|^2$ (see Fig.~\ref{fig:SofQcomponents}). 

\begin{figure*}[!t]
\includegraphics[width=0.33\textwidth]{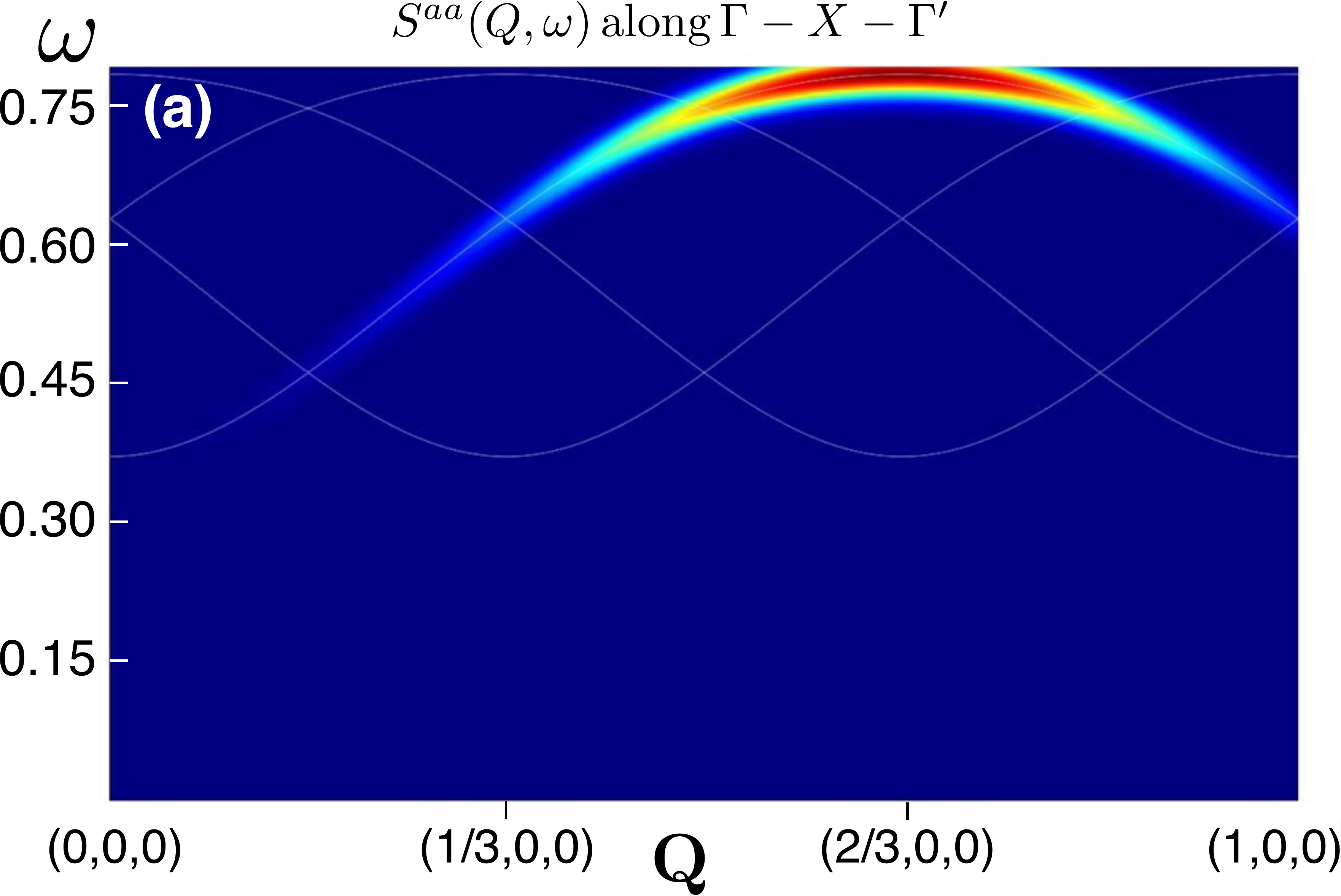}
\includegraphics[width=0.33\textwidth]{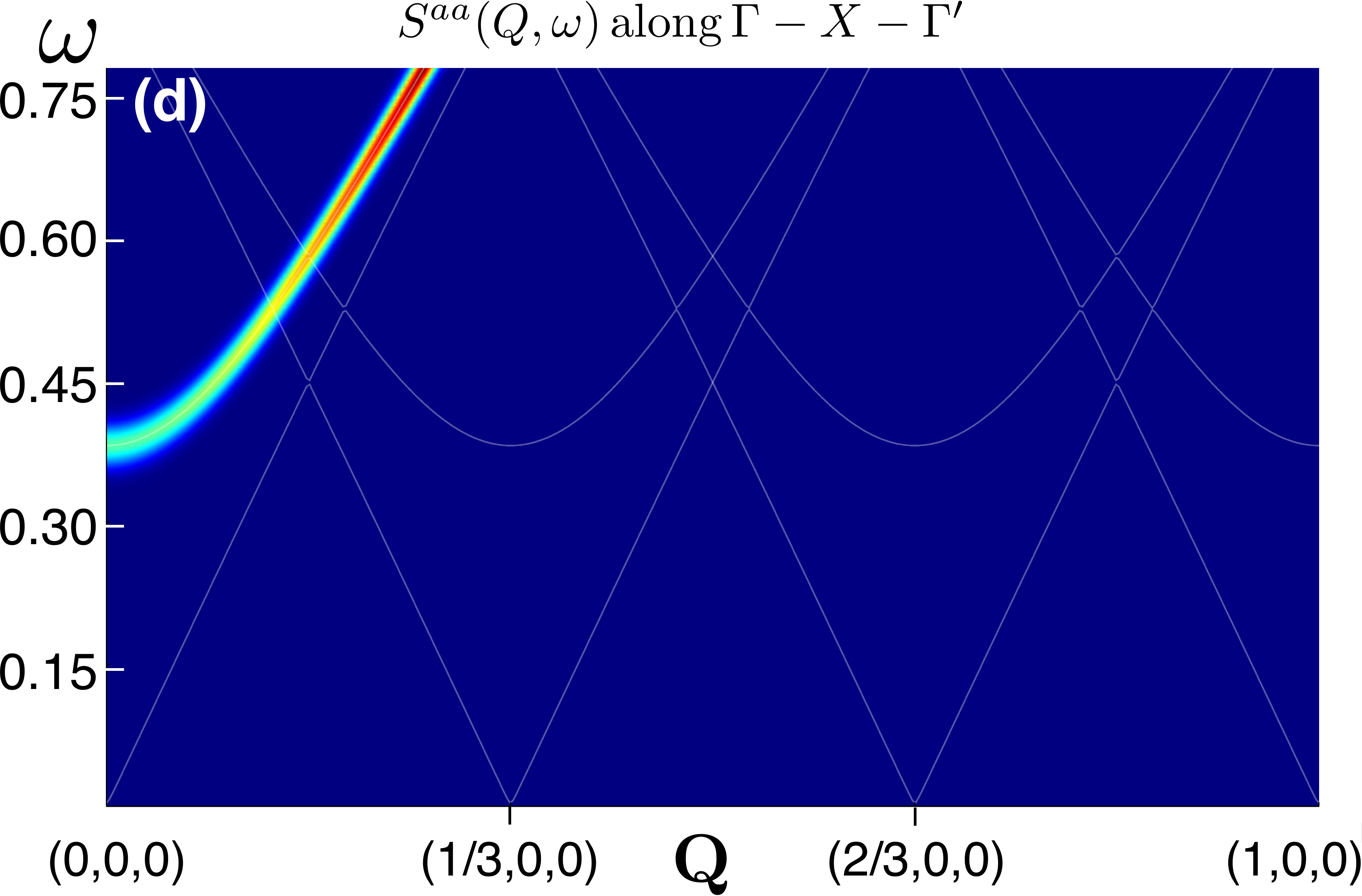}
\includegraphics[width=0.33\textwidth]{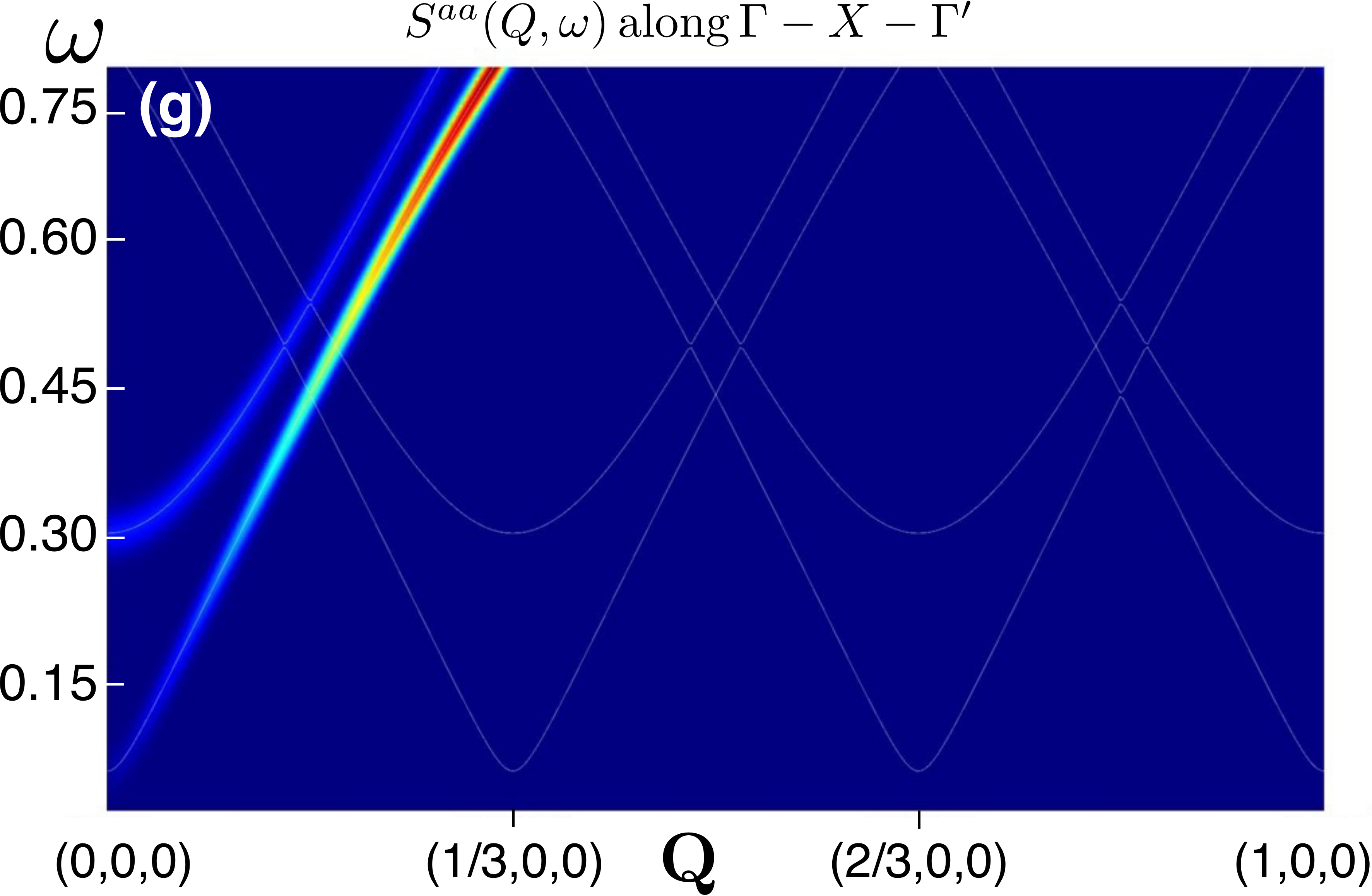}
\includegraphics[width=0.33\textwidth]{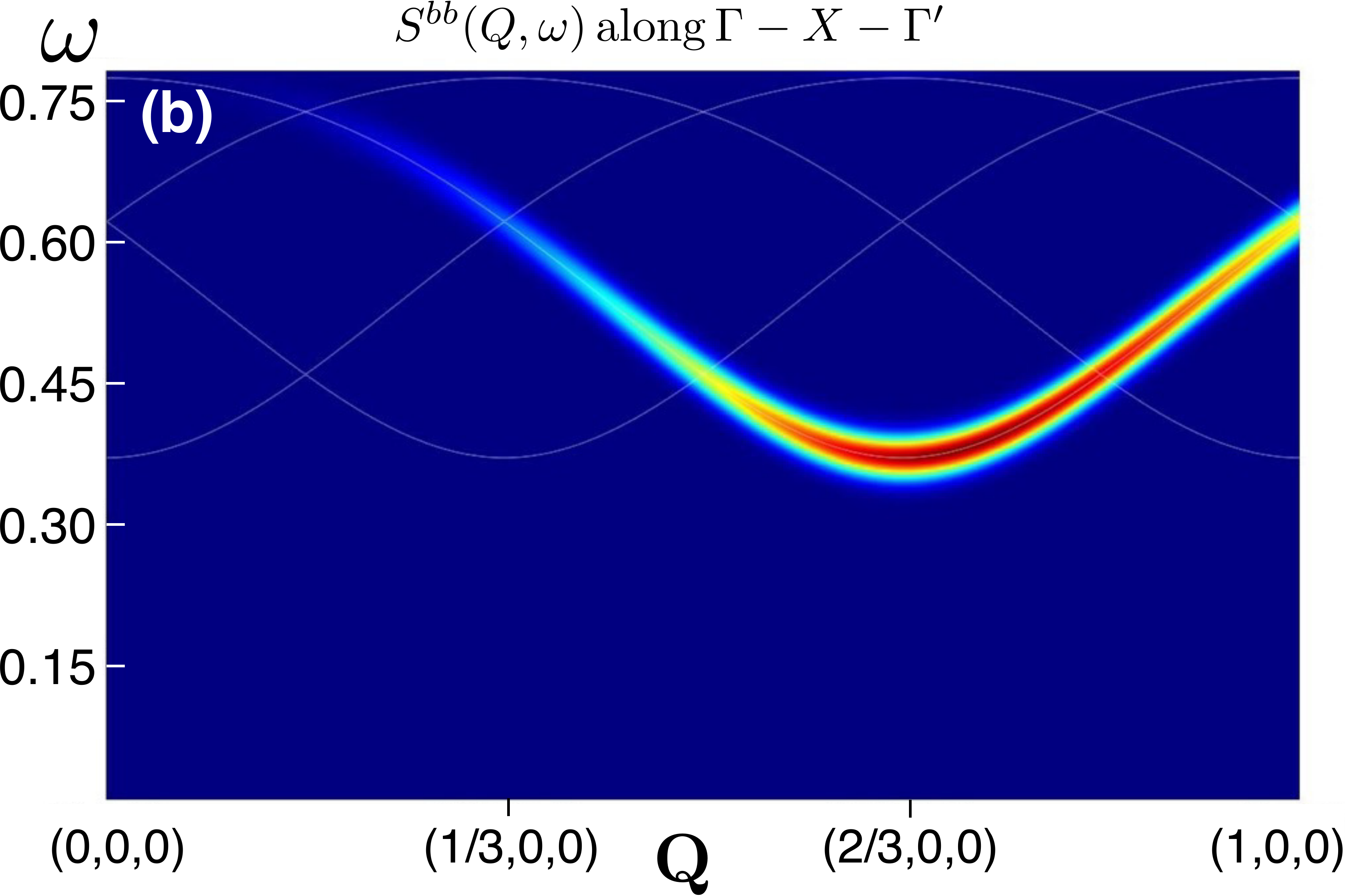}
\includegraphics[width=0.33\textwidth]{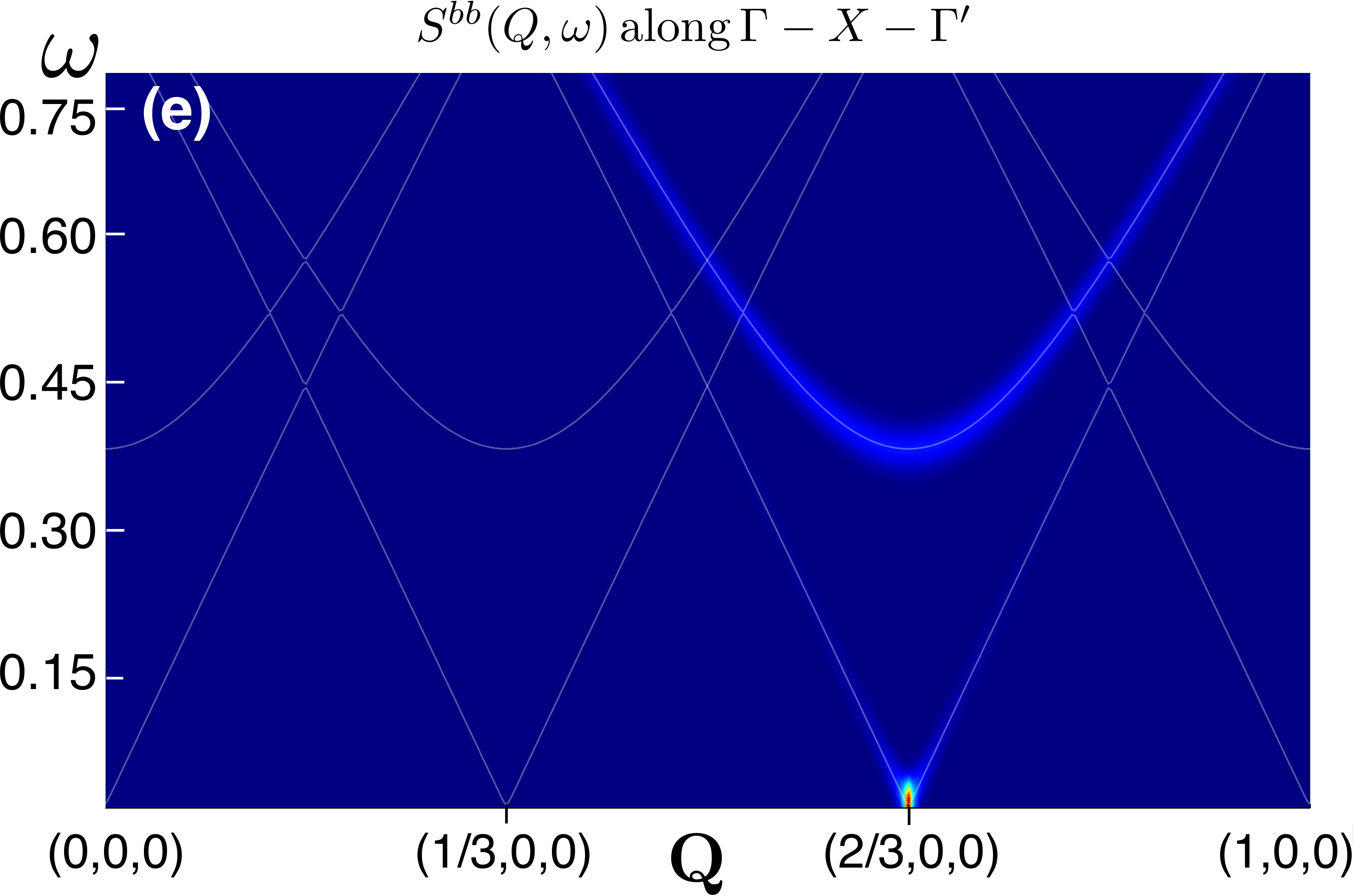}
\includegraphics[width=0.33\textwidth]{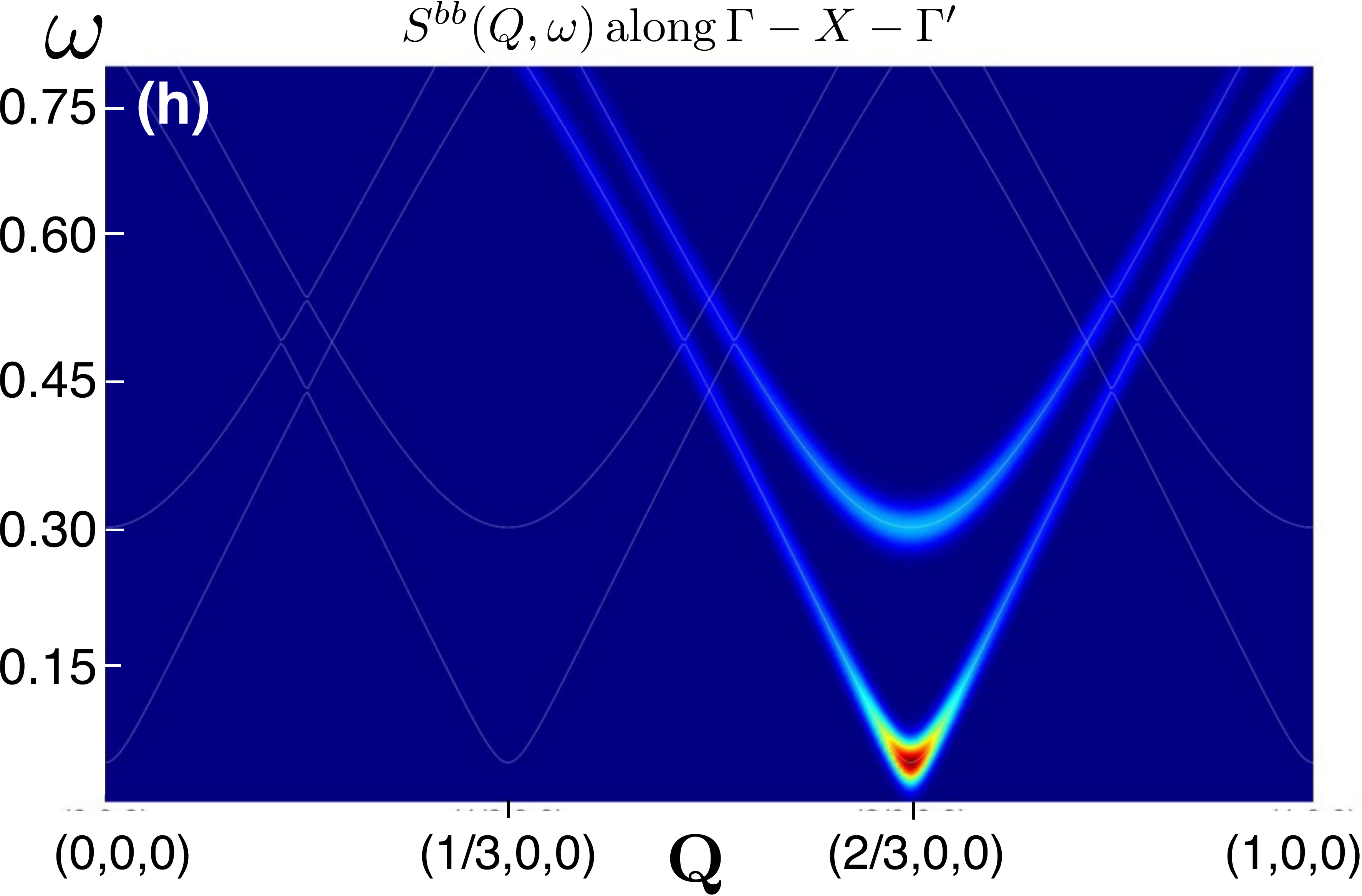}
\includegraphics[width=0.33\textwidth]{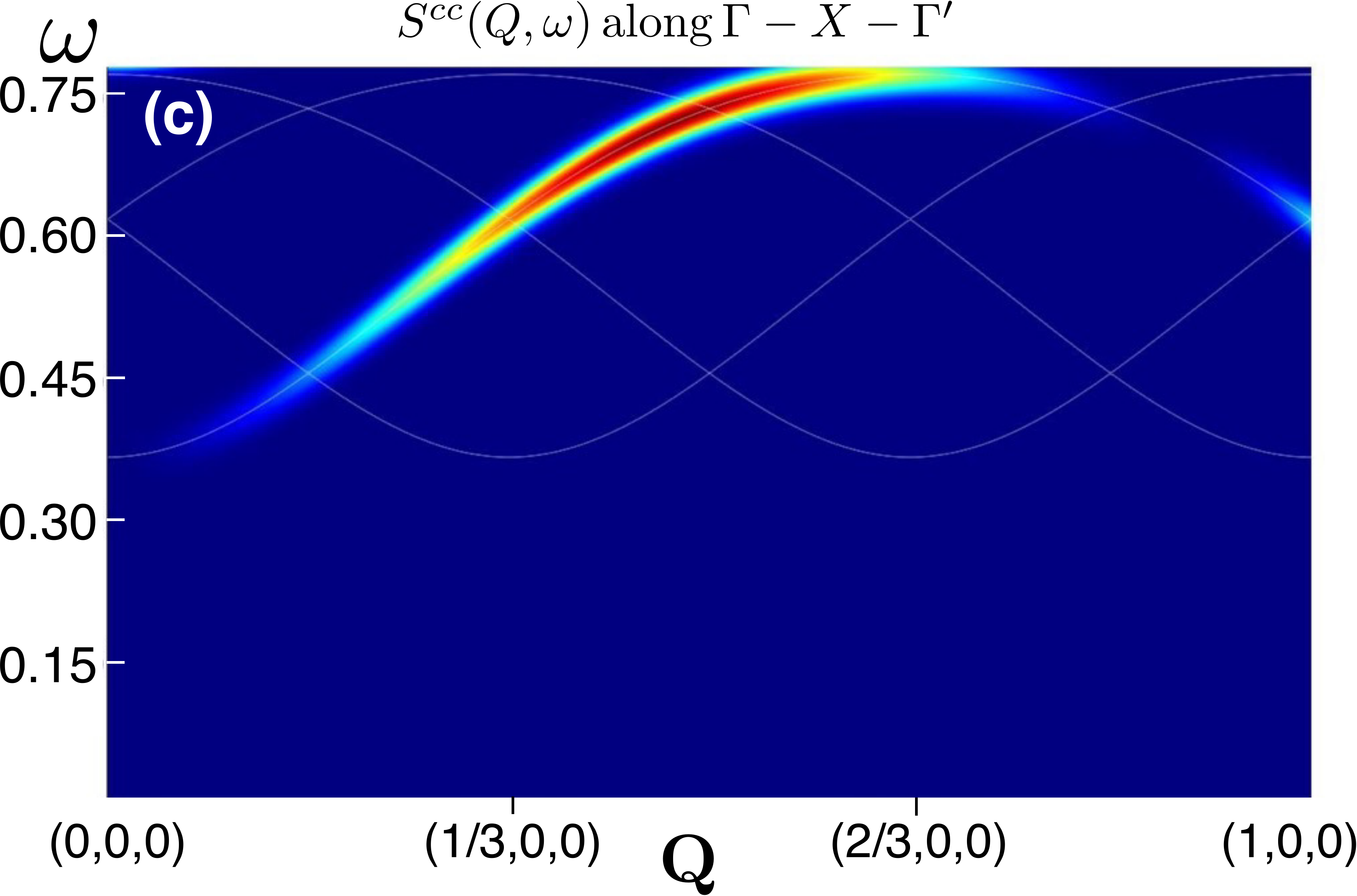}
\includegraphics[width=0.33\textwidth]{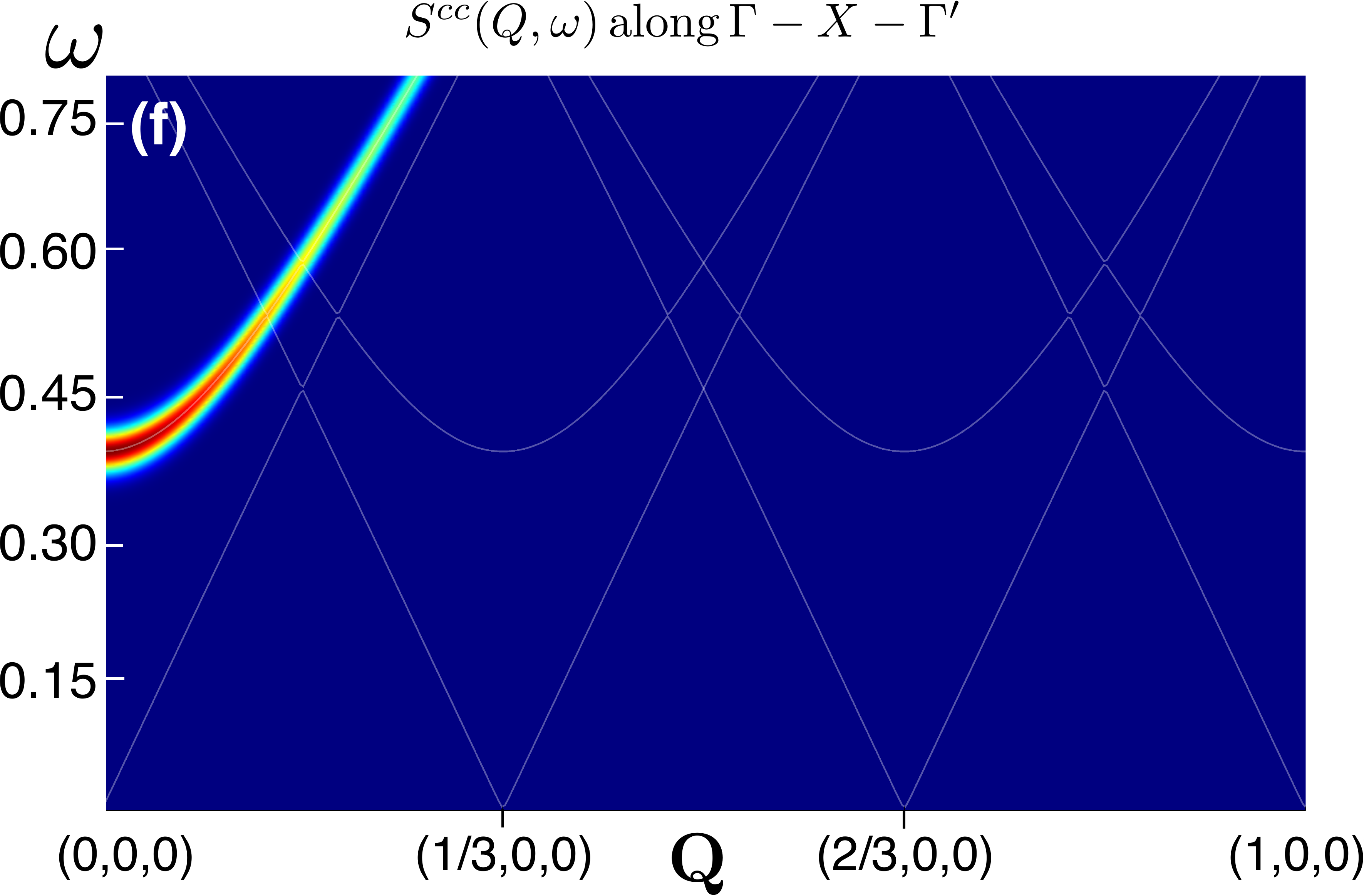}
\includegraphics[width=0.33\textwidth]{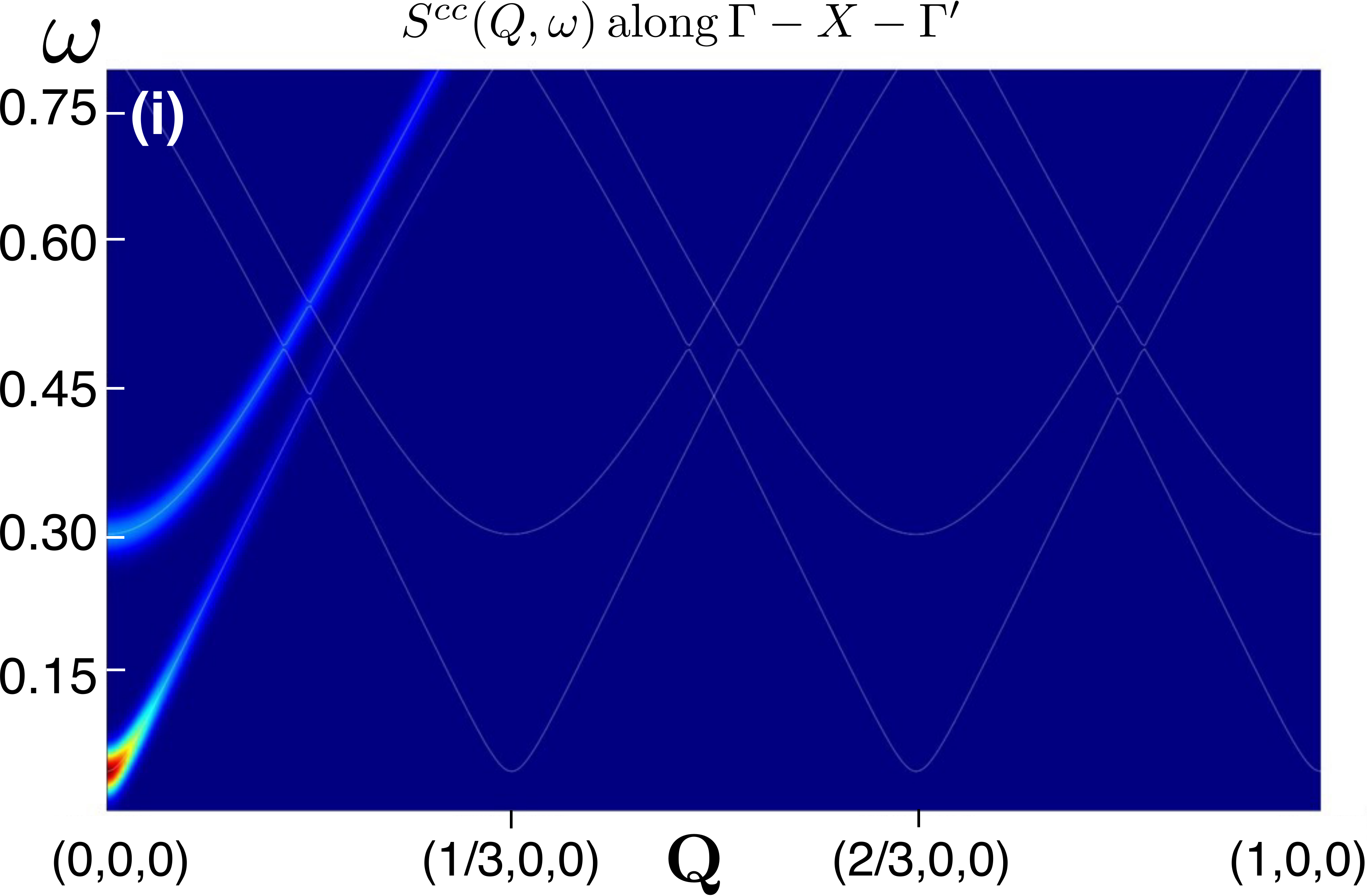}
\caption{The diagonal components of the dynamical spin structure factor $\mc{S}^{aa}({\bf Q}, \omega)$, $\mc{S}^{bb}({\bf Q}, \omega)$ and $\mc{S}^{cc}({\bf Q}, \omega)$, along the $\Gamma-X-\Gamma'$ direction of the orthorhombic BZ, for:
(a, b, c) $(r,\phi)=(\frac{3\pi}{8},\frac{25\pi}{16})$, which is the $P_K$ point of Fig.~\ref{fig:LTphasediagram},
(d, e, f) $(r,\phi)=(0.239\pi,\frac{97\pi}{64})$, which lies inside the $K$-region of Fig.~\ref{fig:LTphasediagram}, but close to the boundary with the $\Gamma$-state,
and 
(g, h, i) $(r,\phi)=(\frac{\pi}{8},\frac{97\pi}{64})$, which is the $P_\Gamma$ point of Fig.~\ref{fig:LTphasediagram}.
 The LSW spectra are shown by white solid
lines.
The colors and the width indicate the magnitude of the associated component after convolving with a gaussian of finite width to emulate finite experimental resolution. The color scale  runs from ``blue" color corresponding to the minimum to  ``red" color corresponding to the maximum of the intensity, and it is  independently normalized for each plot.}\label{fig:polarization}
\end{figure*}

Next, we analyze the polarization dependence of the intensity of the low-energy modes, by plotting the individual components of the dynamical spin structure factor. Fig.~\ref{fig:polarization} shows the calculated  diagonal components, $\mc{S}^{aa}({\bf Q}, \omega)$, $\mc{S}^{bb}({\bf Q}, \omega)$ and $\mc{S}^{cc}  ({\bf Q}, \omega)$, along the direction $\Gamma$-$X$-$\Gamma'$ of the orthorhombic BZ. The off-diagonal components are non-zero (they are subdominant  to the diagonal ones), but we do not show them here because they do not contribute along the direction $\Gamma$-$X$-$\Gamma'$ (where ${\bf Q}\parallel{\bf a}$) due to the vanishing geometrical prefactor $(\delta_{\alpha\beta}$-$\frac{ Q^{\alpha} Q^{\beta}}{Q^2})$ in Eq.~(\ref{intensity}). 
The latter also vanishes for $\mc{S}^{aa}({\bf Q}, \omega)$, so we will focus on the $\mc{S}^{bb}$ and $\mc{S}^{cc}$ channels only. 

In all three parameter points considered in Fig.~\ref{fig:polarization}, the main contribution to the intensity of the ${\bf Q}=(\frac{2}{3},0,0)$ soft mode comes from the $\mc{S}^{bb}$ channel. 
On the contrary,  the intensity of the ${\bf Q}=0$ soft mode of the $\Gamma$-state [panels (g-i)] comes from the $S^{cc}$ channel. 
Given that the ${\bf Q}=0$ component of the static structure factor of the $\Gamma$ state involves a stripy canting along the ${\bf c}$-axis, it follows that the strong intensity of the ${\bf Q}=0$ soft mode implies that the longitudinal modulation of this canting has a large amplitude. This is not however a consequence of a nearby instability toward a stripy phase, because there is no such phase nearby in the phase diagram.~\cite{Lee2015}

\vspace*{-0.3cm} 
\section{Discussion}\label{sec:concl}
\vspace*{-0.3cm}
We have revisited the microscopic $J$-$K$-$\Gamma$ description of $\beta$-Li$_2$IrO$_3$ and have argued that the observed~\cite{Takayama2015,Biffin2014a,Ruiz2017} incommensurate magnetic order can be understood in terms of a long-wavelength deformation of the closest commensurate, period-3 orders in the parameter space. 
The basic working hypothesis of our approach is that irrespective of the details of the actual deformation that takes place at long distances, the period-3 orders should share the same physics at short distances and the same excitation spectrum with the actual incommensurate order above some small energy cutoff.

A comparison of the resulting picture with reported experiment gives strong support to this hypothesis. First, the period-3 states reported here share the same irreducible representation, propagation vector direction and counter-rotation of the moments with the observed incommensurate phase. 
Second, the detailed structure of the $K$-state and its characteristic symmetry properties of the ${\bf Q}=0$ component of the static structure factor is in {\it full agreement} with the ${\bf Q}=0$ Bragg peaks observed in recent scattering experiments under magnetic fields along the ${\bf b}$-axis.~\cite{Ruiz2017}
This shows in particular that $\beta$-Li$_2$IrO$_3$ lies inside the $K$-region of Fig.~\ref{fig:LTphasediagram}, i.e., that $K$ is the dominant interaction, in agreement with {\it ab initio} calculations.~\cite{KimKimKee2016,Katukuri2016}  
Third, a detailed analysis of the magnetization process along the ${\bf b}$-axis, based on the present work (to be presented elsewhere), explains naturally the intensity sum rule observed in Ref.~[\onlinecite{Ruiz2017}] 
Finally, the fact that the uniform ${\bf Q}=0$ components of the structure factor have not been observed in the zero-field scattering experiments of Ref.~[\onlinecite{Biffin2014a}] is in line with $J$ being much weaker than both $K$ and $\Gamma$. 

The distinctive features of the period-3 states reported here (and especially the $K$-state) can be further checked experimentally by local probes such as NMR or $\mu$SR. 
In anticipation of future dedicated INS and RIXS studies on $\beta$-Li$_2$IrO$_3$, we have also provided detailed predictions for the associated spin gaps, the dynamical spin structure factors and INS intensities, which as mentioned above, should follow closely the response of the actual incommensurate order above a small energy cutoff. These predictions can be contrasted with the dynamical response of the counter-rotating spiral of the idealized single-chain model of Ref.~[\onlinecite{Kimchi2015}], and with the response of the exactly solvable Kitaev spin liquid on the hyperhoneycomb lattice.~\cite{Smith2015,Smith2016,RIXSNatalia}

We should further point out that the semiclassical picture presented here should remain qualitatively valid in the fully quantum-mechanical limit, except around the Kitaev point $(r,\phi)\!=\!(\pi/2,3\pi/2)$ and a pocket around the point $r\!=\!0$, where $K$ and $J$ vanish. As argued in Ref.~[\onlinecite{IoannisGamma}], the infinite classical degeneracy of this latter point is eventually lifted by quantum fluctuations, which tend to stabilize a multi-sublattice magnetically ordered state (different from the $\Gamma$-state presented here). However, the associated order-by-disorder energy scale is a very small fraction of $\Gamma$, signifying that the small pocket around $r=0$ will show a correlated {\it classical} spin liquid behavior down to very low temperatures.
This physics appears to become relevant in several experiments under pressure.~\cite{Takayama2015,Breznay2017,Haskel2017,Tsirlin2018} 

Returning to our semiclassical picture, it is noteworthy that both $K$- and $\Gamma$-states can be understood in terms of a simplified, single-chain Hamiltonian, which can be  also identified in the corresponding $J$-$K$-$\Gamma$ model in the 2D honeycomb lattice. This reflects that the physics in the associated parameter regime has universal features. This is exemplified by the universal structure of the classical ground state manifold along the special line $J=0$ (Sec.~\ref{sec:SpecialLine}), which seems to play a central role in several compounds. 
Moreover, the simplicity of the single-chain Hamiltonian (Sec.~\ref{sec:Hc}) suggests a possible route to study the nature of the long-distance deformation of the above commensurate orders and understand e.g. whether this deformation proceeds via domain-wall or soliton-like `discommensurations'.~\cite{DesGennes1975,McMillan1976,Bak1982,Schaub1985,Abrikosov1957,Wright89,Bogdanov1989,Roessler2006,Z2Ioannis2016}
The single-chain Hamiltonian may also allow to deduce in a more tractable way the structure of the actual phase diagram in the relevant regime of interest, and clarify e.g. whether this regime consists of a single phase or a non-trivial cascade of first-order transitions between a multitude of different phases. 
These questions call for further dedicated theoretical studies.

\vspace*{-0.3cm}
\section{Acknowledgments}
\vspace*{-0.3cm}
We are especially grateful to Radu Coldea for several fruitful comments and suggestions and for pointing out the close connection of our results to the recent scattering experiments in finite fields.~\cite{Ruiz2017} We are also grateful to Cristian Batista, Eunsong Choi, Pavel Maksimov and Yuriy Sizyuk for valuable discussions. The authors acknowledge the support from DOE grant 00061911.

\appendix
\vspace*{-0.3cm}
\section{Luttinger-Tisza analysis of the $J$-$K$-$\Gamma$ model}\label{App:A}
\vspace*{-0.3cm}
Here we present the Luttinger-Tisza minimization approach~\cite{LT,Bertaut,Litvin,Kaplan} of the $J$-$K$-$\Gamma$ model, Eq.~(\ref{Ham1}). To this end, we rewrite $\mc{H}$ in the more compact form, 
\begin{equation}\label{Ham2}
\mc{H} =\sum_{ij}\sum_{\mu\nu}\sum_{\alpha\beta} S_{i,\mu}^\alpha H_{i\mu,j\nu}^{\alpha \beta} S_{j,\nu}^\beta ,
\end{equation}
where now the indices $i$ and $j$ label the primitive positions $\vec{R}_i$ and $\vec{R}_j$ of the orthorhombic cells, $\mu,\nu\!=\!1$-$16$ are the sublattice indices inside the orthorhombic cells, and $\alpha,\beta\!=\!x,y,z$. We then switch to momentum space using
\be\begin{array}{l}
S_{i,\mu}^\alpha = \sum_{\vec{Q}}e^{i\vec{Q}\cdot \vec{R}_i}S_{\vec{Q},\mu}^\alpha~,\\
\\
H_{\mu\nu}^{\alpha\beta}(\vec{Q}) = \frac{1}{N_{uc}}\sum_{ij}e^{i\vec{Q}\cdot (\vec{R}_i-\vec{R}_j)}H_{i\mu,j\nu}^{\alpha,\beta}~,
\end{array}
\ee
where the wavevectors $\bf Q$ belong  to the orthorhombic BZ, and $N_{uc}=\frac{N}{16}$ is the number of unit cells ($N$ is the total number of sites). 
The classical energy per site $\epsilon=E/N$ then becomes
\begin{equation}\label{energy}
\epsilon= \frac{1}{16}\sum_{\bf Q}  \sum_{\alpha\beta}\sum_{\mu\nu} S^\alpha_{\mathbf{Q}, \mu} H^{\alpha \beta}_{\mu\nu}({\mathbf{Q}}) S^\beta_{-\mathbf{Q},\nu}.
\end{equation}
The classical ground states minimize $\epsilon$ under the strong spin length constraints, 
\be\label{eq:strong}
\vec{S}_{i,\mu}^2 \!=\!S^2, ~\text{for all}~(i,\mu)~.
\ee 
The Luttinger-Tisza (LT) approach~\cite{LT,Bertaut,Litvin,Kaplan} amounts to replacing these $N$ constraints with a weaker one,
\be\label{eq:soft}
\sum\nolimits_{i,\mu} \vec{S}_{i,\mu}^2 \!=\! N S^2,~~\text{or}~~
\sum\nolimits_{\vec{Q},\mu} \vec{S}_{\vec{Q},\mu}\cdot \vec{S}_{-\vec{Q},\mu} \!=\!16S^2~.
\ee 
Now, let $\{\lambda_\eta(\vec{Q}), \vec{V}_\eta(\vec{Q})\}$, $\eta=1$-$16$, be the set of eigenvalues (ordered such that $\lambda_1\le\lambda_2\le \cdots\lambda_{16}$) and orthogonalized eigenvectors of the matrix ${H}(\vec{Q})$.
Any spin configuration can be expanded in terms of these orthogonal vectors
\be
S_{\vec{Q},\mu}^\alpha = \sum\nolimits_\eta c_{\vec{Q},\eta} V_{\eta,\mu}^\alpha(\vec{Q})~,
\ee
where $c_{\vec{Q},\eta}$ are complex numbers. The weak constraint (\ref{eq:soft}) and the energy per site become
\be
\sum\nolimits_\eta |c_{\vec{Q},\eta}|^2 = 16S^2,~~
\epsilon = \frac{1}{16}\sum\nolimits_{\vec{Q},\eta} \lambda_{\eta}(\vec{Q}) |c_{\vec{Q},\eta}|^2 ~.
\ee
From these relations it follows that in order to saturate the energy minimum we should use a finite value only for the coefficient $c_{\vec{Q}_{\text{min}},\eta=1}$ that corresponds to the lowest eigenvalue $\lambda_1(\vec{Q}_{\text{min}})\equiv\lambda_{\text{min}}$ over the entire BZ. 
The resulting energy from the LT approach is equal to 
\be
\epsilon^{\text{LT}}_{\text{min}} = \lambda_{\text{min}} S^2.
\ee
If the spin configuration corresponding to the associated eigenstate $\vec{V}_{\text{min}}=\vec{V}_1(\vec{Q}_{\text{min}})$ happens to satisfy also the strong constraints (\ref{eq:strong}) then this configuration will be one of the true ground states of the problem.~\cite{LT,Bertaut,Litvin,Kaplan}

Essentially, the LT method corresponds to minimizing the energy over the restricted family of homogeneous states, i.e. states characterized by the the same value of the local mean field exerted at every site. Therefore, this method cannot capture inhomogeneous states with more than one local mean fields, and in particular states described by non-linear incommensurate modulations described by a large number of harmonics ${\mathbf Q}$. For such states, the minimum energy $\epsilon_{\text{min}}^{\text{LT}}$ delivered by the LT approach serves only as a lower energy bound, while the corresponding LT wavevectors may provide useful insights for the actual modulation of the spin structure.

\vspace*{-0.3cm}
\section{Classical ground state from the relaxation dynamics simulations}\label{App:B}
\vspace*{-0.3cm}
Another efficient approach to obtain the classical ground states is  via the so-called overdamped dynamics simulations based on the Landau-Lifshitz-Gilbert (LLG) equations.\cite{Serpico2001,Mochizuki2010,Choi2013,ChoiPRB2013} Since  the LT solution obtained in App.~\ref{App:A} already provides a close approximation to the true classical ground state, such relaxation simulations initiated from the LT state can potentially bring the system to the true ground state.

The LLG equation can be written in the following form:
\begin{equation}\label{LLG}
\frac{\partial \mathbf{S}_{i,\mu}}{\partial t}= \mathbf{S}_{i,\mu} \times (\mathbf{h}_{i,\mu} +   \frac{\alpha_G}{S} \frac{\partial \mathbf{S}_{i,\mu}}{\partial t}),
\end{equation}
where $\bf{ h}_{i,\mu}$ is the effective exchange field given by 
\begin{equation}\label{field}
h_{i,\mu}^{\alpha}=\frac{\partial H}{\partial {S}^\alpha_{i,\mu}} =
\sum_{j,\nu} \sum_{ \beta} H_{i\mu,j\nu}^{\alpha \beta} S_{j,\nu}^\beta~,
\end{equation}
and $\alpha_G$ is a dimensionless damping parameter. The LLG  equations can be integrated numerically by adopting the finite-difference method of Serpico {\it et al}, see Refs. [\onlinecite{Serpico2001}] and [\onlinecite{ChoiPRB2013}].

\begin{figure*}[!t]
\includegraphics[width=0.75\textwidth]{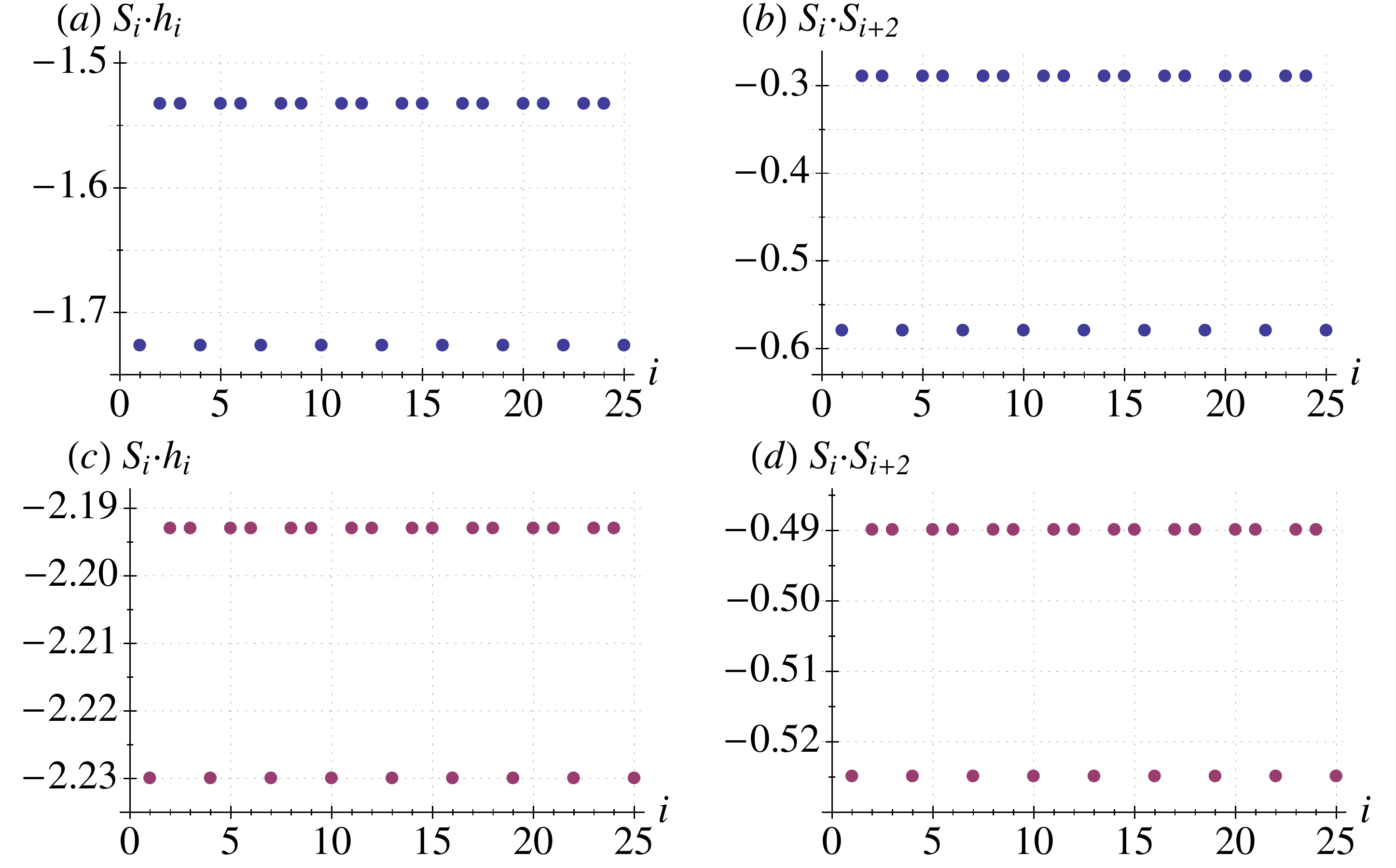}
\caption{(a) and (c): The numerical values of the local energies $\mathbf{S}_i \cdot \mathbf{h}_i$ along individual zigzag chains (with $i$ being the site index along the chain), obtained at $P_K$ and $P_{\Gamma}$ points of Fig.~\ref{fig:LTphasediagram}, respectively. (b, d): Corresponding spin-spin correlations ${\bf S}_i\cdot{\bf S}_{i+2}$ between odd (even) spins of the zigzag chain. 
All results are obtained  from the non-linear LLG simulation  initialized from the commensurate $\mathbf{Q}=\frac{2}{3}\mathbf{a}$ state.}
\label{fig:Sdoth}
\end{figure*}

Here we discuss some aspects of our numerical results for the representative points $P_K$  and   $P_{\Gamma}$ of Fig.~\ref{fig:LTphasediagram}. In our simulations we used a cluster of $240\times 2\times 2$ orthorhombic unit cells  with periodic boundary conditions and $N=15360$ spins.
We first discuss our findings at the $P_K$ point. The LT wavevector that minimizes the classical energy is $\mathbf{Q}_K=( 0.675,0,0)$. The corresponding eigenvalue is $\epsilon_{\text{min}}^{\text{LT}}\simeq-1.58845$. Using the LT result, we construct the initial state for  the non-linear LLG simulations by requiring that  all spins point along the directions determined by the eigenvector ${\mathbf V}_1(\vec{Q}_{K})$ and have unit length. The energy per site in the spin configuration resulting from this simulation is equal to $\epsilon_{\rm LLG}\simeq -1.577718$, which is only  0.6 percent higher than $\epsilon_{\text{min}}^{\text{LT}}$. 

To check if the obtained LLG-state is a local minimum we examine the distribution of the local torques $\mathbf{S}_i \times \mathbf{h}_i$. We find that for the overwhelming majority of the sites the local torques are practically zero, $\mathbf{S}_i \times \mathbf{h}_i\simeq 10^{-7}$, but for some isolated sites the torques are of the order of $10^{-3}$. The presence of these `defected' sites suggests that the LLG-state obtained starting from the optimal LT state is not a true local minimum.

In order to check whether there are any nearby local-minima states, we perform another LLG simulation  initialized from the  commensurate  state described by  the wavevector $\mathbf{Q}=(\frac{2}{3},0,0)$. In this case, the  LLG simulation gives the state with an    energy per site equal to $\epsilon_{\rm LLG} =-1.578237$, which is slightly lower than the energy  obtained in  the  LLG simulation initiated from the LT state. 
In this state, the local torques are below our numerical precision  for all sites, so the state is a local minimum.

Fig.~\ref{fig:Sdoth}~(a) shows the distribution of local energies, $\mathbf{S}_i \cdot \mathbf{h}_i$, along the zigzag chains. We see that  local energies take only two values, approximately equal to -1.533 and -1.727. This means that there are two different local fields acting  on the spins and, thus,  two  different kinds of sites. 
We have checked  that  the same behavior is observed at all 4 zigzag chains of the orthorhombic unit cell  for both even and odd sites. 

Fig.~\ref{fig:Sdoth}~(b) shows the spin-spin correlation function between even (odd) spins along a  single zigzag chain. Here, we also see that  the spin-spin  correlation function  is non-uniform and alternates  between  two  different values with the same periodicity 3. This indicates that the obtained state is not a `homegeneous' counter-rotating spiral described by $+{\mathbf Q}$ on even and $-{\mathbf Q}$ on  odd sites;  for a  homogeneous counter-rotating spiral, the dot product of each pair of spins on  even (odd) sites should  be equal to the same constant given by the pitch of the spiral, and we clearly do not have this case here. 

We also  performed the  LLG-simulations  at the  $P_{\Gamma}$ point  of Fig.~\ref{fig:LTphasediagram}. The results for the local energies and spin-spin correlations are presented in Figs.~\ref{fig:Sdoth} (c) and (d). The  results are very similar to the ones at the $P_K$ point. Starting the  LLG-simulations  form the commensurate  state  we obtain again a state with energy only slightly higher than the lower bound of the energy predicted from the LT analysis and with the periodicity-3 distribution of the local energies and correlation functions. However, in the $\Gamma$-state  the  local fields acting on  two types of  spins, and therefore the local energies and the correlation functions,  are much closer in magnitude than in the $K$-state. Overall, the $\Gamma$-state  is much closer to the  120$^\circ$  order, as discussed in the main text.

\vspace*{-0.3cm}
\section{Modulation of spin components along a single $xy$-chain}\label{app:C}
\begin{figure*}[!t]
\includegraphics[width=0.48 \textwidth]{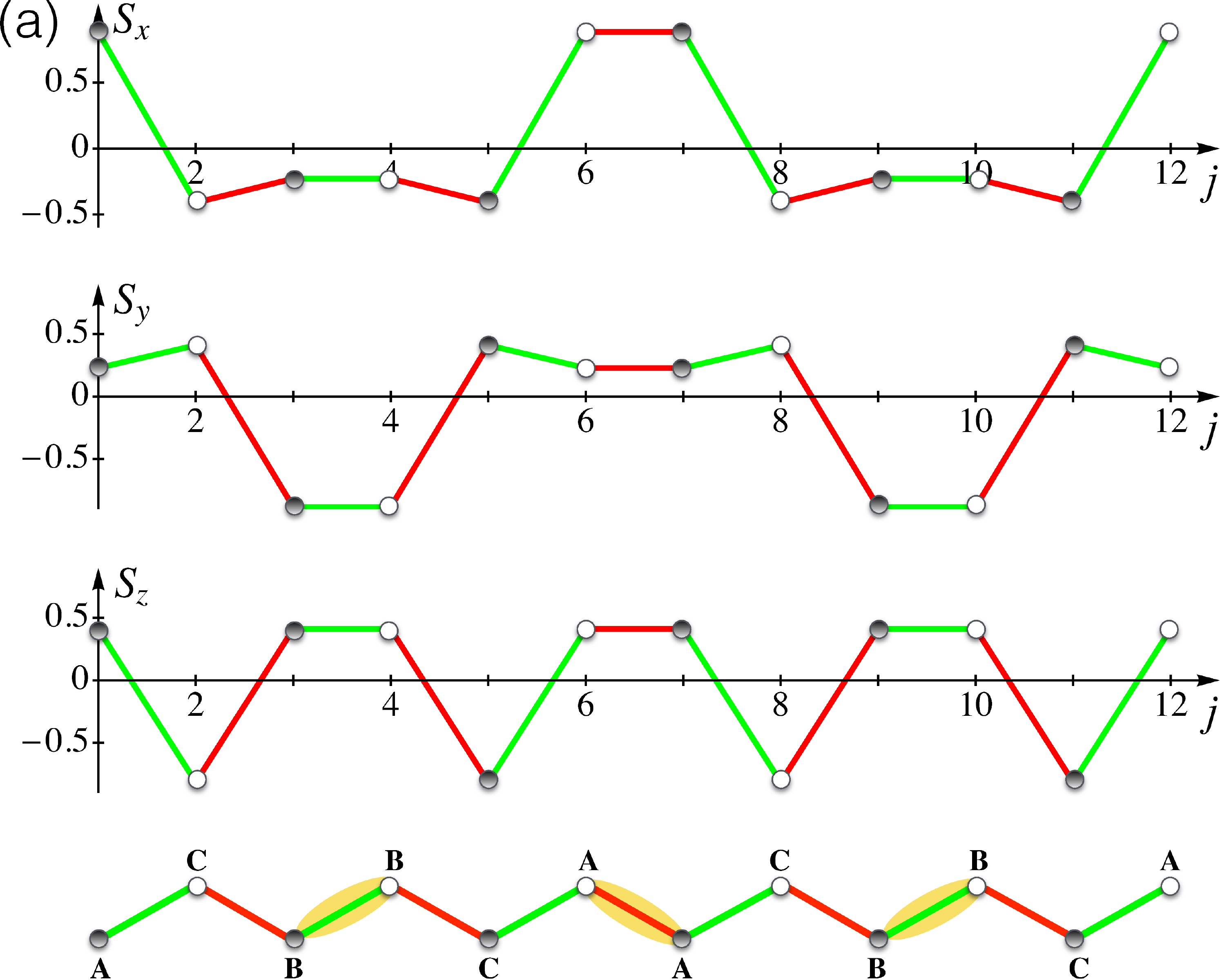}
~~~~
\includegraphics[width=0.48 \textwidth]{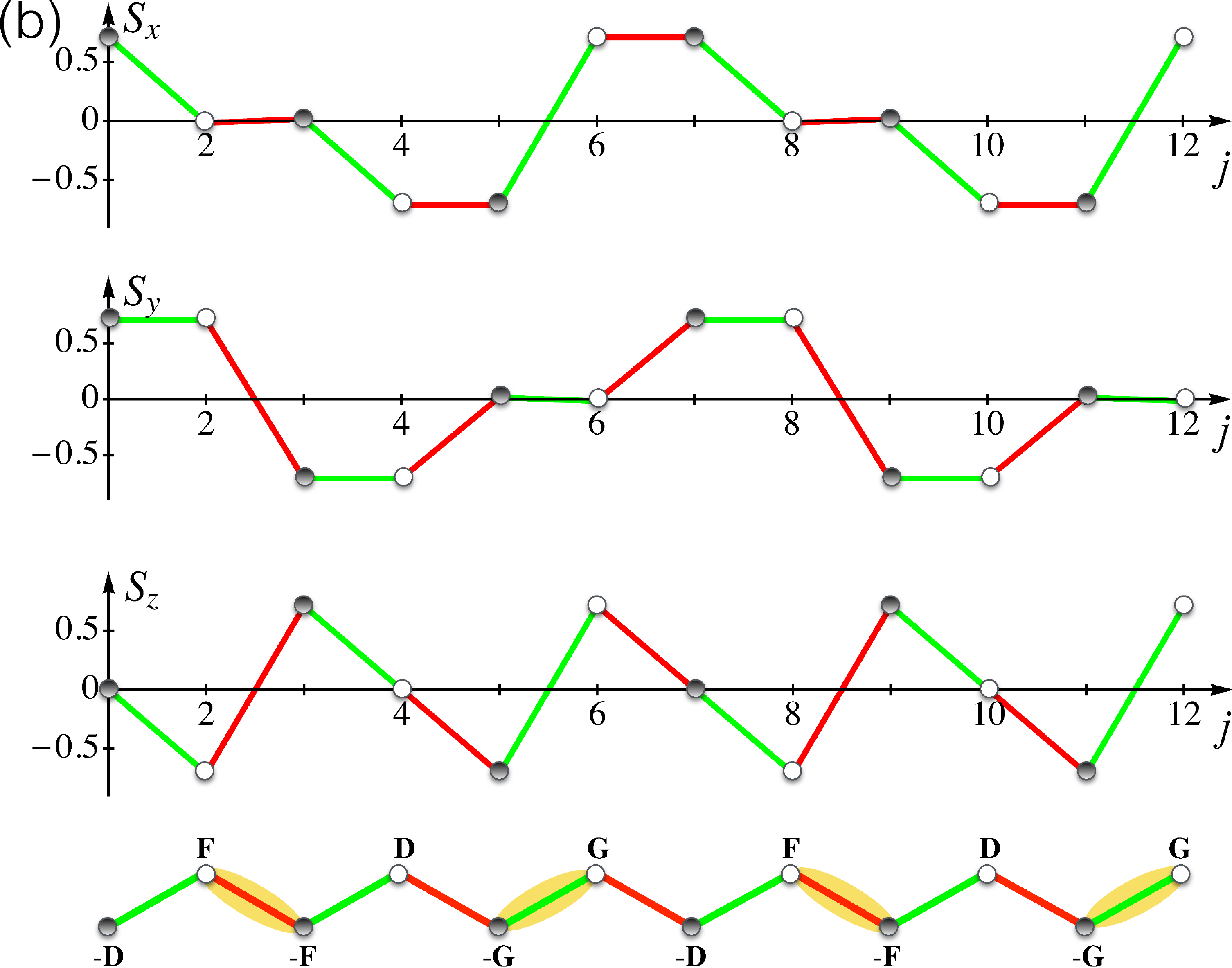}
\caption{The modulation of  $S^x,\, S^y$ and $S^z$ spin components along a single xy-zigzag chain  computed at (a)  $P_K$ and (b) $P_{\Gamma}$ points of Fig.~\ref{fig:LTphasediagram}. White and gray circles represent spins on even and odd sites of the zigzag chain, respectively. Red and green line segments denote $x$- and $y$-type of NN bonds, respectively. The bottom panels show the chain configuration as in Fig.~\ref{fig:KdomGdom}. The FM (AF) dimers in the $K$-state ($\Gamma$-state) are highlighted by yellow ovals.}
\label{fig:CartesianCompsAlongChain}
\end{figure*}

\vspace*{-0.3cm}
Figure~\ref{fig:CartesianCompsAlongChain} shows the modulation of the Cartesian components $S^x$, $S^y$ and $S^z$ along a single xy-zigzag chain at representative points of the shaded region of Fig.~\ref{fig:LTphasediagram}. 

\vspace*{-0.3cm}
\section{Minima of Eqs.~(\ref{eq:EnK}) and (\ref{eq:EnG}) along the line $\phi=3\pi/2$}\label{app:D}
\vspace*{-0.3cm}
Here we discuss the structure of the $K$- and $\Gamma$-states and their degeneracy along the line $\phi=3\pi/2$.
Along this line, $J$ vanishes and the energy of the $K$-state becomes
\be\label{eq:EnK2}
\begin{array}{l}
\frac{E_K}{N} \!=\! \frac{S^2}{6} 
\Big\{ 
-\Gamma (y_1-z_1)^2 -2K (y_1-x_2)^2 \\
+ 3K + 2 \Gamma [ 1 + y_1^2+x_2^2+2x_2z_1+2x_1z_2 ]
\Big\}~,
\end{array}
\ee 
where $x_1^2+y_1^2+z_1^2=1$ and $2x_2^2+z_2^2=1$. Next, we note that the first line of (\ref{eq:EnK2}) is minimized when 
\be\label{eq:CondK}
z_1=x_2=y_1~.
\ee
Imposing these conditions to the second line of (\ref{eq:EnK2}) we get for the total energy per site 
\be\label{eq:EnK3}
\begin{array}{l}
\frac{E_K}{N} \!=\! \frac{S^2}{2} (K+2\Gamma)~,
\end{array}
\ee 
which saturates the lower energy bound from the Luttinger-Tisza method and describe therefore a ground state. 
So the minima of the energy of the $K$-state along the line $\phi=3\pi/2$ obey the conditions (\ref{eq:CondK}).

Let us now do the same for the $\Gamma$-state, whose energy along the line $\phi=3\pi/2$ reads
\be\label{eq:EnG2}
\begin{array}{l}
\frac{E_\Gamma}{N} \!=\! \frac{S^2}{6} 
\Big\{ 
-2\Gamma (z_3-\frac{1}{\sqrt{2}})^2
-2K (x_3-\frac{1}{\sqrt{2}})^2 \\
+4\Gamma (1 +x_3 z_3) + K (3 -4y_3^2)
\Big\}~,
\end{array}
\ee 
where $x_3^2+y_3^2+z_3^2=1$. Here, the first line of (\ref{eq:EnG2}) is minimized when
\be\label{eq:CondG}
x_3=z_3=\frac{1}{\sqrt{2}}, ~~y_3=0~.
\ee
Imposing these conditions to the second line of (\ref{eq:EnG2}) we get for the total energy per site 
\be\label{eq:EnG3}
\begin{array}{l}
\frac{E_\Gamma}{N} \!=\! \frac{S^2}{2} (K+2\Gamma)~,
\end{array}
\ee 
which saturates the lower energy bound from the Luttinger-Tisza method and is therefore a ground state. 
So the minima of the energy of the $\Gamma$-state along the line $\phi=3\pi/2$ obey the conditions (\ref{eq:CondG}).

%\newpage  
%\bibliography{refSam}\end{document}


\end{document}